\documentclass[11pt]{article}
\pdfoutput=1
\usepackage{jheppub}
\usepackage{float, extarrows, tikz-cd}
\usepackage{graphicx}
\usepackage{slashed}
\usepackage{tabularx,ragged2e}
\usepackage{amssymb,fge}
\usepackage{amsmath,amssymb}
\usepackage{slashed}
\usepackage{hyperref}
\usepackage{caption}
\usepackage{dsfont}
\usepackage{verbatim}
\usepackage{subfig}
\usepackage{mathtools,ytableau,amsfonts,tikz}
\usepackage{graphicx}
\usepackage{physics}
\usepackage{amsthm}
\usepackage{tcolorbox}
\usepackage[normalem]{ulem}

\theoremstyle{definition}

\newtheorem{theorem}{Theorem}
\newtheorem{lemma}{Lemma}

\definecolor{colHighlight}{rgb}{0.40,.58,.93}
\definecolor{colHighlight2}{rgb}{0.9,.37,.58}

\newcolumntype{C}{>{\Centering\arraybackslash}X}

\newcommand{\address}[1]{\vbox{\center\em#1}}

\newcommand{\plt}{\text{plateau}}

\ytableausetup{centertableaux,boxsize=.5em}

\usepackage{tikz}
\newcommand*{\boxcolor}{black}
\makeatletter
\renewcommand{\boxed}[1]{\textcolor{\boxcolor}{%
\tikz[baseline={([yshift=-1ex]current bounding box.center)}] \node [rectangle, minimum width=1ex,rounded corners,draw] {\normalcolor\m@th$\;\,\displaystyle#1\;\,$};}}
 \makeatother

\begin{document}

\thispagestyle{empty}

\begin{center}

{\LARGE \bf {Symmetries and spectral statistics in chaotic conformal field theories II:\flushright 
Maass cusp forms and arithmetic chaos}}
\end{center}

\bigskip \noindent

\bigskip

\begin{center}

Felix M.~Haehl,$^a$ Wyatt Reeves,$^b$ and Moshe Rozali$^b$

\address{a) School of Mathematical Sciences and STAG Research Centre,\\ University of Southampton, SO17 1BJ, U.K.}

\address{b) Department of Physics and Astronomy,\\ University of British Columbia, Vancouver, V6T 1Z1, Canada}
\vspace{0.5in}
    
{\tt f.m.haehl@soton.ac.uk, wreeves@phas.ubc.ca, rozali@phas.ubc.ca}

\bigskip

\vspace{1cm}

\end{center}

We continue the study of random matrix universality in two-dimensional conformal field theories. This is facilitated by expanding the spectral form factor in a basis of modular invariant eigenfunctions of the Laplacian on the fundamental domain. The focus of this paper is on the discrete part of the spectrum, which consists of the Maass cusp forms. Both their eigenvalues and Fourier coefficients are sporadic discrete numbers with interesting statistical properties and relations to analytic number theory; this is referred to as `arithmetic chaos'. We show that the near-extremal spectral form factor at late times is only sensitive to a statistical average over these erratic features. Nevertheless, complete information about their statistical distributions is encoded in the spectral form factor if all its spin sectors exhibit universal random matrix eigenvalue repulsion (a `linear ramp'). We `bootstrap' the spectral correlations between the cusp form basis functions that correspond to a universal linear ramp and show that they are unique up to theory-dependent subleading corrections. The statistical treatment of cusp forms provides a natural avenue to fix the subleading corrections in a minimal way, which we observe leads to the same correlations as those described by the [torus]$\times$[interval] wormhole amplitude in AdS${}_3$ gravity.

\newpage

\setcounter{tocdepth}{3}
{}
\vfill
\tableofcontents

\vspace{10pt}
\section{Introduction}

The importance of chaos for conformal field theories and the AdS/CFT correspondence has become increasingly apparent over the years. Quantum chaos is often formulated as a statement about the statistics of the spectrum of energy eigenvalues. Energy levels that are sufficiently close together are expected to show the same statistics as those of an appropriate random matrix ensemble, namely eigenvalue repulsion; the probability of energy levels being nearby decreases as they get closer. This leads to a linear ramp in the spectral form factor, which is the averaged product of partition functions,  at late times. Holographic conformal field theories possess a dense spectrum for large enough energies for any spin, and thus could possibly display random matrix universality.

In theories with symmetries, only the parts of the spectrum that are independent of the symmetries can display random matrix universality. In particular, the spectrum of conformal field theories in two dimensions is subject to translation invariance, Virasoro symmetry, and modular invariance. We can remove the consequences of translation invariance and Virasoro symmetry by focusing on conformal primary operators in fixed spin sectors. This leaves the question of modular invariance, which relates primaries of different energy and spin.

In \cite{Haehl:2023tkr}, we began investigating the relationship between quantum chaos in two-dimensional CFTs and modular invariance. Motivated by the pure gravity wormhole amplitude found by Cotler and Jensen \cite{Cotler:2020hgz,Cotler:2020ugk} (see also \cite{Eberhardt:2022wlc}), we argued that random matrix statistics for the ``near-extremal'' part of the dense spectrum and the corresponding late time linear ramp is an independent feature of each spin sector separately. This is a non-trivial statement because the exact spectrum is fully determined by only the spectrum of spin zero primaries and those of a single non-zero spin. The focus of this analysis was on CFTs where the ramp is encoded solely in the continuous part of the basis of modular invariant functions. There exists a discrete part as well, the Maass cusp forms, that can also encode the ramp.

The cusp forms are interesting objects in their own right: 
\begin{itemize}
    \item Cusp forms arise as bound states for a particle moving in the fundamental domain of $SL(2,\mathbb{Z})$. This is a classically chaotic system, however due to its highly symmetric structure it does not obey random matrix statistics. 
    \item Instead, their spectrum of eigenvalues $R_n^\pm$ is bounded from below and is Poisson distributed, i.e corresponds to independent draws from a known distribution.\footnote{ We label the different cusp forms by an integer $n$ and a sign `$\pm$', which refers to cusp forms of even and odd parity, respectively.}
    \item Their Fourier coefficients $a_m^{(n,\pm)}$ for prime spin $m$ are bounded by $\pm 2$ and are also Poisson distributed independently drawn from known distributions. As a consequence of Hecke relations the Fourier coefficients for non-prime (composite) spins are polynomials of those with prime spins. This implies that the distributions of non-prime spin Fourier coefficients are also determined by the distributions for prime spins.
\end{itemize}
We sometimes refer to the collection of eigenvalues and Fourier coefficients as {\it cusp form data}. Together their statistical properties are sometimes referred to as \textit{arithmetic chaos} \cite{bolteArithmeticalChaosViolation1992,sarnakthesis,Hejhal1992OnTT,BOGOMOLNY1997219}.

A connection between arithmetic chaos and quantum chaos in CFTs is an intriguing possibility (first hinted at in \cite{Benjamin:2021ygh}), particularly since quantum chaos in the wormhole amplitude is encoded solely in the cusp forms for all non-zero spins \cite{DiUbaldo:2023qli}. On the one hand, the fact that objects linked to chaos appear in the spectral decomposition of CFTs suggests that the two (very different) types of chaos may be linked in some way. On the other hand, arithmetic chaos is a property of the general modular invariant basis functions, not of the actual CFT spectrum, so it is also present in integrable CFTs. We intend to clarify the relation in this work.

\begin{figure}
    \centering
    \includegraphics[scale=0.7]{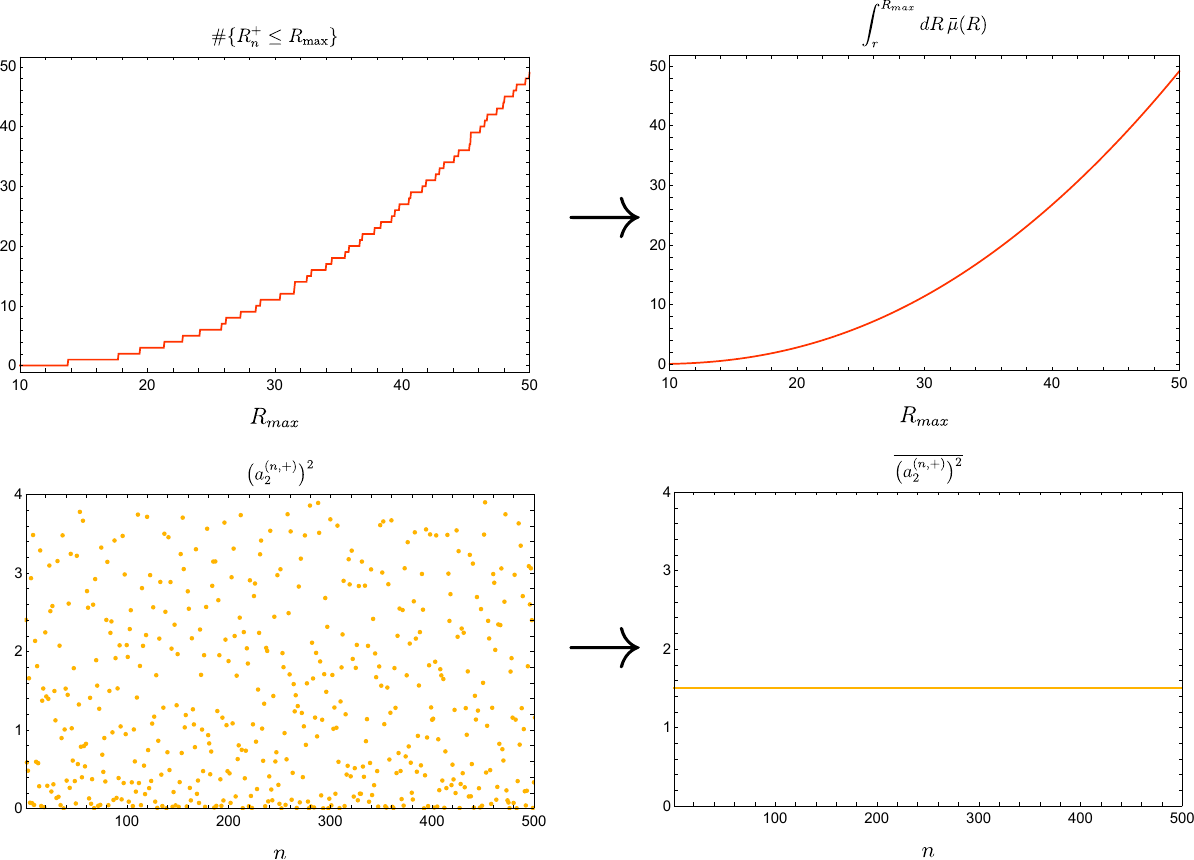}
    \caption{A depiction of the statistical approximation to cusp form data. The ``spectral staircase'' and the erratically distributed Fourier coefficients are replaced with their average values. We will justify this `statistical' coarse-graining in the time regime that is relevant for random matrix universality. It should be distinguished from `microcanonical' coarse-graining, which is always required to discuss correlations in the CFT spectrum.}
    \label{fig:stat-approx}
\end{figure}

\subsubsection*{Summary of results}

In this paper, we extend the techniques developed for the continuous $SL(2,\mathbb{Z})$ spectrum in \cite{Haehl:2023tkr} to the cusp forms and identify the relationship between arithmetic and quantum chaos. We show that taking the near-extremal, late time limit automatically implements a statistical averaging over the cusp form data, in particular over their sporadic eigenvalues and erratic Fourier coefficients. On the one hand, this means that quantum chaos in 2d CFT depends only on statistical features of arithmetic chaos, not the detailed structure of the cusp form data. On the other hand, it is remarkable that full information about (i.e., all statistical moments of) the distributions of the cusp form data is encoded in the spectral form factor, assuming it exhibits a linear ramp in all spin sectors. We find the universal form these statistically averaged correlations in the cusp form sector must take to produce a ramp. As with the continuous sector, modular invariance does not spoil the independence of random matrix universality in each separate spin sector, since the statistical averaging proceeds differently in each spin sector; a linear ramp must be imposed as a separate assumption for each spin sector in order to fully and consistently determine the correlations in cusp form expansion coefficients.

By demanding that there be a linear ramp in every spin sector, we are able to ``bootstrap'' the exact cusp form correlations whose statistical averaging produces random matrix statistics independently in every spin sector. These correlations depend on all moments of the distributions of the Fourier coefficients for prime spin, and are related to well studied number-theoretic objects. In fact, these correlations are essentially universal and unique under some mild assumptions. The gravitational wormhole amplitude exhibits the same universality, while at the same time having the minimal subleading corrections (in the late time limit) to make it consistent with modular invariance \cite{DiUbaldo:2023qli}; under some related minimality assumptions our construction reproduces it exactly.

Our presentation is somewhat pedagogical. For the result on cusp form correlations encoding a linear ramp in all spin sectors, see \eqref{eq:zzRampFinal} and \eqref{eq:KernelDef}. We derive this result by investigating statistical features of the sum over `arithmetically chaotic' cusp forms. We discuss the connection with Euclidean wormholes in section \ref{sec:gravity-ramp}.

\subsubsection*{Outline}

The paper is organized as follows. In section \ref{sec:rampDecomposition} we review the setup of \cite{Haehl:2023tkr}, introducing the fluctuating part of the partition function and decomposing it in the complete basis of modular invariant functions. In section \ref{cusp} we analyze how random matrix statistics appears in the cusp forms, and demonstrate its reliance on only arithmetic chaos. We derive an expression whose statistical average produces a ramp in each spin sector and show that it is unique under mild assumptions. In section \ref{sec:gravity-ramp} we then show that this expression exactly matches a calculation in AdS$_3$ gravity. In the discussion, we put forth some preliminary results on the spectral decomposition of the self-correlations in the spectrum , and how it differs from eigenvalue repulsion. 

Conventions are collected in appendix \ref{app:notation}. We review statistical features of cusp forms in appendix \ref{app:maass} and discuss some important mathematical properties in appendix \ref{app:norms}. Appendix \ref{sec:ramp-independence} concerns the imprint of linear ramps in a given spin sector onto other spin sectors.

\vspace{5pt}
\section{Spectral decomposition of the ramp}
\label{sec:rampDecomposition}

We start by reviewing the $SL(2,\mathbb{Z})$ spectral decomposition of the linear ramp and recollecting the work of \cite{Haehl:2023tkr}. In the next section we extend this analysis to the cusp forms, and in particularly use statistical properties thereof (dubbed ``arithmetic chaos") to simplify the analysis. 

\vspace{5pt}
\subsection{$SL(2,\mathbb{Z})$ spectral theory}

Beginning from the full CFT partition function on a torus with modular parameter $\tau=x+iy$, $Z(x,y)$, in \cite{Benjamin:2021ygh,Haehl:2023tkr} the authors introduce a fluctuating partition function $\widetilde{Z}_\text{P}(x,y)$ by a series of steps intended to account for the symmetries of the problem:

\begin{itemize}
    \item First, the partition function is divided by that of a single non-compact boson, $Z_0 = 1/(y^{1/2}|\eta(x+iy)|^2)$, to remove all Virasoro descendants in a modular invariant fashion.
    \item Then one realizes that the `censored' part of the spectrum (i.e., states with $h$ or $\bar h \leq \frac{c-1}{24}$, equivalently $E\leq 2\pi\left(m-\frac{1}{12}\right)\equiv E_m$) is not typically chaotic. Therefore, one subtracts off this part of the spectrum. In addition, one also removes the part of the dense spectrum ($h,\bar h > \frac{c-1}{24}$, equivalently $E>E_m$) that is determined from the censored spectrum by symmetries (such as modular S-transformations); together, these two parts are called the modular completion of the censored spectrum, $\widehat{Z}_C(x,y)$.
\end{itemize}
The resulting ``fluctuating" partition function $\widetilde{Z}_\text{P}(x,y)$ is the object that can display quantum chaos. Finally, we write this object in terms of a decomposition into sectors with definite spin:
\begin{equation}
\label{eq:ZPtildeDef}
\begin{split}
 \widetilde{Z}_\text{P}(x,y) &=  \sum_{m\in \mathbb{Z}} e^{2\pi i m x} \,\widetilde{Z}^m_{\text{P}}(y)\,.
\end{split}
\end{equation}
This is not quite an ordinary partition function: the density of states it describes corresponds to {\it fluctuations} around the average density of states.

To understand why this is, we have to explain the process of modular completion. There are different ways of performing the modular completion of the censored spectrum $\rho_C (E)$, which are all modular invariant and do not introduce new censored states. We focus on the kind introduced in \cite{Keller:2014xba}, where the modular completion of each censored state results in a continuous density of states in the dense part of the spectrum\footnote{This is in the spirit originally suggested in \cite{Pollack:2020gfa} that the effective disorder average in gravity is related to the conventional one underlying quantum statistical mechanics.}. For example, the modular completion of the vacuum gives a continuous density of states for the dense spectrum that includes the well-known Cardy formula for the leading average density of states at high energies. Other censored states give additional contributions that are subleading at high energy and large central charge.

In effect, the modular completion defines our coarse-graining procedure: $\widehat{\rho}_C(E) = \rho_C(E)+\langle\rho_D(E)\rangle$, where `$D$' refers to the dense part of the full spectrum.\footnote{A similar perspective is established in \cite{DiUbaldo:2023qli}, motivated by the diagonal approximation of semi-classical periodic orbits.} 
With this prescription, subtracting the modular completion of the censored spectrum amounts to eliminating the latter while also removing the {\it average} density of states from the heavy spectrum. Explicitly, 
\begin{equation}
\begin{split}
\widetilde{\rho}_{\text{P}}(E) &\equiv \rho_{\text{P}}(E)-\widehat{\rho}_C(E)=\rho_D(E)-\langle\rho_D(E)\rangle 
\end{split} \label{eq:rhotildeDef}
\end{equation}
is the fluctuating density of states corresponding to the fluctuating partition function, 
\begin{equation}
    \widetilde{Z}^m_{\text{P}}(y) = \frac{\sqrt{y}}{e^{\frac{\pi}{6}y}}\int_{E_m}^\infty dE \, \widetilde{\rho}_{\text{P}}^m(E) e^{-yE} \, , 
\end{equation}
where we include the normalization factors from the non-compact boson. Thus, \eqref{eq:ZPtildeDef} is the modular-invariant object that can display quantum chaos. It is `fluctuating' in the sense that $\widetilde{\rho}_\text{P}(E)$ has a vanishing microcanonical average (in particular it has both positive and negative contributions).

\paragraph{The linear ramp:} We are interested in the universal correlations that this fluctuating spectrum exhibits due to quantum chaos. In particular, a quantum chaotic CFT will have a universal asymptotic contribution to the variance of $\widetilde{Z}_\text{P}$, describing eigenvalue repulsion. This is often called the `linear ramp' and corresponds to analytically continuing $y_1\rightarrow \beta+iT$ and $y_2 \rightarrow \beta - iT$ and taking the large $T$ limit of the spectral form factor. 
A linear ramp is captured in $y_i$ variables by the following limiting behavior:
\begin{equation}
\label{eq:ZPZPuniv}
\begin{split}
 \big{\langle} \widetilde{Z}_\text{P}^{m_1}(y_1) \, \widetilde{Z}_\text{P}^{m_2}(y_2) \big{\rangle}_{\text{ramp}} &=\delta_{m_1m_2} \left[\frac{1}{\pi} \frac{y_1 y_2}{y_1+y_2}\, e^{-2\pi |m_1|(y_1+y_2)} \right] + \ldots \quad\;\; \Big( y_i \gg 1, \; \frac{y_1}{y_2} = \text{fixed}\Big)
 \end{split}
\end{equation}
where  `$\ldots$' denotes subleading terms in the large $y_i$ limit. This should be read as a statement about each spin sector separately. We choose here (and henceforth) a normalization for the ramp corresponding to the GOE universality class, which matches the discussion in \cite{DiUbaldo:2023qli,Yan:2023rjh}. The normalization would be different for other universality classes, in particular it would differ by a factor $\frac{1}{2}$ for GUE as in \cite{Cotler:2020ugk} and our previous work \cite{Haehl:2023tkr}.\footnote{It was shown in \cite{Yan:2023rjh} that every CFT contains an anti-linear, anti-unitary $RT$ symmetry, implying that the relevant universality class for two-dimensional CFTs is GOE.}.

\paragraph{Spectral decomposition:} It is useful to expand the fluctuating partition functions in a complete basis of normalizable modular invariant functions on the fundamental domain ${\cal F}$ \cite{Benjamin:2021ygh} (see also \cite{Collier:2022emf,DiUbaldo:2023qli,DiUbaldo:2023hkc,DHoker:2022dxx}). Such eigenfunctions consist of a continuous spectrum of  Eisenstein series $E_{s}(y)$ with $s \in \frac{1}{2} + i \mathbb{R}$, and a discrete spectrum of Maass cusp forms $\nu_{n,\pm}(y)$:
\begin{equation}
    \begin{split}
    \Delta_{_{\cal F}} E_{\frac{1}{2} + i\alpha}(\tau) = \left( \frac{1}{4} + \alpha^2 \right) E_{\frac{1}{2} + i\alpha}(\tau) \,,\qquad
    \Delta_{_{\cal F}} \nu_{n,\pm}(\tau) = \left( \frac{1}{4} + \big(R_n^\pm\big)^2 \right) \nu_{n,\pm}(\tau)\,.
    \end{split}
\end{equation}
in addition to a constant function, $\Delta_{_{\cal F}} \nu_{0}(\tau)=0$.
Note that there are both even ($+$) and odd ($-$) cusp forms, so there are two sets of eigenvalues $\{R_n^\pm\}$. These are randomly distributed, which we will quantify later.
After expanding the fluctuating part of the partition function in this modular invariant basis, the {\it expansion coefficients} are then unconstrained by modular invariance and their statistical properties are a good diagnostic of chaos. To write this, we refine the decomposition \eqref{eq:ZPtildeDef}:
\begin{equation}
\label{eq:ZPtildeDefDiscCon}
\begin{split}
 \widetilde{Z}_\text{P}(x,y) &= \widetilde{Z}^{0}_\text{P}(y) + 2\sum_{m>0} \left\{ \cos(2\pi m x)\left[\widetilde{Z}^m_{\text{P,disc.,}+}(y) + \widetilde{Z}^m_{\text{P,cont.}}(y) \right] + \sin(2\pi m x) \widetilde{Z}^m_{\text{P,disc.}-}(y) \right\} 
\end{split}
\end{equation}
where the spectrum consists of the following pieces:\footnote{ In writing the first line we imposed $\Lambda(i\alpha) z_{\frac{1}{2} + i\alpha} = \Lambda(-i\alpha) z_{\frac{1}{2} - i \alpha}$, which is a symmetry of the Eisenstein series.}
\begin{equation}
\label{eq:ZmPtildeDecom}
\begin{split}
\text{spin 0, continuous:} \quad\qquad\qquad\;\; \widetilde{Z}^{0}_\text{P}(y) &= \text{vol}({\cal F})^{-\frac{1}{2}} \; {\color{colHighlight}z_0} 
+ 2\sqrt{y} \int_{\mathbb{R}} \frac{d\alpha}{4\pi}  \; {\color{colHighlight}z_{\frac{1}{2}+i\alpha}}\, y^{i\alpha} \,,
\\
\text{spin}>0\text{, discrete:} \quad\qquad \widetilde{Z}^{m> 0}_{\text{P,disc.,}\pm}(y) &= \sum_{n\geq 0} \, {\color{colHighlight}z_{n,\pm}} \,   \nu_{n,\pm}^{m}(y)\,,\\
\text{spin}>0\text{, continuous:} \quad\qquad\;\; \widetilde{Z}^{m> 0}_{\text{P,cont.}}(y) &=  \int_{\mathbb{R}} \frac{d\alpha}{4\pi}  \,  {\color{colHighlight}z_{\frac{1}{2} + i\alpha} }\,   E_{\frac{1}{2} + i \alpha}^m(y) \,,
\end{split}
\end{equation}
with $\text{vol}({\cal F}) = \frac{\pi}{3}$ and the norm of cusp forms refers to the Petersson norm. The modular invariant expansion coefficients are $\{{\color{colHighlight} z_0},{\color{colHighlight} \, z_{n,\pm}} , \, {\color{colHighlight}z_{\frac{1}{2} + i\alpha}} \}$, and we are interested in their variance and how it encodes the linear ramp. Using the explicit basis functions for $m>0$,\footnote{Our conventions for Fourier coefficients are consistent with \cite{Benjamin:2021ygh} and \cite{Haehl:2023tkr}, but differ from \cite{DiUbaldo:2023qli}. For comparison, we give the translation:
\begin{equation}
    \big(\mathtt{a}_{j\equiv m}^{(s\equiv \frac{1}{2}+i\alpha)}\big)_{\text{there}} = \big(a_{m}^{(\alpha)} \big)_\text{here} \,,\qquad
    \big(\mathtt{b}_{j\equiv m}^{(n)}\big)_{\text{there}} = \frac{|\!|\nu_{n,\pm}|\!|}{2 } \, \big(a_{m}^{(n,\pm)} \big)_\text{here} \,.
\end{equation}}
\begin{equation}
\label{eq:aAlphaDef}
    \begin{split}
       \nu_{n,\pm}^{m}(y) &= a_m^{(n,\pm)} \, \sqrt{y} K_{iR_n^\pm} (2\pi m y) \,,\\
        E_{\frac{1}{2} + i \alpha}^{m}(y) &=   \frac{2\,a_m^{(\alpha)}}{\Lambda(-i\alpha)} \, \sqrt{y} K_{i\alpha}(2\pi m y)
        \,,  \qquad a_m^{(\alpha)} = \frac{2\sigma_{2i\alpha}(m)}{ m^{i\alpha}}  \,,
    \end{split}
\end{equation}
where $\Lambda(s)  \equiv \Lambda(\frac{1}{2}-s)\equiv \pi^{-s} \Gamma(s) \zeta(2s)$.

We also trivially obtain a decomposition into bases of Bessel functions (both with continuous order, $K_{i\alpha}$, and sporadic discrete order, $K_{iR_n^\pm}$) by defining the {\it spin-dependent spectral overlap coefficients}:
\begin{equation}
    \begin{split}
{\color{colHighlight2}z^m_{n,\pm}} &\equiv a_m^{(n,\pm)}\, {\color{colHighlight}z_{n,\pm}} \,,\\
{\color{colHighlight2}z^m(\alpha)} &\equiv  \frac{2\,a_m^{(\alpha)}}{ \Lambda(-i\alpha)}\,{\color{colHighlight}z_{\frac{1}{2}+i\alpha}} \,. 
    \end{split}
    \label{eq:zzzzDef}
\end{equation}

The fact that $\{{\color{colHighlight} z_0}, \, {\color{colHighlight}z_{n,\pm}} , \, {\color{colHighlight}z_{\frac{1}{2} + i\alpha}} \}$ are independent of spin leads to {\it spectral determinacy} \cite{Benjamin:2021ygh}: full knowledge of $\widetilde{Z}_{\text{P, cont./disc.,},\pm}^m(y)$ for only $m=0$ and a single non-zero spin determines the partition function for every other spin.\footnote{For partition functions that arise as Poincar\'{e} series, further constraints follow \cite{DiUbaldo:2023qli}. However, we do not assume this here.} That the coefficients must be independent of spin will prove to be important.

\vspace{5pt}
\subsection{Linear ramp from correlations in spectral overlap coefficients} 

We wish to discuss how the ramp \eqref{eq:ZPZPuniv} translates into specific universal correlations between the coefficients of the spectral decomposition. This discussion should a priori be had for each spin sector individually.

\paragraph{Ramp for spin $0$:} For spins $m_1=m_2=0$, the ramp is encoded in $\big{\langle}  \widetilde{Z}^{0}_\text{P}(y_1)\widetilde{Z}^{0}_{\text{P}}(y_2)  \big{\rangle}$, and in particular it is fully determined by correlations in the overlap coefficients with Eisenstein series 
\begin{equation}
\label{eq:spin0Ramp}
\begin{split}
\big\langle \widetilde{Z}_\text{P}^0(y_1) \widetilde{Z}_\text{P}^0(y_2) \big\rangle &=  4\sqrt{y_1y_2} \int_\mathbb{R} \frac{d\alpha_1 d\alpha_2}{(4\pi)^2} \, \big{\langle} {\color{colHighlight}z_{\frac{1}{2}+i\alpha_1} \, z_{\frac{1}{2}+i\alpha_2}} \big\rangle_{\text{spin }0\text{ ramp}} \; y_1^{i\alpha_1} y_2^{i\alpha_2}  + \ldots
\\
\big{\langle}{\color{colHighlight} z_{\frac{1}{2}+i\alpha_1} \, z_{\frac{1}{2}+i\alpha_2}} \big\rangle_\text{spin 0 ramp} &\sim \frac{1}{2\cosh(\pi \alpha_1)} \times 4\pi\delta(\alpha_1+\alpha_2)  \qquad (|\alpha_i| \rightarrow \infty)\,.    
\end{split}
\end{equation} 
where terms subleading in the large $y_i$ limit (denoted as `$\ldots$') are required to obtain a modular invariant expression; these correspond to deviations from the asymptotic form given in the second line of \eqref{eq:spin0Ramp}.\footnote{We thank E.\ Perlmutter for pointing out the importance of the large $|\alpha_i|$ limit, see \cite{DiUbaldo:2023qli} and \cite{Haehl:2023tkr} for more details.} While the correlations in ${\color{colHighlight} z_{\frac{1}{2}+i\alpha}}$ will in general contain more information, the above should be understood as the {\it universal} contribution that is due to a ramp in the spin 0 sector.\footnote{Note that the correlation \eqref{eq:spin0Ramp} is manifestly {\it diagonal} in the spectral parameters $\alpha_i$. Such diagonality was proposed in \cite{DiUbaldo:2023qli} as a natural constraint analogous to Berry's diagonal approximation in the theory of periodic orbits. It is also a distinctive feature exhibited by the pure gravity result for the $\mathbb{T}^2 \times I$ amplitude \cite{Cotler:2020ugk}.}

\paragraph{Ramp for non-zero spins:} 

For non-zero spins, decomposing the ramp using \eqref{eq:ZPtildeDefDiscCon} and noting that the leading term shown in \eqref{eq:ZPZPuniv} is even in spin, it is clear that the ramp could apriori be encoded in cross-correlations between any of the terms in \eqref{eq:ZPtildeDefDiscCon} (subject to producing the correct parity).
We will now discuss the form of the correlations that can encode a ramp in a single spin sector. However we will later stress the limitations of this approach when asking for linear ramps in more than one spin sector simultaneously, as these are not independent of each other and additional consistency conditions must be imposed.

Using the spectral decomposition of the partition function, we can write its even and odd parts for spin $m\geq 1$ in a basis of Bessel functions, where each individual term is still modular invariant by construction:
\begin{equation}
\begin{split}
    \widetilde{Z}^{m}_{\text{P},+}(y) &\equiv   \sum_{{\small n> 0}} {\color{colHighlight2}z^m_{n,+}} \, \sqrt{y} K_{iR_n^+}(2\pi my) + \int_{\mathbb{R}} \frac{d\alpha}{4\pi} \, {\color{colHighlight2}z^m(\alpha)} \sqrt{y} K_{i\alpha}(2\pi m y)\,,\\
    \widetilde{Z}^{m}_{\text{P},-}(y) &\equiv \sum_{{\small n> 0}} {\color{colHighlight2}z^m_{n,-}} \, \sqrt{y} K_{iR_n^-}(2\pi my) \,.
\end{split}\label{eq:z-m-alpha-def}
\end{equation}

Let us now briefly review how the ramp could be encoded the Eisenstein series correlations (see \cite{Haehl:2023tkr}).
For the continuous part of the spectral decomposition, we can use the orthogonality of Bessel functions
to invert the $\alpha$-integral in \eqref{eq:z-m-alpha-def}:
\begin{equation}
    {\color{colHighlight2}z^m(\alpha)} =  \frac{2}{\pi}\,\alpha \sinh(\pi \alpha) \int_0^{\infty} \frac{dy}{y^{3/2}}\,  K_{i \alpha}(2\pi m y) \widetilde{Z}^{m}_\text{P,cont.}(y)\,.\label{eq:z-m-alpha}
\end{equation}
This allows us to translate the universal expression for RMT eigenvalue repulsion, \eqref{eq:ZPZPuniv}, into an expression for the correlations of ${\color{colHighlight2}z^m(\alpha)}$ coefficients by performing two correlated $y$-integrals of the form \eqref{eq:z-m-alpha}:
\begin{equation}
\begin{split}
 &   \langle {\color{colHighlight2}z^{m_1}(\alpha_1)z^{m_2}(\alpha_2)}\rangle_\text{ramp} = 
 2\alpha_1 \tanh(\pi \alpha_1) \,\delta_{m_1m_2}\, \left[ \delta(\alpha_1-\alpha_2) +\delta(\alpha_1+\alpha_2) \right]   \,.
 \end{split}
\label{eq:z-m-alpha-result}
\end{equation}
This shows how a ramp for specific spin $m$ can be encoded in the coefficients of the Eisenstein series, and the required correlations are again diagonal in the spectral parameter.\footnote{ Since $z^m(\alpha)$ is even in $\alpha$ by definition, we refer to the presence of the symmetrized sum of delta-functions in \eqref{eq:z-m-alpha-result} as diagonal.}
It is straightforward to verify this result explicitly by transforming \eqref{eq:z-m-alpha-result} back to $y$-variables, which reproduces \eqref{eq:ZPZPuniv}. 
We can equivalently write \eqref{eq:z-m-alpha-result} as:
\begin{equation}
\label{eq:z-m-result-early}
\begin{split}
    \big\langle {\color{colHighlight}z_{\frac{1}{2}+i\alpha_1} z_{\frac{1}{2}+i\alpha_2} }\big\rangle_{\text{spin }m\text{ ramp}}
    &=
     \frac{\Lambda(-i\alpha_1) \Lambda(-i\alpha_2)}{2\big(a_m^{(\alpha_1)}\big)^2}\,
     \alpha_1 \tanh(\pi \alpha_1) \, \left[ \delta(\alpha_1-\alpha_2) +\delta(\alpha_1+\alpha_2) \right] 
\end{split}
\end{equation}
The fact that this relation depends explicitly on spin might be understood as follows: the existence of a ramp in each spin sector gives partial information about the correlation of the modular invariant coefficients in different regimes, roughly organized by scale of oscillation as function of $\alpha$. The different regimes are spin-dependent, so \eqref{eq:z-m-result-early} is to be understood as being valid only in the regime informed by the asymptotic form of the spin-$m$ partition function.\footnote{To organize the information conveniently and discuss the relationship between all the statements implied by the ramp in different spin sectors, ref.\ \cite{Haehl:2023tkr} introduced a conjugate variable $\xi$; the existence of a ramp in each spin sector then is localized in that variable, in a different location for different spin sectors. The transformation to the $\xi$ variables is roughly a Fourier transform, so localization in that variable translates to a definite scale of oscillatory behavior in the $\alpha$ variables.}

Note that \eqref{eq:z-m-result-early} is not consistent with the asymptotic condition \eqref{eq:spin0Ramp}. This means that once we impose the spin 0 ramp on the correlations in the Eisenstein sector, the spinning ramps must be encoded in the cusp form correlations.


\vspace{5pt}
\section{Ramp from cusp forms -- the statistical approximation}
\label{cusp}

In this section we extend the above analysis to the cusp forms. The main tool we use is a continuous approximation to the sum over the cusp forms, which utilizes statistical information, known as ``arithmetic chaos". We introduce the approximation in the context of a simple ansatz that yields ramps in a single spin sector from cusp form sums, and show how this is reproduced to good accuracy by the statistical approximation. This sets the stage for an improved ansatz, discussed in the next section, where we ``assemble" ramps for {\it all} spin sectors simultaneously.

We would like to explore the type of correlations in the overlap coefficients, $\langle {\color{colHighlight}z_{n_1,\pm} z_{n_2,\pm}}\rangle_\text{ramp}$, which yield a linear ramp through a sum over cusp forms:
\begin{equation}
\begin{split}
   \Big{\langle}  \widetilde{Z}^{m_1}_{\text{P,disc.,}\pm}(y_1) \widetilde{Z}^{m_2}_{\text{P,disc.,}\pm}(y_2) \Big{\rangle}_\text{ramp} 
     = \sum_{n_1,n_2>0}
    \langle {\color{colHighlight}z_{n_1,\pm} z_{n_2,\pm}}\rangle_\text{ramp}\,
    \nu_{n_1,\pm}^{m_1}(y_1) \nu_{n_2,\pm}^{m_2}(y_2)\,, 
\end{split}
\label{eq:discSumRes0}
\end{equation}
where the l.h.s.\ takes the universal form \eqref{eq:ZPZPuniv}, and we recall $\nu_{n,\pm}^m(y) 
\equiv a_{m}^{(n,\pm)}\sqrt{y} K_{iR_{n}^\pm}(2\pi m y)$. Note that the sum is over erratic eigenvalues $R_n^\pm$ and erratic Fourier coefficients $a_m^{(n,\pm)}$. The first few eigenvalues are:
\begin{equation}
\label{eq:fourierEx}
\begin{split}
 R_n^+ &= 13.7798.. , \quad 17.7386..,\quad 19.4235.., \quad 21.3158..,\quad 22.7859.., \quad 24.1124.., \quad 25.8262.., \; \ldots \\
 R_n^- &=\;\; 9.5337.., \quad 12.1730.., \quad 14.3585.., \quad 16.1381.., \quad 16.6443.., \quad 18.1809..,\quad 19.4847.., \; \ldots
\end{split}
\end{equation}
These are sporadically distributed and become increasingly dense. The cusp form Fourier coefficients take  a similarly erratic form (for fixed spin), for example:
\begin{equation}
    \begin{split}
    a^{(n,+)}_{m=2} &= +1.5493..,\quad -0.7655..,\quad -0.6928..,\quad +1.2875..,\quad +0.2677..,\quad +1.7124.., \; \ldots \\
    a^{(n,-)}_{m=2} &= -1.0683..,\quad +0.2893..,\quad -0.2309..,\quad +1.1619..,\quad -1.5402..,\quad +0.3741.., \; \ldots
    \end{split}
\end{equation}
where we normalized such that $ a^{(n,\pm)}_{m=1} = 1$. The Fourier coefficients are distributed according to a Wigner semi-circle for prime spins $m \rightarrow \infty$.
Studying the nearest neighbor spacings reveals that both the eigenvalues and the Fourier coefficients (for fixed spin) are Poisson distributed -- a fact we shall refer to as {\it arithmetic chaos}; see Appendix \ref{app:maass} for details and plots. In a sense, arithmetic chaos is more akin to an {\it integrable} rather than a chaotic structure. One of our goals is to elucidate the relationship between this randomness in the structure of the Maass cusp form expansion and the genuine {\it quantum chaos} described by the linear ramp in the spectral form factor. 
To reproduce the ramp from a sum over cusp forms, we will have to address this interplay.

A central ingredient in our analysis is a certain continuum approximation to the discrete sum over cusp forms; relatedly, we will argue that all cusp form data can be replaced with its statistical average, which we will explain in turn. Before giving details, let us summarize the steps we will follow:
\begin{enumerate}
    \item To find $\langle{\color{colHighlight} z_{n_1,\pm} z_{n_2,\pm}}\rangle$ such that the sum \eqref{eq:discSumRes0} yields a ramp, we first note that the linearly increasing density of eigenvalues $R_n^\pm$ allows us to approximate the sum by an integral over a continuous eigenvalue density. We will argue that this approximation becomes arbitrarily good for large $y_i$. Equivalently, we can think of the large $y_i$ limit as implementing a statistical averaging over eigenvalues.
    \item While less obvious, we will show that the large $y_i$ limit also acts as a statistical averaging over the Fourier coefficients $a_m^{(n,\pm)}$. Since they appear squared in the spectral form factor, the cusp form sum is effectively only sensitive to their statistical variance. Thanks to certain Hecke relations, the information contained in the variances of Fourier coefficients for all spins $m$ is equivalent to the information contained in the full distribution of those with prime $m$.
    \item Using these statistical properties, we illustrate what kind of correlations $\langle{\color{colHighlight} z_{n_1,\pm} z_{n_2,\pm}}\rangle$ can yield a ramp in a given spin sector. We then show how to get a ramp in {\it every} spin sector in a very constrained way. The correlations thus obtained come with a certain amount of freedom. We show that fixing this freedom in the simplest possible way leads to a result that matches the pure gravity wormhole amplitude \cite{Cotler:2020ugk}.
\end{enumerate}

Throughout this section we make extensive use of a database of $5832$ even and $6282$ odd Maass cusp forms (corresponding to eigenvalues $R_n^\pm < 400$), computed in \cite{Then_2004} (see also \cite{lmfdb} for a subset). We also assume the non-degeneracy of cusp forms, which is a widely believed but unproven conjecture.

\vspace{5pt}
\subsection{Statistical treatment of the sum over eigenvalues $R_n^\pm$}
\label{sec:StatRn}

A ramp can be encoded in the coefficients of Maass cusp forms; to extract this, we need to invert the discrete part of \eqref{eq:z-m-alpha-def}. This requires an appropriate regularization of the integrals over Bessel functions to ensure their orthogonality in the discrete solution space. We avoid this technical point for the moment by working with an approximate continuous representation. This will allow us to derive the solution. We will see that this representation utilizes many of the statistical properties of the cusp forms, thus connecting arithmetic chaos to the expansion of the ramp in the cusp forms.

To start we define the density of cusp forms by $\mu_\pm(R)$, defined by
\begin{equation}
\widetilde{Z}^m_{\text{P,disc.,}\pm}(y) =  \int_{r_\pm}^\infty dR \, \mu_\pm(R)\, z^m_{R,\pm} \, \sqrt{y} K_{iR}(2\pi m y)\,, \qquad \mu_\pm(R) = \sum_{n\geq 1} \delta(R-R_n^\pm) \,,
\label{eq:Zapproximate}
\end{equation}
where $z^m_{R,\pm}$ is a smooth function of $R$ such that $z^m_{R_n^\pm,\pm} \equiv {\color{colHighlight2}z^m_{n,\pm}}$. We will justify by construction that this is consistent with a ramp.

The asymptotic density of cusp forms can be approximated by a continuous function, using the `Weyl law' (see for example \cite{Steil:1994ue,PhysRevA.44.R7877}):
\begin{equation}
\label{eq:muRapprox}
\begin{split}
    \mu_+(R) &\approx \bar{\mu}_+(R) = \frac{1}{12} \, R - \frac{3}{2\pi} \, \log R + \frac{\log(\pi^4/2)}{4\pi} + {\cal O}\left(\frac{\log R}{R^2}\right)\,, \\
    \mu_-(R) &\approx \bar{\mu}_-(R) = \frac{1}{12} \, R - \frac{1}{2\pi} \, \log R - \frac{\log 8}{4\pi} + {\cal O}\left(\frac{\log R}{R^2}\right)\,.
\end{split}
\end{equation}
The lower cutoff $r_\pm>0$ in \eqref{eq:Zapproximate} is chosen appropriately such as to avoid over-counting of the constant cusp form. We review this approximation and various other statistical properties of the Maass cusp forms in appendix \ref{app:maass}, see in particular figure \ref{fig:appWeylLaw}. For the purpose of our analysis, the smooth approximation to the density of eigenvalues $R_n^\pm$ sometimes allows us to replace sums by integrals:
\begin{equation}
\label{eq:sumnApprox}
    \sum_{n>0} f(R_n^\pm) \stackrel{?}{\approx} \int_{r_\pm}^\infty dR \, \bar{\mu}_\pm(R) \, f(R) \,,
\end{equation}
which one might expect to hold for sufficiently smooth functions $f$. Clearly the approximation is better for functions $f$ with support at larger values of $R$, since the eigenvalue density increases linearly with $R$; thus, more precisely, for any $\varepsilon>0$ and sufficiently smooth functions $f$, there is a sufficiently large $n_0$ such that
\begin{equation}
\label{eq:sumnApprox2}
   \left| \sum_{n>n_0} f(R_n^\pm) - \int_{R_{n_0}}^\infty dR \, \bar{\mu}_\pm(R) \, f(R) \right| < \varepsilon\,.
\end{equation}
\begin{figure}
    \centering
\includegraphics[width=.49\textwidth]{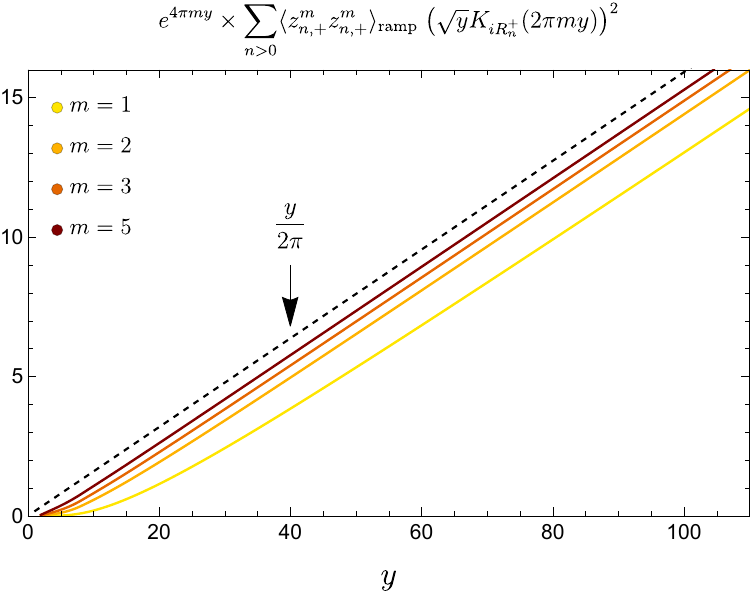}$\,$
\includegraphics[width=.49\textwidth]{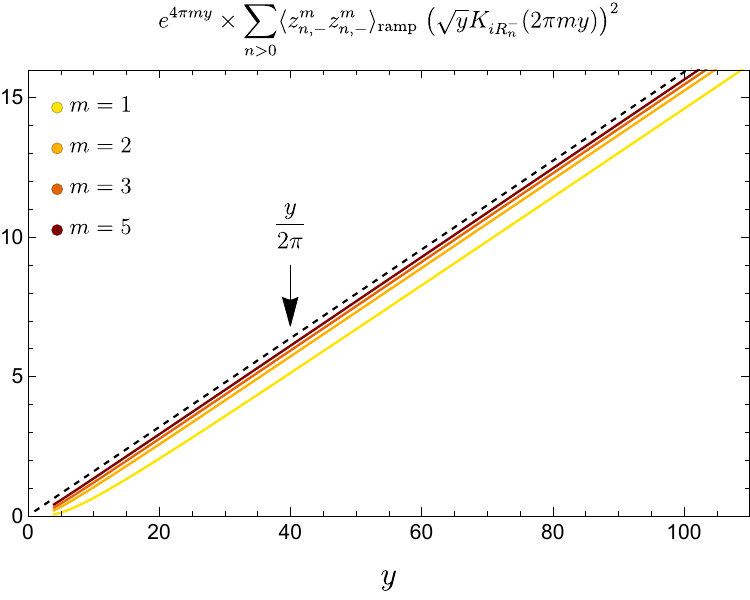}
    \caption{ Numerical verification of the encoding of a linear ramp in correlations of even (left) and odd (right) Maass cusp forms, according to \eqref{eq:discSumRes}, for $y_1=y_2\equiv y$. The summation over $n$ is performed up to some cutoff such that convergence is achieved within the displayable accuracy. The plots show that the sum converges to the ramp linear $y/(2\pi)$ up to an $m$-dependent constant that is subleading as $y\rightarrow \infty$.}
    \label{fig:rampCusp}
\end{figure}
Working with this continuous approximation, a calculation identical to \eqref{eq:z-m-alpha-result} gives: 
\begin{equation}
\begin{split}
 &  \big\langle z^{m_1}_{R_1,\pm}z^{m_2}_{R_2,\pm}\big\rangle_\text{ramp} 
 \approx \frac{2R_1 \tanh(\pi R_1)}{\pi^2\,\bar{\mu}_\pm(R_1)^2} \,\delta_{m_1m_2}\, \delta(R_1 - R_2)   \,.
 \end{split}
\label{eq:z-m-R-result}
\end{equation}
One can immediately see that this approximate continuous expression translates into the following correlations for the discrete coefficients:
\begin{equation}
 \big\langle {\color{colHighlight2}z^{m_1}_{n_1,\pm}\,z^{m_2}_{n_2,\pm}}\big\rangle_\text{ramp} 
 \approx \frac{2R_{n_1}^\pm \tanh(\pi R_{n_1}^\pm)}{\pi^2\,\bar{\mu}_\pm(R_{n_1}^\pm)} \,\delta_{m_1m_2}\, \delta_{n_1n_2}   \,.
\label{eq:z-m-R-result-disc}
\end{equation}
or, equivalently, the cusp form sum \eqref{eq:discSumRes0} encoding a linear ramp should be of the form
\begin{equation}
\begin{split}
    &\Big{\langle}  \widetilde{Z}^{m_1}_{\text{P,disc.,}\pm}(y_1) \widetilde{Z}^{m_2}_{\text{P,disc.,}\pm}(y_2) \Big{\rangle}_\text{ramp} \\
    &\qquad
     \approx \sum_{n_1,n_2>0}
    \left( \frac{2R_{n_1}^\pm \tanh(\pi R_{n_1}^\pm)}{\pi^2\,\bar{\mu}_\pm(R_{n_1}^\pm)} \,\delta_{m_1m_2} \delta_{n_1n_2} \right) \sqrt{y_1} K_{iR_{n_1}^\pm}(2\pi m_1 y_1)\,\sqrt{y_2} K_{iR_{n_2}^\pm}(2\pi m_2 y_2)
    \end{split}
    \label{eq:discSumRes}
\end{equation}
This is approximate in the following sense.
To evaluate the sum we can proceed in two ways: $(i)$ analytically, we can approximate the sum by an integral as in \eqref{eq:sumnApprox}, which in turn recovers the {\it exact} ramp in {\it every} spin sector (by construction). The approximation is then due to replacing the sum by an integral. This approximation becomes increasingly good for $y_i \rightarrow \infty$ because the support of the Bessel functions becomes peaked shifts to larger values of $R_n^\pm$ where these are more dense. Indeed, the sum receives most of its support from a window $n_1=n_2\in [n_\text{min},n_\text{max}]$, where both $n_\text{min}$ and $n_\text{max}$ increase with $y_i$. $(ii)$ Numerically, we can confirm directly that the discrete sum \eqref{eq:discSumRes} (cut off at an appropriate $n_\text{max}$) does also reproduce the ramp up to an error (a subleading constant shift) that goes to zero as $y_i \rightarrow \infty$.\footnote{ The constant shift is the error introduced by the summands with small values of $n$, where the continuum approximation is worse. It is strictly subleading to the linear ramp for large $y_i$.} Figure \ref{fig:rampCusp} illustrates the result (both for even and odd parity cusp forms). We see that the numerical evaluation of the Maass cusp form sum asymptotes to the expected linear ramp for large values of $y_i$ (we only show the case $y_1=y_2\equiv y$, but other cross sections of the $(y_1,y_2)$ plane were checked similarly). In appendix \ref{app:errorEstimate} we give more details on these approximations.

\vspace{5pt}
\subsection{Statistical treatment of the Fourier coefficients $a_m^{(n,\pm)}$}\label{sec:stat-fourier-coeff}

Let us return to the sum over cusp forms, \eqref{eq:discSumRes0}. We wish to address the following question: {\it what form of correlations $\langle {\color{colHighlight}z_{n_1,\pm} z_{n_2,\pm}}\rangle$ yields the ramp \eqref{eq:discSumRes}?} Naively, it seems that we have already answered this question in \eqref{eq:z-m-R-result-disc}. However, that expression, taken literally, would via \eqref{eq:zzzzDef} give a different, {\it spin-dependent} form of $\langle {\color{colHighlight}z_{n_1,\pm} z_{n_2,\pm}}\rangle$ for {\it every} spin, which clearly cannot be correct. So how is \eqref{eq:discSumRes} consistent with spin-independent correlations $\langle {\color{colHighlight}z_{n_1,\pm} z_{n_2,\pm}}\rangle$? To resolve this conundrum, we take a detour to discuss properties of the Fourier coefficients $a_m^{(n,\pm)}$ of the cusp forms.\\

What does the erratic nature of the Fourier coefficients mean for the validity of our continuous approximation to the eigenvalues?
We argued that the sum over $n$ is dominated by a window of $R_n^\pm \in [R_\text{min}, R_\text{max}]$.
For any desired error in the evaluation of the cusp form sum, the corresponding $R_\text{min}$ and $R_\text{max}$ increase indefinitely as $y_i \rightarrow \infty$ (see appendix \ref{app:errorEstimate}), so the relevant {\it density} of eigenvalues $R_n^\pm$ increases as well. Summing over an increasingly dense set of $R_n^\pm$ acts as a {\it statistical coarse-graining} over the $n$-dependent summands. In particular, the product of the Fourier coefficients appearing in the sum and the correlations $\langle {\color{colHighlight}z_{n_1,\pm} z_{n_2,\pm}}\rangle$ get averaged over. We therefore expect to be able to replace the discrete erratic Fourier coefficients by their statistical distribution.

\paragraph{Distribution of Fourier coefficients:}
The statistical distribution of the Fourier coefficients is a well-known topic of mathematical research, and we review it in some detail in appendix \ref{app:maass}. Let us only point out the most crucial aspects. First, the asymptotic distribution of $a_m^{(n,\pm)}$ for fixed {\it prime} spins $m\equiv p$ is well known \cite{sarnakStatisticalPropertiesEigenvalues1987}:
\begin{equation}
    \boldsymbol{\mu}_p(x)=
    \begin{cases}
        \frac{(p+1) \sqrt{4-x^2}}{2\pi \left( \left(p^{1/2}+p^{-1/2}\right)^2-x^2\right)} & \text{if } |x|<2 \\
        0 & \text{otherwise}
    \end{cases}
    \label{eq:coefficient_mu}
\end{equation}
For large prime spins, this approaches a Wigner semicircle $(2\pi)^{-1}\sqrt{4-x^2}$. Another notable feature is that the distribution suggests that $|a_p^{(n,\pm)}| < 2$ for all $n$, a property known as the Ramanujan-Petersson conjecture \cite{sarnakStatisticalPropertiesEigenvalues1987}. We are interested in moments of these distributions. Since the sum \eqref{eq:discSumRes0} features the squares of Fourier coefficients, a statistical feature of particular interest is their variance
\begin{equation}
\label{eq:varianceDef}
    {\cal N}_m^\pm \equiv \overline{\big(a_m^{(n,\pm)}\big)^2 } \equiv \lim_{n_0\rightarrow \infty}\; \frac{1}{n_0}\sum_{n=1}^{n_0} \big( a_m^{(n,\pm)} \big)^2 \,,
\end{equation}
which has the following exact value for prime spins:\footnote{
It is interesting to note that since the Fourier coefficients for prime spins are Poisson distributed, as shown in appendix \ref{app:maass}, the variance in \eqref{eq:varianceDef} already implies delta-functions in spin and eigenvalue index,
\begin{equation}
   p_1,p_2\, \text{ prime:}\qquad  \overline{a_{p_1}^{(n_1,\pm)} a_{p_2}^{(n_2,\pm)} } = \overline{\big(a_{p_1}^{(n_1,\pm)}\big)^2}\; \delta_{n_1,n_2} \delta_{p_1,p_2}  = {\cal N}_{p_1}^\pm \; \delta_{n_1,n_2} \delta_{p_1,p_2} \,.
\end{equation}
This suggests that arithmetic chaos is linked to the diagonal approximation in the periodic orbit picture of \cite{DiUbaldo:2023qli}. The delta-function in the eigenvalue indices persists even for non-prime spins and is therefore tied to the effective statistical averaging implemented by the correlated cusp form sums \eqref{eq:discSumRes0}. Note, however, that we will later average over summands involving higher moments of Fourier coefficients, which complicates the picture.\label{foot:diag}} 
\begin{equation}
    {\cal N}_p^\pm = \frac{p+1}{p} \qquad \text{(}p\text{ prime; exact).}
\end{equation}
See \eqref{eq:momentsFormula} for higher moments. We use the notation $\overline{(\cdots)}$ to denote {\it statistical} averaging (over $n$). This is independent of the {\it microcanonical} averaging, denoted by $\langle \cdots \rangle$, which we always use to discuss correlations in the coarse-grained CFT spectrum.

For non-prime spins $m$, the variances ${\cal N}^\pm_m$ are determined by the distributions for prime spins. Importantly, not only the variances of the distributions for prime spins, but also their {\it higher moments} are needed. The reason is that the Fourier coefficients themselves are determined as non-linear polynomials of those for prime spins by a certain Hecke algebra, see \eqref{eq:HeckeExamples} for some examples. Statistical averaging over such polynomials requires knowledge of higher moments of the prime distributions. In summary, the following three pieces of information are equivalent: 
    \begin{gather*}
    \text{variances }  {\cal N}_m^\pm \equiv \overline{\big(a_m^{(n,\pm)}\big)^2 } \text{ of distributions of {\it all} spins } m \\
    \Leftrightarrow \\
    \text{{\it all} moments } \overline{\big(a_p^{(n,\pm)}\big)^k } \text{ of distributions of {\it prime} spins $p$} \\
   \Leftrightarrow \\
    \text{distributions \eqref{eq:coefficient_mu} of {\it prime} spins}
    \end{gather*}
We review these statements in appendix \ref{app:maass} and give examples in \eqref{eq:HeckeExamples}. Using the first 5832 even and 6282 odd Fourier coefficients, we find numerically for their variances as a function of spin $m$:
\begin{equation}
\label{eq:variancesNum}
    \begin{split}
{\cal N}_m^+  \; &\approx \; 
    {\bf 1}\,,\; {\bf 1.46}\,,\; {\bf 1.27} \,,\; 1.65\,,\; {\bf 1.13}\,,\; 1.84\,,\; {\bf 1.07} \,,\; 1.72 \,,\; 1.32 \,,\; 1.63 \,,\; {\bf 1.02} \,,\ldots \\
    {\cal N}_m^- \; &\approx \; 
    {\bf 1}\,,\; {\bf 1.47}\,,\; {\bf 1.30} \,,\; 1.68\,,\; {\bf 1.16}\,,\; 1.89\,,\; {\bf 1.09} \,,\; 1.76 \,,\; 1.36 \,,\; 1.68 \,,\; {\bf 1.04} \,,\ldots
    \end{split}
\end{equation}
where values for prime $m$ are printed in boldface (see tables \ref{tab:table1} and \ref{tab:table2} for more details).

\paragraph{Statistical averaging in the spectral form factor:} 
Whenever the statistical averaging over $n$ applies to our cusp form sum over $n$, it means that we can replace discrete erratic expressions by their statistical average. This amounts to a significant simplification for evaluating sums such as \eqref{eq:discSumRes0}: for large $y_i$ the {\it exact} squared Fourier coefficients (which oscillate erratically) can be replaced by their {\it mean value}, i.e., the variance of their distribution \eqref{eq:coefficient_mu}, thus `forgetting' about the detailed sporadic values and only keeping track of statistical information. This explains how it was possible that the correlations $\langle {\color{colHighlight}z_{n_1,\pm} z_{n_2,\pm}}\rangle$ that follow from \eqref{eq:z-m-R-result-disc} could depend on spin in such a fine tuned way as to cancel all erratic Fourier coefficients $a_{m_1}^{(n_1,\pm)}a_{m_2}^{(n_2,\pm)}$: the correlations $\langle {\color{colHighlight}z_{n_1,\pm} z_{n_2,\pm}}\rangle$ do not actually need to cancel the Fourier coefficients exactly, but only {\it on average}. As we will see, this is indeed possible in a spin-independent way.

Focusing on a single spin sector, the fact that the Fourier coefficients only need to cancel on average means we would expect to reproduce the linear ramp in the spin $m$ sector from
\begin{equation}
\begin{split}
    &\Big{\langle}  \widetilde{Z}^{m}_{\text{P,disc.,}\pm}(y_1) \widetilde{Z}^{m}_{\text{P,disc.,}\pm}(y_2) \Big{\rangle}_\text{ramp naive} \\
    &\qquad
     \equiv \frac{1}{{\cal N}_m^\pm} \sum_{n>0}
    \left( \frac{2R_{n}^\pm \tanh(\pi R_{n}^\pm)}{\pi^2\,\bar{\mu}_\pm(R_{n}^\pm)} \right) \big(a_{m}^{(n,\pm)}\big)^2 \, \sqrt{y_1} K_{iR_{n}^\pm}(2\pi m y_1)\,\sqrt{y_2} K_{iR_{n}^\pm}(2\pi m y_2)
    \end{split}
    \label{eq:z-m-R-result-disc-zn3}
\end{equation}
This corresponds to correlations of the form\footnote{ The second approximation, $\frac{2R_{n}^\pm \tanh(\pi R_{n}^\pm)}{\pi^2\,\bar{\mu}_\pm(R_{n}^\pm)} \approx \frac{24}{\pi^2}$, is valid asymptotically for very large $n$, i.e., for very large $y_i$. For all numerical results in this paper, this approximation is not good enough and is not used.} 
\begin{equation}
\label{eq:znznApproxDef}
    \big\langle {\color{colHighlight}z_{n,\pm}\,z_{n,\pm}}\big\rangle_{\text{spin }m\text{ ramp naive}}  \equiv \frac{1}{{\cal N}_m^\pm} \, \frac{2R_{n}^\pm \tanh(\pi R_{n}^\pm)}{\pi^2\,\bar{\mu}_\pm(R_{n}^\pm)} \approx   \frac{24}{\pi^2 \, {\cal N}_m^\pm}   \quad\;\;\; (m\geq 1,\, n\gg 1)
\end{equation}

\begin{figure}
    \centering
\includegraphics[width=.48\textwidth]{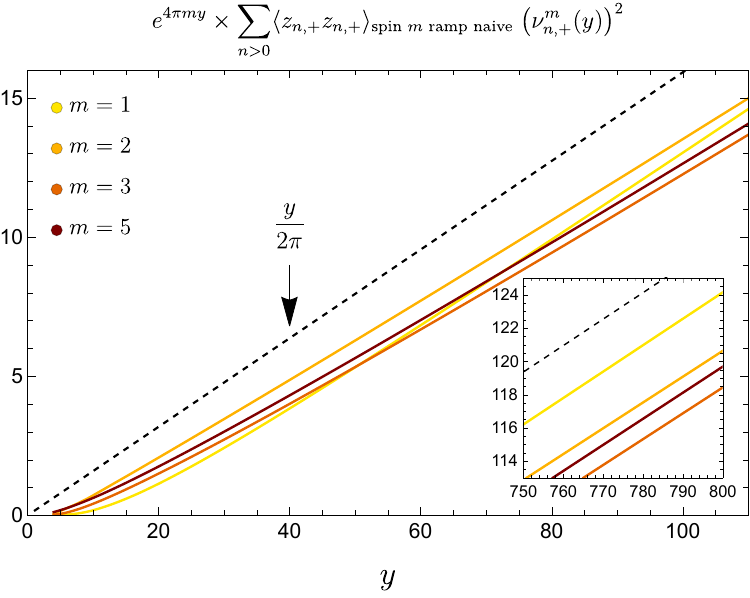}$\,$
\includegraphics[width=.49\textwidth]{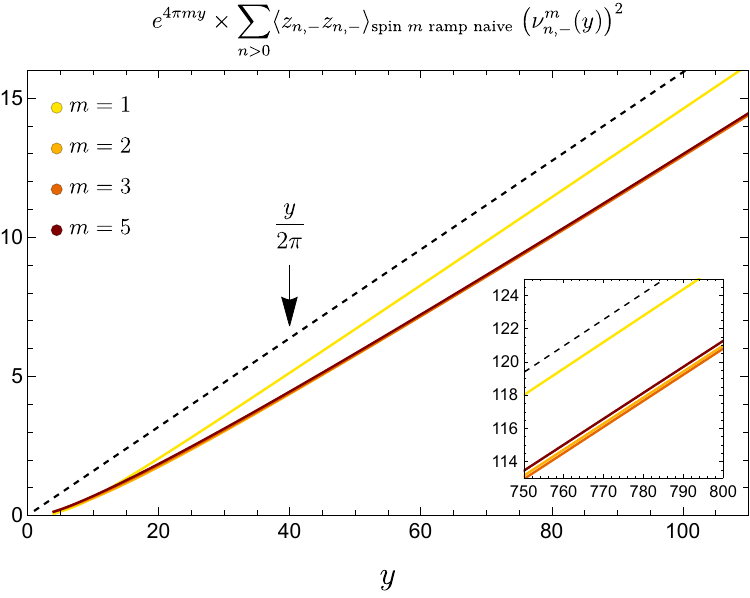}
    \caption{We compute the Maass cusp form sum using the variance of the Fourier coefficients instead of their exact values in \eqref{eq:z-m-R-result-disc-zn3}. For large $y$ increasingly many Fourier coefficients contribute to the sum over $n$, which means that their square can be increasingly well approximated by their variance. We therefore reproduce the linear ramp asymptotically (up to a subleading constant shift), c.f.\ figure \ref{fig:rampCusp}. The left (right) shows the case of even (odd) parity cusp forms. In the odd case the ramps for different spins lie almost on top of each other. Insets show larger values of $y$.}
    \label{fig:rampConvergence}
\end{figure}

We can check the validity of this claim numerically by computing the sum \eqref{eq:z-m-R-result-disc-zn3} and comparing it with the true form of the ramp. As can be seen in figure \ref{fig:rampConvergence}, for large $y$ the correct linear ramp is approached, again up to a constant which is subleading for $y \rightarrow \infty$.\\

Evidently, \eqref{eq:znznApproxDef} still depends on the spin $m$ via the normalization ${\cal N}_m^\pm$, albeit much more weakly than had we tried to cancel the erratic Fourier coefficients in \eqref{eq:z-m-R-result-disc-zn3} exactly (term by term). It is therefore still not a good candidate for correlations $\langle {\color{colHighlight}z_{n,\pm} z_{n,\pm} }\rangle$ that yield linear ramps independent of the choice of spin. And indeed, correlations of the form \eqref{eq:znznApproxDef} only yield a ramp with the correct slope in the spin $m$ sector. In other spin sectors $m'$, we would need a similar form of correlations, but with a different normalization $1/{\cal N}_{m'}^\pm$. We will remedy this situation in the following subsection.

\subsection{Ramps in all spin sectors: number theory and uniqueness}

As we have seen, \eqref{eq:znznApproxDef} only encodes the ramp in the spin $m$ superselection sector, but it `contaminates' the slope of any putative ramp in other spin sectors. The basic assumption of quantum chaos, however, would be a linear ramp with the correct slope in {\it all} spin sectors. To achieve this, let us now take the statistical averaging one step further and improve the naive ansatz \eqref{eq:znznApproxDef} such that it works {\it on average} for every spin sector, i.e., in a spin-independent way. We wish to write:
\begin{equation}
\label{eq:zzRampFinal}
\boxed{ \;\;   \langle {\color{colHighlight}z_{n_1,\pm} z_{n_2,\pm}} \rangle_{\text{ramp}} \approx 
    \frac{2R_{n_1}^\pm \tanh(\pi R_{n_1}^\pm)}{\pi^2 \bar{\mu}_\pm(R_{n_1}^\pm)}\,\delta_{n_1n_2} \, f^{(n,\pm)}
    \approx 
    \frac{24}{\pi^2 }\,\delta_{n_1n_2} \, f^{(n,\pm)}
    \;\;}
\end{equation}
with a {\it spin-independent} function $f^{(n,\pm)}$ such that
\begin{equation}
    \overline{ \big( a_m^{(n,\pm)} \big)^2 f^{(n,\pm)}} = 1 \quad \text{ for all } m\geq 1.
    \label{eq:fCond}
\end{equation}
The effective averaging over $n$ will then guarantee that in the limit $y_i \rightarrow \infty$, we recover the ramp {\it for all spins $m$}:
\begin{equation}
\begin{split}
&\sum_{n_1,n_2>0} \langle {\color{colHighlight}z_{n_1,\pm} z_{n_2,\pm} }\rangle_\text{ramp} \;  \nu_{n_1,\pm}^m(y_1) \nu_{n_2,\pm}^m(y_2) \\
    &\quad \stackrel{y_i\rightarrow\infty}{\longrightarrow} \sum_{n >0} \frac{24}{\pi^2}\, \overline{\big( a_m^{(n,\pm)} \big)^2 f^{(n,\pm)}} \, \sqrt{y_1} K_{iR_n^\pm}(2\pi m y_1)\,\sqrt{y_2} K_{iR_n^\pm}(2\pi m y_2) + \ldots \\
     &\qquad\;\;= \frac{1}{\pi} \frac{y_1 y_2}{y_1+y_2}\, e^{-2\pi m(y_1+y_2)}  + \ldots
\end{split}
\end{equation}
where we replaced $\big( a_m^{(n,\pm)} \big)^2 f^{(n,\pm)}$ by its average according to \eqref{eq:fCond} in the second line and then simply applied \eqref{eq:discSumRes}. We denote subleading terms by `$\ldots$'.

We will refer to $f^{(n,\pm)}$ as the {\it arithmetic kernel} associated with the cusp form $\nu_{n,\pm}$.
This name is inspired by the fact that any function satisfying \eqref{eq:fCond} must obviously depend on all Fourier coefficients for all spins in a fine-tuned way such that it produces just the right normalization for the ramp in every spin sector. It must, in a sense, encode all the information loosely referred to as {\it arithmetic chaos}, such as Hecke relations \eqref{eq:HeckeDef} and the statistical distribution of Fourier coefficients \eqref{eq:coefficient_mu}. Note that the ansatz \eqref{eq:zzRampFinal} assumes diagonality in $n_i$. We will justify by construction that this is a consistent assumption. Note further that the condition \eqref{eq:fCond} really only needs to hold asymptotically as a statement about the average over terms in the spectral form factor with large $n$.\footnote{ For example, we can imagine performing a `moving average' over large but finite windows of $n$, which determine the cusp form sum over corresponding `batches' of cusp forms, then for small $n$ it is certainly allowed that the average fluctuates around $1$.} Deviations for small $n$ will only affect subleading terms in the late-time spectral form factor. We fix this ambiguity in the minimal way, i.e., by imposing \eqref{eq:fCond} as an average over all $n$ as written.

Given all the information encoded in `arithmetic chaos', it is remarkable that such a function exists. We will now first write down this function, then explain why it works, and then derive it, showing that it is essentially unique (under the above assumptions). The arithmetic kernel satisfying \eqref{eq:fCond} is given by
\begin{equation}
\label{eq:KernelDef}
\boxed{\;\;
    f^{(n,\pm)} = \prod_{p \text{ prime}} \left[ \frac{p+1}{p} - \frac{1}{p+1} \, \big( a_p^{(n,\pm)} \big)^2 \right] \,.
    \;\;}
\end{equation}
Let us first confirm that this function satisfies \eqref{eq:fCond}. We do this in three steps:
\begin{enumerate}
    \item {\bf If $m \equiv p$ is prime:} Since the Fourier coefficients for prime spins are independently distributed, we only need to know the second and fourth moments of the distributions \eqref{eq:coefficient_mu}, which are easy to calculate. We immediately find:
    {\small
    \begin{equation}
        \begin{split}
            \overline{ \big( a_p^{(n,\pm)} \big)^2 f^{(n,\pm)}}
            &= \left[ \frac{p+1}{p} \,\overline{ \big( a_p^{(n,\pm)} \big)^2}- \frac{1}{p+1}\, \overline{ \big( a_p^{(n,\pm)} \big)^4}\right]\prod_{\substack{p' \text{ prime}\\ p' \neq p}} \left[ \frac{p'+1}{p'} - \frac{1}{p'+1} \overline{ \big( a_{p'}^{(n,\pm)} \big)^2} \right] =1\,,
        \end{split}
        \label{eq:avCalc}
    \end{equation}
    }
    where every factor is individually 1 due to the following statistical facts:
    \begin{equation}
        p \text{ prime:} \qquad 
        \overline{ \big( a_p^{(n,\pm)} \big)^2} = \frac{p+1}{p} \,,\qquad \overline{ \big( a_p^{(n,\pm)} \big)^4} = \frac{2p^2+3p+1}{p^3} \,.
    \end{equation}
    \item {\bf If $m = p^k$ is a prime power:} For prime power spins, we can analogously show that every factor in an expression similar to \eqref{eq:avCalc} is 1. For the first factor ($p'=p$) we need some more non-trivial facts about the Fourier coefficients, which follow from the Hecke multiplicativity rules \eqref{eq:HeckeDef}. The required properties are (see appendix \ref{app:norms} and in particular Lemma \ref{lemma:averages}):
    \begin{equation}
        p \text{ prime:} \quad\;\; \overline{ \big( a_{p^k}^{(n,\pm)} \big)^2} = \frac{p-p^{-k}}{p-1} \,,\qquad 
        \overline{ \big( a_{p^k}^{(n,\pm)} \big)^2 \big( a_{p}^{(n,\pm)} \big)^2} = \frac{2(p+1) - p^{-k}(p+2+p^{-1})}{p-1} 
    \end{equation}
    Note that these properties encode {\it all} information about the distributions \eqref{eq:coefficient_mu}.
    \item {\bf Arbitrary $m$:} For any general integer $m$, there is a prime factorization $m=p_1^{k_1} \cdots p_r^{k_r}$. The Hecke multiplicativity rules \eqref{eq:HeckeDef} imply 
    \begin{equation}
    \label{eq:primeFac}
        m=p_1^{k_1} \cdots p_r^{k_r} \qquad \Rightarrow \qquad
\big(a_m^{(n,\pm)}\big)^2 = \left(a_{p_1^{k_1}}^{(n,\pm)} \right)^2\cdots \left(a_{p_r^{k_r}}^{(n,\pm)}\right)^2 \,.
    \end{equation}
    The property \eqref{eq:fCond} follows factor by factor.
\end{enumerate}

The arithmetic kernel $f^{(n,\pm)}$ has a deep number theoretical meaning in terms of Hecke $L$-functions. We elaborate on these fascinating mathematical properties in appendix \ref{app:norms}. We can also {\it derive} $f^{(n,\pm)}$ from physical requirements, i.e., by merely imposing \eqref{eq:fCond} in all spin sectors. We sketch the derivation below, delegating details to appendix \ref{app:fDerivation}.

\paragraph{Uniqueness of the arithmetic kernel:}
We will now derive the arithmetic kernel \eqref{eq:KernelDef} by arguing that the requirement \eqref{eq:fCond} fixes it {\it uniquely} (within an ansatz class). First recall that Fourier coefficients have multiplicative properties due to them being eigenvalues of Hecke operators. In particular, if the spin has a prime factor decomposition as in \eqref{eq:primeFac}, since $a_{p^k}^{(n,\pm)}$ are independently distributed for different primes $p$ it is useful to first solve the problem \eqref{eq:fCond} for prime power spins, $m=p^k$. Consider an ansatz of the form
\begin{equation}
    f^{(n,\pm)}_p
    =  \sum_{r\geq 0} c_{p,r} \big( a_p^{(n,\pm)} \big)^{2r} 
\end{equation}
for prime $p$. Since we already assumed diagonality in eigenvalues $R_{n_i}^\pm$ in \eqref{eq:zzRampFinal}, odd powers of Fourier coefficients will average to zero, and we discard them in our ansatz. Such terms would not affect the construction of the universal ramp, but they would change the subleading behavior of the late time spectral form factor. Discarding odd powers in the ansatz can thus be viewed as a {\it minimality assumption} about the ansatz. It would be interesting to constrain such ambiguities further, using input from the off-diagonal sector.

The condition \eqref{eq:fCond} yields an infinite linear system constraining the parameters $c_{p,r}$ in terms of moments of the distribution of Fourier coefficients. After some investigation (see appendix \ref{app:fDerivation}), this system can be written as follows:
\begin{equation}
\label{eq:systemT}
     \sum_{r\geq 0} c_{p,r}  \overline{\big( a_p^{(n,\pm)} \big)^{2(k+r)}} = \frac{(2k)!}{k!(k+1)!} \,.
\end{equation}
Making extensive use of $(i)$ Hecke relations and $(ii)$ all moments of the distributions of prime Fourier coefficiens, the solution of this system for $m=p^k$ is unique:
\begin{equation}
    c_{p,0} = \frac{p+1}{p} \,,\quad c_{p,1} = - \frac{1}{p+1} \,,\quad c_{p,r\geq 2} = 0 \,.
\end{equation}
Using \eqref{eq:primeFac}, the condition \eqref{eq:fCond} for all $m$ is then solved by
\begin{equation}
   f^{(n,\pm)} = \prod_{p \text{ prime}} f^{(n,\pm)}_p = \prod_{p \text{ prime}} \left[ \frac{p+1}{p} - \frac{1}{p+1} \big( a_p^{(n)} \big)^2 \right] \,.
\end{equation}
While we have made simplifying assumptions in the derivation of this kernel (see the discussion after \eqref{eq:fCond}), its uniqueness within a large class of possibilities is remarkable. We show in the next subsection that the structure of the result \eqref{eq:zzRampFinal}, \eqref{eq:KernelDef} is more than just a mathematical curiosity; it has a number theoretical interpretation and its simplicity is in fact intimately tied to a calculation in AdS$_3$ pure gravity.

\vspace{5pt}
\section{Matching universal correlations to the AdS${}_3$ wormhole}\label{sec:gravity-ramp}

Our `bottom-up' construction of the spectral overlap coefficients encoding the linear ramp was based on minimal assumptions about quantum chaos in all spin sectors and consistency with the symmetries of CFTs. We also assumed a certain minimality in the ansatz for the arithemtic kernel $f^{(n,\pm)}$, which then allowed us to fully determine it. In this section we compare this `minimally consistent' arithmetic kernel with the wormhole amplitude found in AdS$_3$ pure gravity, which also exhibits such linear ramps. We find detailed agreement.

Demanding universal eigenvalue repulsion (i.e., a linear ramp) in every spin sector of the CFT, and assuming that for $m>0$ this property is encoded in the cusp form sector alone, we constructed the following form of spectral correlations as the simplest consistent possibility:
\begin{equation}
\label{eq:summaryRamp}
    \begin{split}
    \big\langle {\color{colHighlight} z_{\frac{1}{2}+i\alpha_1} \, z_{\frac{1}{2}+i\alpha_2} } \big\rangle_{\text{spin }0 \text{ ramp}} &= \frac{1}{2\cosh(\pi \alpha_1)} \times 4\pi\delta(\alpha_1+\alpha_2) \,,\\
    \big\langle {\color{colHighlight} z_{n_1,\pm} \, z_{n_2,\pm}} \big\rangle_{\text{spin }m>0 \text{ ramps}} &= \frac{24}{\pi^2}  \,f^{(n,\pm)}\times \delta_{n_1,n_2} \,, \qquad  f^{(n,\pm)}  \equiv \prod_{p \text{ prime}} \left[ \frac{p+1}{p} - \frac{1}{p+1} \, \big( a_p^{(n,\pm)} \big)^2 \right]
    \end{split}
\end{equation}
By virtue of being spin-independent, these correlations provide a manifestly modular invariant encoding of a linear ramp in all spin sectors. (Of course, the `bare' asymptotic ramp is not modular invariant by itself, so the subleading corrections produced by \eqref{eq:summaryRamp} are important.)

Let us now turn to gravity. The spectral decomposition of the $\mathbb{T}^2 \times I$ wormhole amplitude in AdS$_3$ pure gravity \cite{Cotler:2020ugk,Cotler:2020hgz} was given in \cite{DiUbaldo:2023qli}, and provides an explicit example of a modular invariant spectral form factor that contains a ramp in the large $y_i$ limit.\footnote{We  thank Scott Collier for private conversation on this result.} In our notation it corresponds to the following non-zero variances:\footnote{ To compare with \cite{DiUbaldo:2023qli}, note that $\frac{\pi}{\cosh(\pi \alpha)} = \Gamma(\tfrac{1}{2}+i\alpha) \Gamma(\tfrac{1}{2}-i\alpha)$. To compare with \cite{Cotler:2020ugk}, note that we introduced an additional factor of 2 in the wormhole amplitude to match the GOE universality class, c.f., \cite{Yan:2023rjh,DiUbaldo:2023qli}.}
\begin{equation}
    \begin{split}
     \big\langle {\color{colHighlight}z_{\frac{1}{2}+i\alpha_1} z_{\frac{1}{2}+i\alpha_2}} \big\rangle_\text{wormhole} &=  \frac{1}{2\cosh (\pi \alpha_1)}\times 4\pi \delta(\alpha_1+\alpha_2) \,,\\
     \big\langle {\color{colHighlight}z_{n_1,\pm} \, z_{n_2,\pm} }\big\rangle_\text{wormhole} &=  \frac{1}{2\cosh(\pi R_{n_1}^\pm)}\,\frac{1}{ |\!|\nu_{n,\pm}|\!|^2} \times \delta_{n_1n_2} \,,
    \end{split}
    \label{eq:wormholeResult}
\end{equation}
where the cusp form norms are computed with respect to the Petersson inner product (see appendix \ref{app:norms} for more details, and figure \ref{fig:norms} for concrete values).
The second line is meant to indicate that both the even and odd correlations as indicated give a ramp with correct normalization. In a CFT with parity symmetry, the even and odd ramps describe chaos in different parity superselection sectors.

Now compare our result \eqref{eq:summaryRamp} with \eqref{eq:wormholeResult}. The continuous part of the correlations, which encodes the spin 0 ramp, matches immediately (which is by construction). More interestingly, the discrete correlations, which we constructed by imposing quantum chaotic universality consistently across spin sectors, also match the gravity result. To see this, we need an important fact from arithmetic number theory, which is derived and explained in appendix \ref{app:norms}. The central observation is that our arithmetic kernel $f^{(n,\pm)}$ is a particular meromorphic {\it symmetric square $L$-function} $L_{\nu\times\nu}^{(n,\pm)}(s)$ evaluated at $s=1$. For every single cusp form, this function provides a generalization of the Riemann zeta-function that encodes all the statistical properties and Hecke relations between different spin Fourier coefficients. The precise statement is:
\begin{equation}
    f^{(n,\pm)} = \frac{\zeta(2)}{L_{\nu\times\nu}^{(n,\pm)}(s=1)}  
    =  \frac{\pi^2}{48  \cosh(\pi R_n^\pm) |\!|\nu_{n,\pm} |\!|^2} \,.
    \label{eq:fLsym}
\end{equation}
The intermediate steps in this equation are reviewed in appendix \ref{app:norms}. This establishes equality of, on the one hand, the correlations found from demanding a `bare' linear ramp in all spin sectors (taking into account the mechanism of statistical averaging over cusp forms and constructing a minimal spin-independent arithmetic kernel) in \eqref{eq:summaryRamp}, and, on the other hand, the pure gravity result, \eqref{eq:wormholeResult}.

It is interesting to note that the spin-0 ramp, encoded in the Eisenstein sector, can similarly be expressed in terms of a suitable $L$-function:
\begin{equation}
\begin{split}
    \big\langle {\color{colHighlight} z_{\frac{1}{2}+i\alpha_1} \, z_{\frac{1}{2} + i\alpha_2} }\big\rangle_\text{spin 0 ramp} &=
 \frac{ \Lambda(i\alpha_1)\Lambda(i\alpha_2)}{2L_E^{(2\alpha_1)}(1)} \times 4\pi\delta(\alpha_1+\alpha_2) 
 \,,
\end{split}
\label{eq:LEis}
\end{equation}
where $L_E^{(\alpha)}(s) = \zeta(s+i\alpha) \zeta(s-i\alpha) $ is the meromorphic continuation of a sum over Fourier coefficients. See appendix \ref{app:LfuncCont} for more details.

To summarize, we have found that a linear ramp in {\it all spin sectors} $m\geq 1$ is encoded in the following sum over cusp forms in the near-extremal limit:
\begin{equation}
\label{eq:fnormRel}
     \sum_{n>0} \frac{1}{2\cosh(\pi R_n^\pm) }\,  \frac{\nu_{n,\pm}^m(y_1)}{|\!|\nu_{n,\pm}|\!|} \,\frac{\nu_{n,\pm}^m(y_2)}{|\!|\nu_{n,\pm}|\!|}
= \frac{1}{\pi} \frac{y_1 y_2}{y_1+y_2}\, e^{-2\pi |m|(y_1+y_2)}  + \ldots \quad \text{(for all }m\text{)}
\end{equation}

The agreement of the wormhole amplitude with the `minimal' realization of quantum chaos across spin sectors was called the {\it MaxRMT principle} in \cite{DiUbaldo:2023qli}. It amounts to the statement that the gravity amplitude is the {\it minimal} modular completion of a spectral form factor exhibiting linear ramps. More precisely, ref.\ \cite{DiUbaldo:2023qli} shows that the wormhole amplitude is the minimal extension of the `bare' ramp, after imposing `diagonal' and `Hecke' projections onto correlated eigenvalues and eigenfunctions in the spectral decomposition of the spectral form factor.\footnote{The Hecke projection of \cite{DiUbaldo:2023qli} refers to demanding that the spectral decomposition features no mixed correlations between Eisenstein series and cusp forms. It is then proven that Hecke symmetric wormhole amplitudes must have an identical functional form of correlations in the continuous and discrete sectors, which is indeed a remarkable feature of \eqref{eq:wormholeResult} after absorbing the cusp form norms into the normalization of Fourier coefficients. We explore this feature in more detail in \cite{Haehl:2023mhf}.} 

Our investigation similarly imposed some minimality requirements: the main assumptions were the realization of quantum chaos in all spin sectors and modular invariance; we argued that these assumptions required a spin-independent form of the arithmetic kernel and then constructed the simplest consistent kernel from an ansatz \eqref{eq:zzRampFinal} by solving the {\it statistical constraints}. The main assumptions in this analysis concern the nature of these statistical constraints: by discarding from the ansatz any terms that would be invisible to our statistical condition \eqref{eq:fCond}, we fixed it fully and recovered the wormhole amplitude. Recall also that we demanded the averaging condition \eqref{eq:fCond} to hold exactly for all $n$. This assumption extrapolates the linear ramp beyond the asymptotic regime in the simplest way, i.e., by discarding fluctuations from the statistical average. Relaxing these assumptions would give the flexibility to change the subleading corrections to the ramp such as to encode spectra not described by the wormhole. This provides a statistical perspective based on arithmetic chaos on the MaxRMT principle of \cite{DiUbaldo:2023qli}.

\vspace{10pt}
\section{Discussion}
\label{sec:discussion}

To summarize, we note again that the Euclidean wormhole amplitude \eqref{eq:wormholeResult} describes a universal part of the spectral correlations in any individual chaotic CFT, which dominates the late time near-extremal limit. We constructed the same object `bottom-up' by imposing quantum chaos (in the form of a linear ramp) in every spin sector separately and consistently balancing the imprints ramps in any given spin sector have on the slope of ramps in other spin sectors. We delineated the way in which a solution can be constructed based on statistical considerations of Maass cusp forms. A crucial role was played by the effective statistical averaging over erratic data defining the modular invariant Maass cusp forms. It is due to this averaging that, on the one hand, all statistical information about `arithmetic chaos' is encoded in the collection of linear ramps, while, one the other hand, detailed erratic features of cusp forms are washed out and a single spin-independent form of chaotic correlations could be bootstrapped. There is some freedom in the construction of the solution, which would affect subleading corrections to the spectral form factor; the match with the gravitational result was established by not making use of any of this freedom, i.e., fixing it in the minimal and most symmetric way, which we quantified.
We conclude with some further comments.

\vspace{5pt}
\subsubsection*{Spectral determinacy}

It was found in \cite{DiUbaldo:2023qli} that the spectral decomposition of the AdS$_3$ wormhole amplitude is such that the correlations in the Eisenstein series coefficients and those in the Maass cusp form coefficients are identical. We found the same result by imposing statistical universalities (quantum chaos) in all spin sectors and implementing them in a minimal way through a sum over cusp forms. This strengthens the spectral determinacy property of general two-dimensional CFTs \cite{Benjamin:2021ygh}, as in these examples {\it all} spin sectors exhibit identical correlations (`strong spectral determinacy' \cite{DiUbaldo:2023qli}). How is this consistent with one of the basic assumptions of quantum chaos, i.e., the {\it independence} of spectral universalities in each symmetry superselection sector? We take the following perspective: even though the statistical approximation required that our result \eqref{eq:summaryRamp} for spin $m>0$ linear ramps had to be the same for all spins, it nevertheless encodes separate input from all spin sectors. This is manifest when we consider the arithmetic kernel $f^{(n,\pm)}$: it contains all squared Fourier coefficients for all spin sectors in a highly fine-tuned way such as to ensure the correct statistical property \eqref{eq:fCond} for all spin sectors. For example, had we only imposed the ramp in some particular spin sector, then the naive ansatz \eqref{eq:znznApproxDef} would have been sufficient. But this would have impacted the slope of the ramp in all other spin sectors. Finding the universal kernel that yields the correct slope for all spins required us to separately assume the existence of a ramp for all spins and input the corresponding information into the construction of $f^{(n,\pm)}$ in a correlated way.\footnote{ Note also that we did not assume additional structures in the CFT partition function, such as it being a Poincar\'{e} sum over images of a seed function, which would lead to further constraints; see \cite{DiUbaldo:2023qli}.}

We can summarize this as follows: imposing random matrix universality in just one given spin sector leaves a lot of freedom for the choice of the cusp form correlations $\langle z_{n,\pm} z_{n,\pm}\rangle$, thanks to statistical coarse-graining in the late time limit. It does by no means imply a linear ramp with the correct slope for any other independent spin sector. But imposing random matrix universality in {\it all} spin sectors, leads to enough constraints to determine a universal, spin-independent form for the correlations describing the leading order linear ramp. Further, the statistical conditions we investigated naturally led to a `minimal' solution of this problem, which agrees with the gravity result.

\subsubsection*{Deriving chaos} 
A first-principles, bottom-up derivation of chaos in holographic CFTs still eludes us. While we now understand the relationship between quantum chaos and modular invariance better, quantum chaos is still a basic assumption that we show is consistent with other features of the 2d CFTs. This is contrasted with the wormhole amplitude in gravity, which can be derived from first principles. Some standard properties of holographic CFTs might be sufficient for such a derivation, in particular the assumptions that yield a dense spectrum above the extremal limit (large central charge, no conserved currents, and a twist gap). A promising path towards this would be the construction of an Efetov sigma model as in \cite{Altland:2020ccq,Belin:2021ibv}, similar to how chaos is derived in the SYK model \cite{Altland:2017eao}. It would be fascinating to see if such an approach can be adopted using recent discussions of random matrix ensembles for 2d CFT operator data and OPE coefficients, which furnish approximate solutions to the bootstrap equations \cite{Belin:2020hea,Chandra:2022bqq,Belin:2023efa}.

\vspace{5pt}
\subsubsection*{The plateau}

In chaotic quantum mechanics the universal form of eigenvalue correlations is expected to take the random matrix form for sufficiently close energy levels, depending on the universality class (see, e.g., \cite{Mirlin:2000cla}). For the GUE universality class, this is
\begin{equation}
\label{eq:RMTDensityCorrelator}
\langle \rho(E+\omega/2) \rho(E-\omega/2) \rangle= \langle \rho(E)\rangle^2+\langle \rho(E)\rangle \delta(\omega) - \frac{\sin^2 \left(\pi \omega \langle \rho(E)\rangle\right)}{(\pi \omega)^2}\,.
\end{equation}
The first term describes the disconnected part, the third the famous sine-kernel which gives rise to the ramp in the time domain. We now wish to discuss the second term, i.e., the tautological ``self-correlations'',  to provide comparison with the chaotic case -- eigenvalue repulsion and the ramp -- discussed before. In quantum chaotic systems this term gives rise to the eventual plateau for sufficiently long times (or equivalently for sufficiently close eigenvalues), but this term is even more universal as it also exists in integrable systems. Systems with Poissonian statistics are completely described by the first two terms in \eqref{eq:RMTDensityCorrelator}, up to non-universal terms at early times.\footnote{If we discuss multiple independent Hamiltonians, the self-correlations exist for identical matrices (tautologically) but are absent for distinct random matrices.}

While the ramp appears to a natural object in the spectral decomposition, and can be described as analogous to the  ``diagonal approximation'' \cite{DiUbaldo:2023qli} in a periodic orbit expansion (i.e., the correlations in spectral eigenvalues, $\alpha$ and $R_n^\pm$, are diagonal), we will see that the plateau is perhaps less natural. This is consistent with the analogy with the semi-classical periodic orbits, for which the plateau is non-perturbative. Note also that in gravity calculations, the plateau is much more difficult to obtain than the ramp; in JT gravity it arises from an infinite sum of wormhole geometries \cite{Blommaert:2022lbh,Saad:2022kfe}. We will now offer a few comments on the spectral decomposition of the plateau, leaving a full analysis for future work.

\paragraph{Self-correlations:}
First, we comment on the expected height of the plateau in a quantum chaotic system, and see how this is reproduced in our language with the fluctuating partition function $\widetilde{Z}_\text{P}^m(y)$. Recall from \eqref{eq:rhotildeDef} that the density of states for the fluctuating partition function is just the density of states for the dense spectrum minus its average, $\widetilde{\rho}_\text{P}(E)=\rho_D(E)-\langle\rho_D(E)\rangle$.
This means that the second and third terms in \eqref{eq:RMTDensityCorrelator} have to do with the correlation of the fluctuating partition function. Thus the height of the plateau we expect 
from considering the  fluctuating partition function is just that of a standard partition function (multiplied by $\frac{\sqrt{y_1 y_2}}{e^{\frac{\pi}{6}(y_1+y_2)}}$ from the definition of $\widetilde{Z}_\text{P}$). Focusing on the second term in \eqref{eq:RMTDensityCorrelator}:
\begin{equation}
    \begin{split}
     \langle \widetilde{\rho}_\text{P}^{\,m_1}(E_1) \widetilde{\rho}_\text{P}^{\,m_2}(E_2)\rangle_{\plt} &= \langle \rho_D^m(E_1) \rangle \delta(E_1-E_2) \delta_{m_1m_2}
    \\
    \Rightarrow \quad {\langle} \widetilde{Z}^{m_1}_\text{P}(y_1) \, \widetilde{Z}^{m_2}_\text{P}(y_2) {\rangle}_{\plt}&= \frac{\sqrt{y_1 y_2}}{e^{\frac{\pi}{6}(y_1+y_2)}}{\langle} Z^{m_1}_\text{P,D}(y_1+y_2){\rangle}\delta_{m_1m_2} \\
    &\equiv \frac{\sqrt{y_1 y_2}}{e^{\frac{\pi}{6}(y_1+y_2)}}\,\delta_{m_1m_2}\int_{E_{m_1}}^{\infty}dE\, \langle\rho_D^{m_1}(E)\rangle e^{-(y_1+y_2)E} \,,
    \end{split}
\end{equation}
where $\langle \rho_D^m(E_1) \rangle$ is the average density of spin $m$ Virasoro primaries.\footnote{
We use the average density from \cite{Mukhametzhanov:2020swe}, given by 
\begin{equation}
    \langle \rho_D^m(E_1) \rangle \approx\frac{1}{2\pi}\frac{2}{1+\delta_{m_1,0}}\frac{1}{\frac{E_1}{2\pi}+\frac{c}{12}}\exp{2\pi\sqrt{\frac{c-1}{3}\left(\frac{E_1}{2\pi}+\frac{c}{12}\right)}} \, . \label{eq:average-density-spin-m}
\end{equation}
}
Note that  the plateau coefficient is given by $Z_{\text{P,D}}^{m}$, which is {\it not} the modular invariant, fluctuating, dense partition function. It is just the standard partition function for the dense primaries of spin $m$; in particular it is not modular invariant. 

By taking $y_i\rightarrow \infty$, we can estimate the plateau height:\footnote{This is obtained via Laplace's method; the integral is dominated by the global maximum at $E_i=E_{m_i}$, as the local maximum for large $y_i$ lies outside the region of integration as long as $y_1+y_2 \gtrsim c\gg1$.}
\begin{equation}
\begin{split}
\label{eq:plateau-sff}
    \big\langle \widetilde{Z}_\text{P}^m(y_1) \widetilde{Z}_\text{P}^m(y_2) \big\rangle_{\plt}&  \approx \langle \rho_D^m(E_m) \rangle \frac{\sqrt{y_1 y_2}}{y_1+y_2}e^{-2\pi |m|(y_1+y_2)}.
\end{split}
\end{equation}
Comparing \eqref{eq:plateau-sff} to the ramp, we see that the ramp and plateau become equal to each other when $\sqrt{y_1 y_2} \sim T =  \langle \rho_D^m(E_m)\rangle \equiv \Delta(E_m)^{-1} $, i.e., when at times of order the inverse mean-level spacing at threshold energy, as expected from general considerations.

\paragraph{Spectral decomposition:}
We can now analyze how the plateau appears in the Eisenstein series; the cusp forms come with new technical issues, and we relegate their discussion to appendix \ref{sec:plateau-cusp-forms}. For the spin $0$ case, we plug \eqref{eq:plateau-sff} into the usual integral transform (similar to \eqref{eq:z-m-alpha}):
\begin{equation}
\begin{split}
    \big\langle {\color{colHighlight}z_{\frac{1}{2}+i\alpha_1}\, z_{\frac{1}{2}+i\alpha_2}} \big\rangle_{\text{spin }0\text{ plateau}}
   & \approx 2i \pi^2 \langle \rho_D^0(E_0) \rangle \frac{1}{\sinh(\pi \alpha_1)}\delta(\alpha_1+\alpha_2-i) \,.
 \end{split} \label{eq:alphaPlateau-spin0}
\end{equation}
The most interesting feature of this expression is that it is not diagonal in $\alpha_i$.\footnote{The appearance of unfamiliar delta function of a complex argument is due to our function space including functions that grow exponentially in $y$, see for example the discussion in \cite{Maxfield:2019hdt}.}

For the spin $m_i>0$ case, we find similarly:\footnote{This should be understood as a distribution, i.e., 
\begin{equation}
\int {\cal D}\frac{1}{x^2} \phi(x) \equiv \int \frac{1}{x^2}\left(\phi(x)-\phi(0)-x\phi'(0)\right) = \int -\log|x| \phi''(x) \, .
\end{equation}
Such a procedure is necessary as the Fourier transform of $|\xi|$ only makes sense as a distribution i.e. when integrated against test functions, and the resulting distribution cannot be defined without some method of regularizing the singularity at $\alpha_1 \pm \alpha_2=0$. The first method is by subtracting the first two terms in the Taylor series so that the singularity becomes removable; the second is to use integration by parts and discard boundary terms, which makes the singularity integrable.}
\begin{equation}
\label{eq:alphaPlateau-cont}
    \langle {\color{colHighlight2} z^{m_1}(\alpha_1)z^{m_2}(\alpha_2)}\rangle_\text{plateau} \approx -4\pi^2 m \langle \rho_D^m(E_m) \rangle {\cal D}\left( \frac{1}{\left(\alpha_1-\alpha_2\right)^2}+\frac{1}{\left(\alpha_1+\alpha_2\right)^2}\right)\delta_{m_1 m_2} \quad (\alpha_i \rightarrow \infty)
\end{equation}
Again, the correlations for the plateau in any spin sector are not diagonal (there is no delta-function imposing $\alpha_1 = \pm \alpha_2$). From the perspective of \cite{DiUbaldo:2023qli}, this means the plateau does not come from the diagonal approximation analogous to the semi-classical periodic orbits, as one would expect.

Similar to our discussion of the ramp, we can ask if \eqref{eq:alphaPlateau-cont} should be improved by imposing a plateau consistently across all spin sectors. We leave such an analysis to the future, but discuss the question of the imprint of a plateau in a given spin sector onto other spin sectors, using numerical evidence, in appendix \ref{sec:plateau-independence}.

\vspace{10pt}
\section*{Acknowledgments}

We thank Scott Collier and especially Eric Perlmutter for enlightening discussions and comments.
We are also grateful to Holger Then for sharing extensive numerical data on Maass cusp forms with us.
F.H.\ is supported by the UKRI Frontier Research Guarantee Grant [EP/X030334/1]. F.H.\ is grateful for the hospitality of Perimeter Institute, where part of this work was finalized. M.R.\ and W.R.\ are supported by a Discovery Grant from NSERC.

\vspace{10pt}
\appendix

\newpage
\section{Notation and conventions}
\label{app:notation}

In this appendix we collect some conventions and useful formulae. 
We consider the spectral decomposition of the Laplacian on the fundamental domain ${\cal F} = \{ \tau = x+iy\,,\; y>0 \} / SL(2,\mathbb{Z})$, which admits continuous and discrete solutions:
\begin{equation}
    \Delta_{_{\cal F}} E_s(\tau) = s(1-s) E_s(\tau) \,,\qquad \Delta_{_{\cal F}} \nu_{n,\pm}(\tau) = \left( \frac{1}{4} + (R_n^\pm)^2 \right) \nu_{n,\pm}(\tau) \,,
\end{equation}
where the Eisenstein series and Maass cusp forms have the following Fourier decomposition:
\begin{equation}
\begin{split}
E_{s}(\tau=x+iy) &= \left[ y^{s} + \frac{\Lambda(1-s)}{\Lambda(s)} \, y^{1-s} \right] + \sum_{m\geq 1} \cos(2\pi m x)\, \frac{4\,\sigma_{2s-1}(m)}{m^{s-\frac{1}{2}}\Lambda\left( s\right)} \, \sqrt{y} K_{s-\frac{1}{2}}(2\pi m y)\,,\\
\nu_{n,\pm}(\tau=x+iy) &= \sum_{m\geq 1} \left\{ \begin{aligned}\cos(2\pi  m x) \\ \sin(2\pi m x) \end{aligned}\right\}\, a_m^{(n,\pm)} \, \sqrt{y} K_{iR_n^\pm} (2\pi m y) \,.
\end{split}
\end{equation}
The continuous eigenvalues are $s \equiv \frac{1}{2} + i \alpha$ with $\alpha\in  \mathbb{R}$, while $R_n^\pm > 0$ are discrete randomly distributed real numbers (see appendix \ref{app:maass} for details). We work with unnormalized cusp forms, satisfying $a_1^{(n,\pm)}=1$. We also define Fourier coefficients for the Eisenstein series, via $a_m^{(\alpha)} = 2m^{-i\alpha} \sigma_{2i\alpha}(m)$. (The Hecke eigenvalues are $\frac{1}{2} a_m^{(\alpha)}$.)

The spectral decomposition of a normalizable modular invariant function takes the form 
\begin{equation}
 f(\tau) = \int_{-\infty}^\infty \frac{d\alpha}{4\pi} \, {\color{colHighlight}\big( f,\, E_{\frac{1}{2}+i\alpha} \big)} \, E_{\frac{1}{2}+i\alpha}(\tau) + \sum_\pm\sum_{n\geq 0} {\color{colHighlight}\frac{(f,\,\nu_{n,\pm})}{|\!|\nu_{n,\pm}|\!|^2} }\, \nu_{n,\pm}(\tau) \,. 
\end{equation}
where the Petersson inner product is $(f,g) \equiv \int_{\cal F} dxdy \, y^{-2} \, f \, \bar{g}$.
In particular:
\begin{equation}
  (f,E_{\frac{1}{2}+i\alpha}) = \int_{\cal F} \frac{dxdy}{y^2} \, f(x+iy) E_{\frac{1}{2}-i\alpha}(x-iy) = \int_0^\infty dy \, y^{-\frac{3}{2} - i \alpha} \, f^{m=0}(y)\,.
\end{equation}

\vspace{5pt}
\section{Dominant regime of eigenvalues in \eqref{eq:discSumRes}}
\label{app:errorEstimate}

In this appendix we elaborate on the dominance of large $R_n^\pm$ as $y_i \rightarrow \infty$ in the evaluation of the cusp form sum \eqref{eq:discSumRes}. 
As functions of $y$, the Bessel functions $K_{iR_n^\pm}(2\pi m y)$ have strong oscillations for $0<2\pi m y \lesssim R_n^\pm$, with amplitude of order $\sqrt{2\pi/R_n^\pm} \,e^{-\pi R_n^\pm/2}$, and subsequently decay exponentially like $\sqrt{1/(4my)}\,e^{-2\pi m y}$, independent of $R_n^\pm$. First, the exponential decay implies that the sum over $n$ converges and can thus be truncated in numerical evaluation. More non-trivially, the sum is dominated by terms with increasingly large values of $R_n^\pm$.
\begin{figure}
    \centering
\includegraphics[width=.49\textwidth]{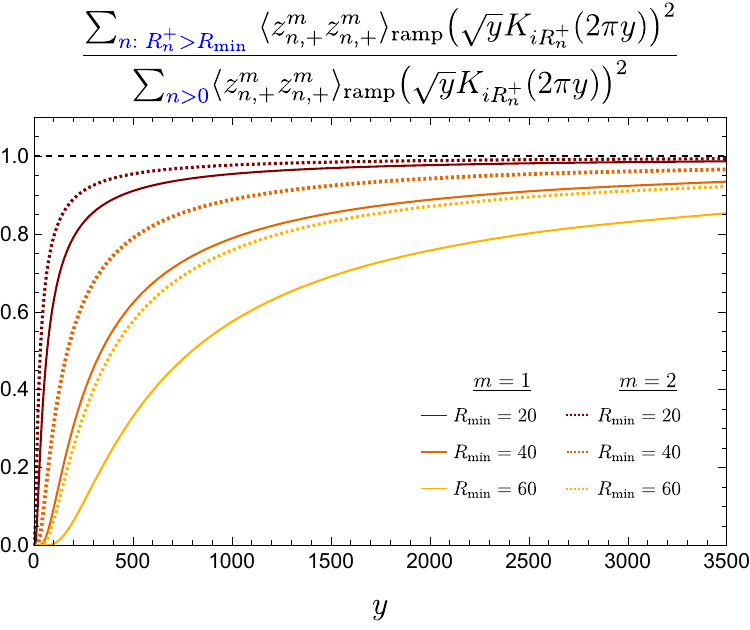}$\,$
\includegraphics[width=.49\textwidth]{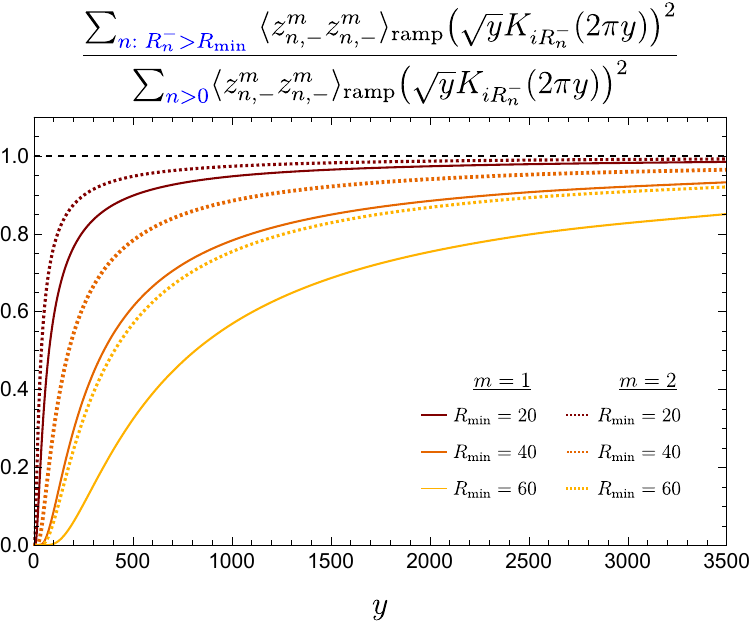}
    \caption{We quantify how the sum over cusp forms indexed by $n$ depends on the terms with small $n$ (and hence small $R_n^\pm$). We plot the ramp in the spectral form factor computed using only values of $n$ for which $R_n^\pm > R_\text{min}$ and normalize it by the complete result. Asymptotically as $y \rightarrow \infty$ this ratio converges to $1$, no matter how many low-lying values of $R_n^\pm$ we exclude. We show the cases of even (left) and odd (right) cusp forms separately (they are almost indistinguishable). Solid lines correspond to spin $m=1$, dashed lines to $m=2$.}
    \label{fig:convergence}
\end{figure}
We verify this numerically in figure \ref{fig:convergence}: we compare the ramp computed using all relevant terms in the sum with the partial result obtained by dropping all terms with $0<R_n^\pm<R_\text{min}$. We observe that the ratio of these two quantities approaches $1$ as $y \rightarrow \infty$, for any choice of $R_\text{min}$. Equivalently, any partial sum over only low-lying $R_n^\pm$ converges to a finite constant (times the usual $e^{-2\pi m (y_1+y_2)}$) as $y_i \rightarrow \infty$, as the Bessel functions become independent of $R_n^\pm$. This is therefore subleading to the ramp:
\begin{equation}
\label{eq:large-Rn-dominance-ramp}
\begin{split}
    \lim_{y_i \rightarrow \infty} e^{2\pi m (y_1+y_2)} &\;\sum_{n=1}^{n_{\text{max}}} \langle {\color{colHighlight2}z_{n,\pm}^m z^m_{n,\pm}}\rangle_\text{ramp} \,  \sqrt{y_1} K_{iR_n^\pm} (2\pi m y_1)\sqrt{y_2}K_{iR_n^\pm} (2\pi m y_2) 
    \\
    =&\; \frac{1}{4m}\sum_{n=1}^{n_{\text{max}}} \langle {\color{colHighlight2}z_{n,\pm}^m z^m_{n,\pm}}\rangle_\text{ramp} 
    \sim \frac{1}{4 \pi^2 m} \, \left( R_{n_\text{max}}^\pm\right)^2 \,,
\end{split}
\end{equation}
which holds for any $n_\text{max}$ as long as $y_i \gg \frac{1}{2\pi m} \, R_{n_\text{max}}^\pm$.
Effectively the sum over $n$ is dominated by a window $R_\text{min} \lesssim R_n^\pm \lesssim R_\text{max}$ where both $R_{\text{min}}$ and $R_\text{max}$ increase with $y_i$. This justifies using the continuous approximation, i.e., treating the eigenvalues $R_n^\pm$ statistically for large $y_i$.

To summarize, the sum \eqref{eq:discSumRes} can be (roughly) split into three pieces, which qualitatively contribute as follows to the spectral form factor:
\begin{equation}
    \begin{split}
    (1)&\;\; 0<R_n^\pm \lesssim R_\text{min}(y_i): \qquad\qquad\quad\; \text{subleading constant} \\
    (2)&\;\;  R_\text{min}(y_i)\lesssim R_n^\pm \lesssim R_\text{max}(y_i): \qquad \text{linear ramp} \\
    (3)&\;\;  R_\text{max}(y_i) \lesssim R_n^\pm : \qquad\qquad\qquad\quad \text{exponentially small} \\
    \end{split}
    \label{eq:sumSplit}
\end{equation}
To understand the dependence of $R_\text{min}$ and $R_\text{max}$ on $y_i$, we study the dominant contributions to the sum \eqref{eq:discSumRes}. 
The integrand in the continuous approximation of the cusp form sum is $\bar{\mu}_\pm (R)\langle {\color{colHighlight2}z_{n,\pm}^m z^m_{n,\pm}}\rangle_\text{ramp}$, which grows monotonically with $n$, while the Bessel functions decay very slowly as functions of $R$ until $R \gtrsim 2 \pi m y_i$. This leads to an integrand that peaks at a value of $R$ that increases with $y_i$, in turn making the continuous approximation better.
This is best seen by approximating \eqref{eq:discSumRes} as an integral:
\begin{equation}
\begin{split}
    &\Big{\langle}  \widetilde{Z}^{m}_{\text{P,disc.,}\pm}(y_1) \widetilde{Z}^{m}_{\text{P,disc.,}\pm}(y_2) \Big{\rangle}_\text{ramp} 
     \approx \int_{R_{n_0}}^\infty dR\, 
     \frac{2R \,\tanh(\pi R)}{\pi^2}\,  \sqrt{y_1}K_{iR}(2\pi m y_1)\,\sqrt{y_2} K_{iR}(2\pi m y_2)
    \end{split}
\label{eq:discSumResInt}
\end{equation}
Instead of evaluating this analytically, we consider the integrand ${\cal I}$ as a function of $R$ for fixed $y_i$. Initially the integrand grows linearly, ${\cal I}(R) \sim \frac{1}{2m \pi^2} \,R\, e^{-2\pi m (y_1+y_2)}$. The integrand reaches a maximum at $R_*\sim \frac{\pi}{2} \sqrt{\frac{2my_1y_2}{y_1+y_2}}$ where its value scales as ${\cal I}(R_*)\sim \frac{1}{2m \pi^2} \, R_* \, e^{-2\pi m (y_1+y_2)}$. The integrand then decays to zero polynomially and becomes negligible for values of $R$ greater than $R_\text{max}\sim 2\pi \sqrt{\frac{2my_1y_2}{y_1+y_2}}$. The choice of $R_\text{min}(y_i)$ corresponds to dropping a finite number of terms in the regime of linear growth. Since both the maximum of the integrand as well as the upper region of integration grow as $\sqrt{\frac{y_1y_2}{y_1+y_2}}$, we can drop terms with small $R$ less than $R_\text{min} \sim \sqrt{\frac{y_1y_2}{y_1+y_2}}$.\footnote{ This estimate ensures an error less than about $1\%$.}

\vspace{5pt}
\section{Statistics of Maass cusp forms: arithmetic chaos}
\label{app:maass}

We review some statistical facts about the Maass cusp forms, along with clarifying aspects that (to our knowledge) do not appear in the literature. Some of this information can also be found in the main text, repeated here for convenience. We use the first 5832 even and 6282 odd cusp forms for all numerics (this corresponds to $R_n^\pm < 400$) \cite{Then_2004}.

\begin{figure}
    \centering
\includegraphics[width=.6\textwidth]{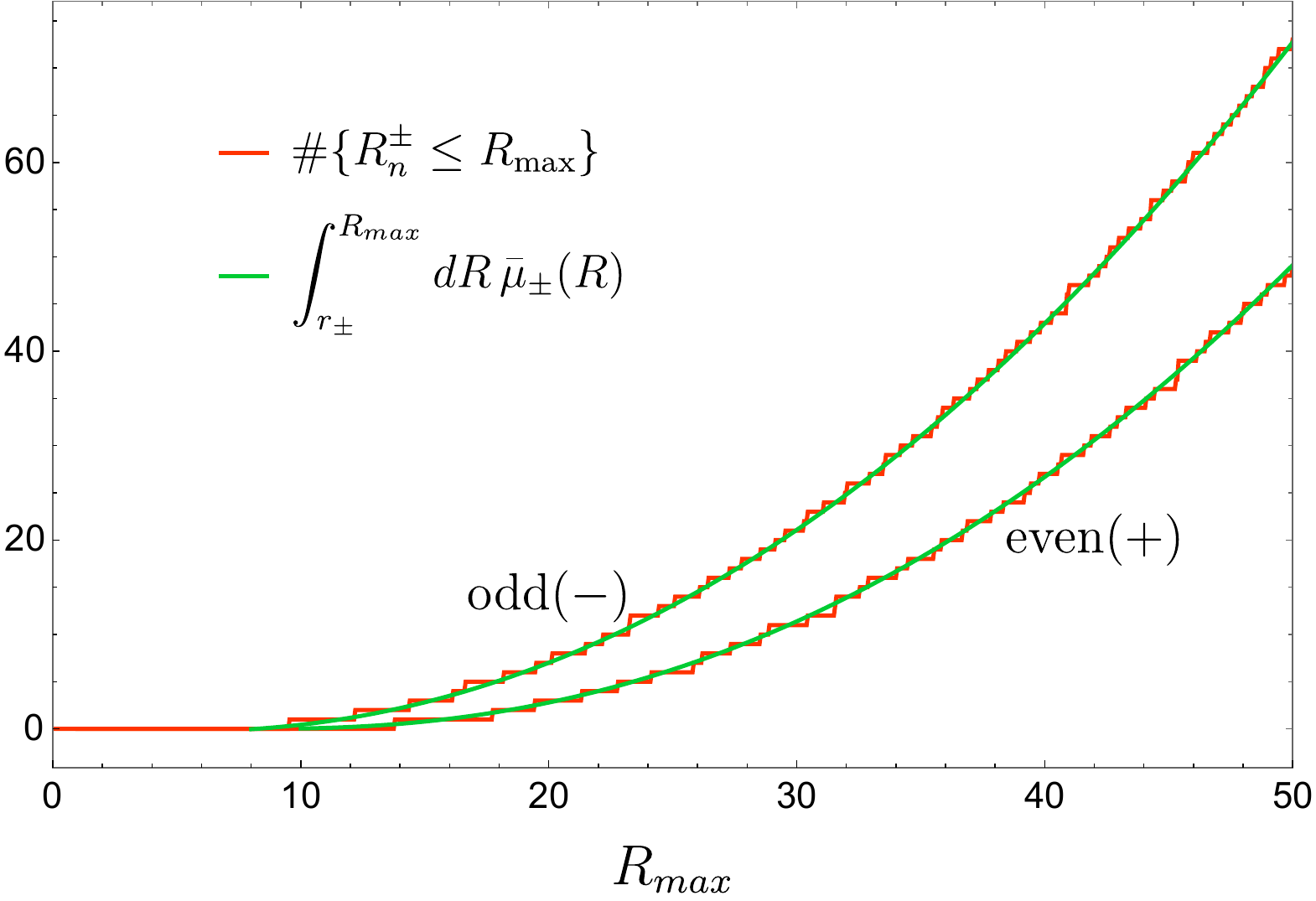}
    \caption{Comparison of the exact counting function of discrete eigenvalues of the Laplacian with the Weyl law approximation as given in \eqref{eq:muRapproxapp}.}
    \label{fig:appWeylLaw}
\end{figure}

The eigenvalues of cusp forms are distributed according to the {\it Weyl law} \cite{Steil:1994ue,PhysRevA.44.R7877}:
\begin{equation}
\label{eq:muRapproxapp}
\begin{split}
    \mu_+(R) &\approx \bar{\mu}_+(R) = \frac{1}{12} \, R - \frac{3}{2\pi} \, \log R + \frac{\log(\pi^4/2)}{4\pi} + {\cal O}\left(\frac{\log R}{R^2}\right)\,, \\
    \mu_-(R) &\approx \bar{\mu}_-(R) = \frac{1}{12} \, R - \frac{1}{2\pi} \, \log R - \frac{\log 8}{4\pi} + {\cal O}\left(\frac{\log R}{R^2}\right)\,.
\end{split}
\end{equation}
We illustrate this in figure \ref{fig:appWeylLaw}.

\begin{figure}
    \centering
\includegraphics[width=.48\textwidth]{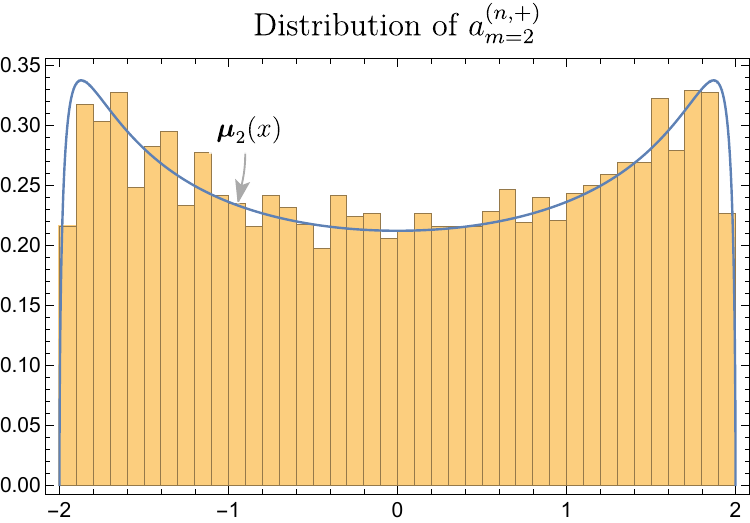}$\,$
\includegraphics[width=.48\textwidth]{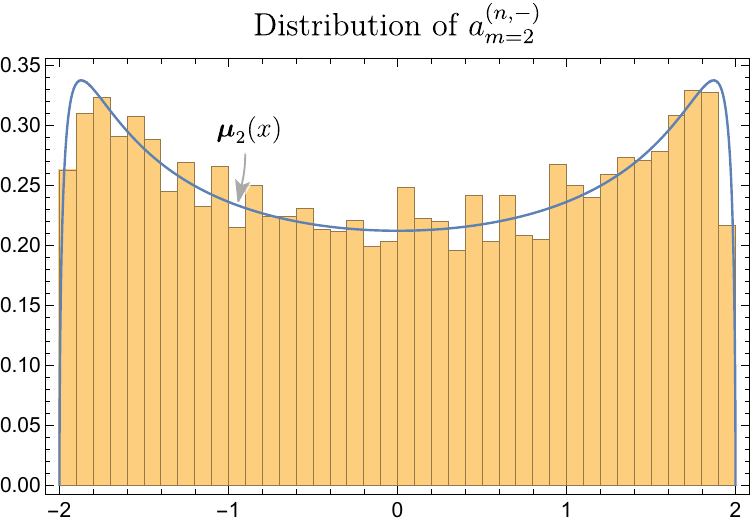}$\,$
\includegraphics[width=.48\textwidth]{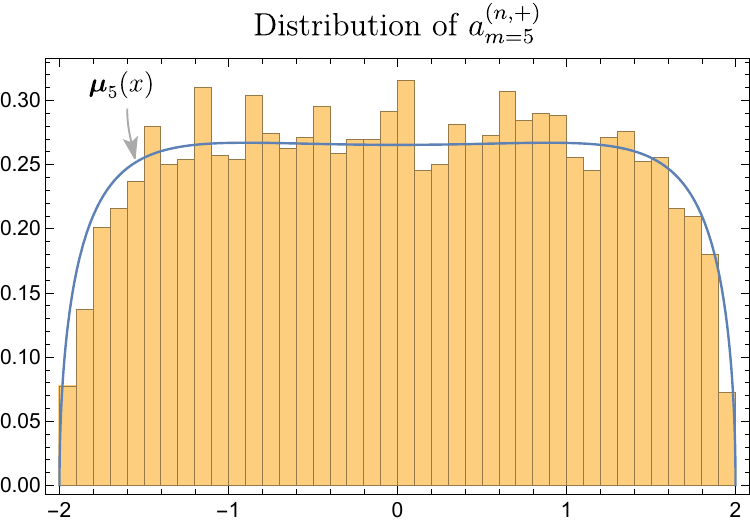}$\,$
\includegraphics[width=.48\textwidth]{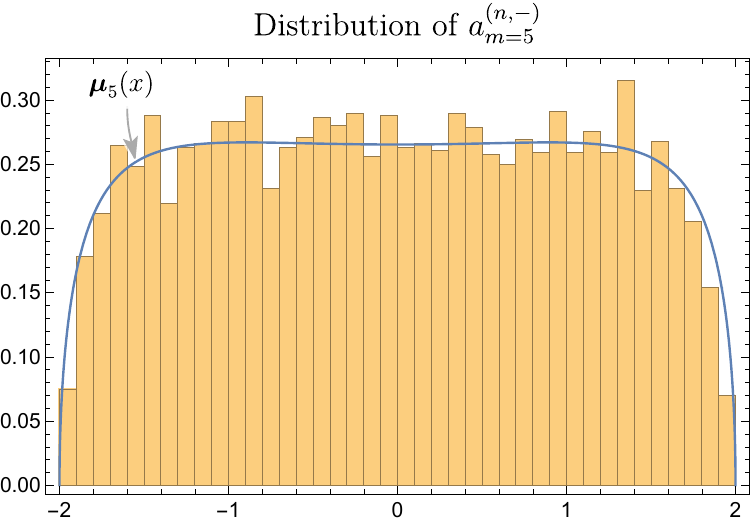}$\,$
\includegraphics[width=.48\textwidth]{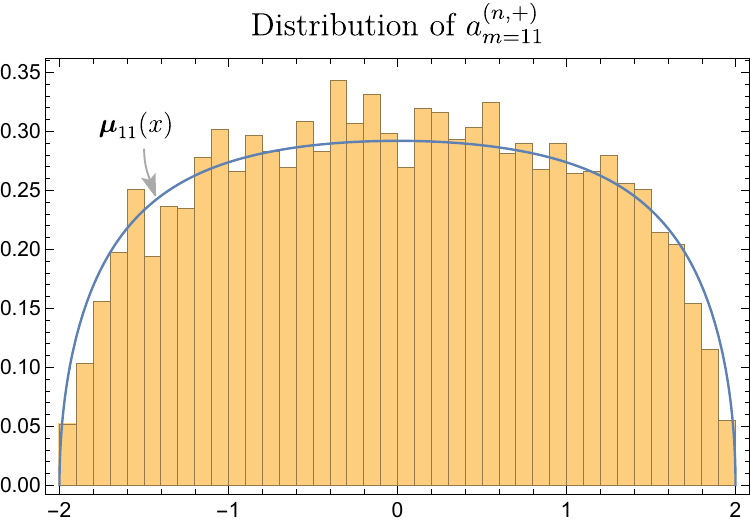}$\,$
\includegraphics[width=.48\textwidth]{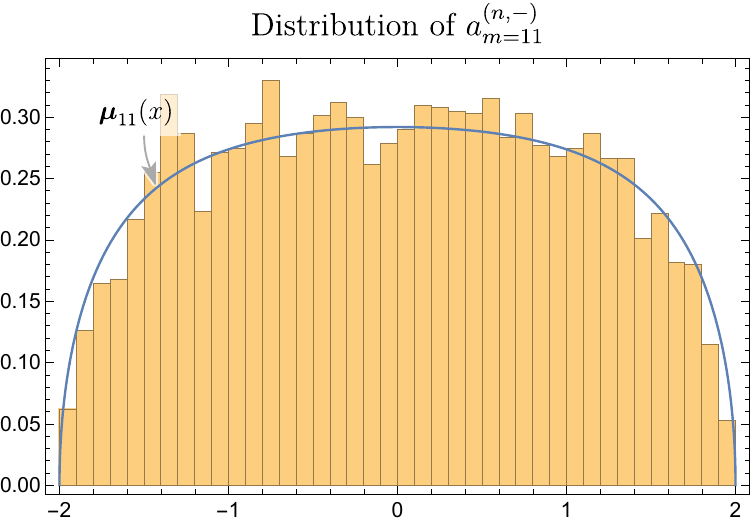}$\,$
    \caption{The distribution of the first 5832 even and 6282 odd Fourier coefficients for prime spins, compared to the distribution they are equidistributed with respect to.}
    \label{fig:distfouriercoeff}
\end{figure}

The Fourier coefficients $\{a_p^{(n,\pm)}\}$ for fixed prime spin $p$ and ordered by increasing corresponding eigenvalue $R_n^\pm$ are equidistributed according to the distributions \cite{sarnakStatisticalPropertiesEigenvalues1987,Steil:1994ue}
\begin{equation}
    \boldsymbol{\mu}_p(x)=
    \begin{cases}
        \frac{(p+1) \sqrt{4-x^2}}{2\pi \left( \left(p^{1/2}+p^{-1/2}\right)^2-x^2\right)} & \text{if } |x|<2 \\
        0 & \text{otherwise}
    \end{cases}
    \label{eq:coefficient_mu2}
\end{equation}  
which approaches the Wigner semi-circle $\frac{1}{2\pi}\sqrt{4-x^2}$ as $p\rightarrow \infty$. Equidistribution means that averages over all cusp forms for a given spin can be replaced with averages over the distribution, i.e.,
\begin{equation}
    \lim_{n_0\rightarrow \infty} \;\sum_{n=1}^{n_0} f\left(a_p^{(n,\pm)}\right) = \int dx \, \boldsymbol{\mu}_p(x) f(x).
\end{equation}
This is illustrated in figure \ref{fig:distfouriercoeff}.

We now investigate the nearest-neighbour level spacing, both for the eigenvalues $R_n^\pm$ and for the Fourier coefficients $a_m^{(n,\pm)}$. This provides a more numerically tractable way of analyzing the correlations than the density of states two-point function. After ``unfolding'' the spectrum,\footnote{Unfolding the spectrum corresponds to replacing each member $R_n^\pm \rightarrow x_n^\pm=\langle N^\pm(R_n^\pm)\rangle$. This yields $\langle N^\pm(R^\pm) \rangle = \int_{-\infty}^{R^\pm} dR'^\pm \, \overline{\mu}_\pm(R'^\pm)=\int_{-\infty}^{x^\pm} dx'^\pm  = x^\pm$, i.e., the spectrum has constant unit density in $x$ variables \cite{bohigasChaoticMotionRandom1984}.} 
we calculate the distribution of the difference between nearest-neighbour levels:
\begin{equation}
 \begin{split}
     P_{R_n^\pm}(s) \equiv \#\{x_n: \quad x_{n+1}-x_n = s\} \, .
 \end{split}    
\end{equation}
An integrable spectrum is distributed according to Poissonian statistics, $P_P(s)=e^{-s}$, which means level {\it attraction}: $P_P(s)\rightarrow 1$ as $s\rightarrow 0$. Chaotic spectra, on the other hand, are distributed according to the ensemble with appropriate symmetries, e.g., the Gaussian orthogonal ensemble with $P_\text{GOE}(s)=\frac{1}{2}\pi s e^{-\pi s^2/4}$, which exhibits level {\it repulsion}: $P_\text{GOE}(s)\rightarrow 0$ as $s\rightarrow 0$.

\begin{figure}
    \centering
\includegraphics[width=.48\textwidth]{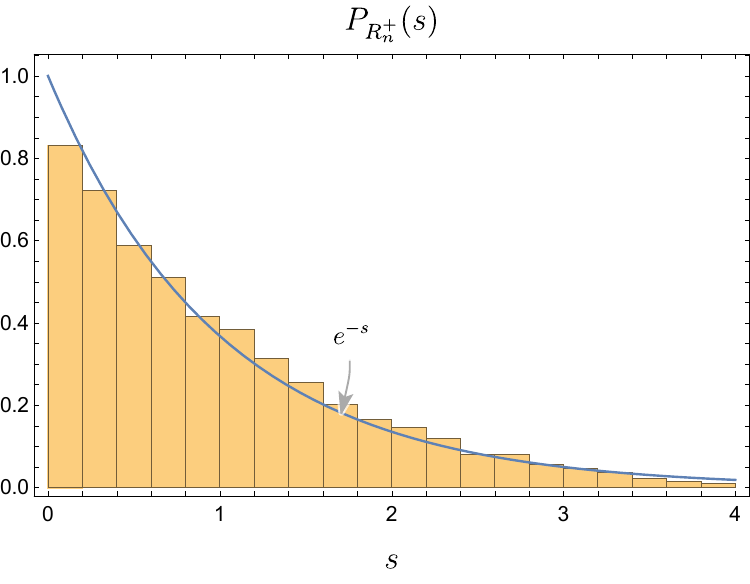}$\,$
\includegraphics[width=.48\textwidth]{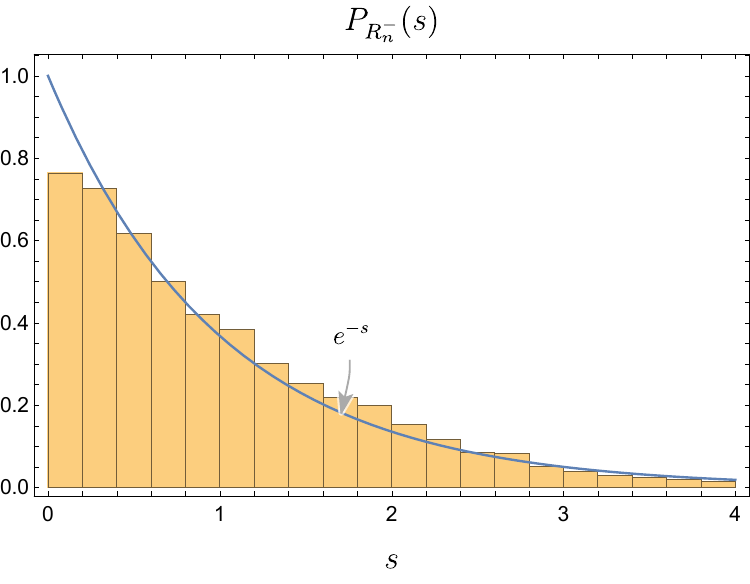}$\,$
\includegraphics[width=.48\textwidth]{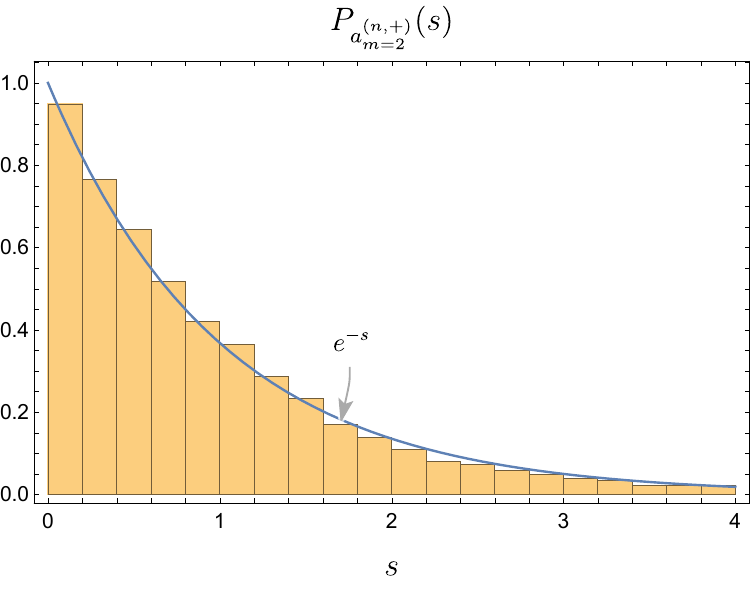}$\,$
\includegraphics[width=.48\textwidth]{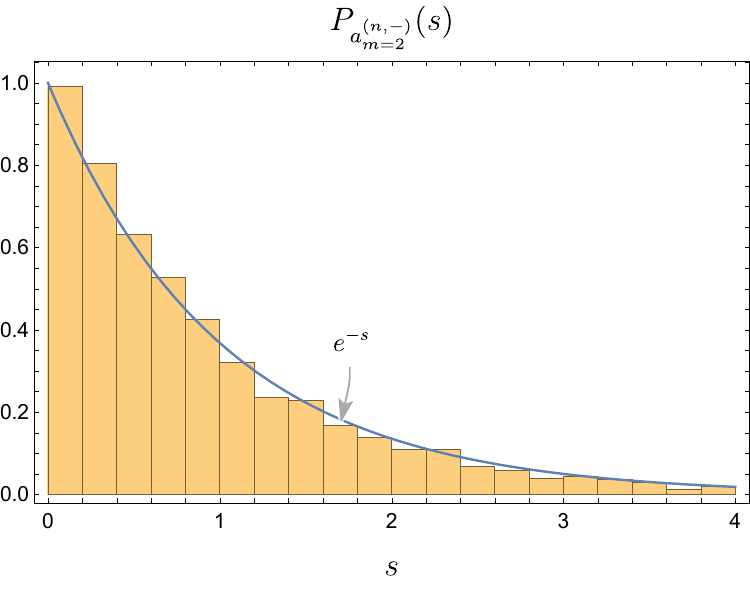}$\,$
    \caption{The distribution of the nearest neighbor spacing of the first 5832 even and 6282 odd eigenvalues and spin $2$ Fourier coefficients, compared to the Poissonian expectation. Note that the statistics for all other prime spins is similar.}
    \label{fig:distspacing}
\end{figure}

The eigenvalues of the cusp forms are known to obey Poissonian statistics \cite{bolteArithmeticalChaosViolation1992}, and we find that the Fourier coefficients for prime spin obey the same, shown in figure \ref{fig:distspacing}; hence, both the eigenvalues and Fourier coefficients are distributed {\it randomly but not chaotically}. Effectively, for any given spin $m$, the Fourier coefficients for different $n$ are independent random variables.\\

We can equivalently characterize the distributions \eqref{eq:coefficient_mu2} through their moments. For the distributions of prime spin Fourier coefficients, \eqref{eq:coefficient_mu2}, the odd moments are zero and the even moments are:\footnote{This formula can be easily derived by computing arbitrary moments of \eqref{eq:coefficient_mu2} and realizing that they correspond to an integral representation of the hypergeometric function.}
\begin{equation}
\boxed{
    p \text{ prime:}\qquad \overline{\big(a_p^{(n,\pm)}\big)^{2k}} = \frac{p}{p+1} \;\frac{(2k)!}{k!(k+1)!} \; {}_2F_1\left( 1,\,k+\tfrac{1}{2} , \, k+2 ,\, \tfrac{4p}{(p+1)^2} \right) \,.
    }
    \label{eq:momentsFormula}
\end{equation}
For example:
\begin{equation}
    \overline{\big(a_p^{(n,\pm)}\big)^{2}} = \frac{p+1}{p} \,,\qquad \overline{\big(a_p^{(n,\pm)}\big)^{4}} = \frac{(p+1)(2p+1)}{p^2}\,,\qquad \overline{\big(a_p^{(n,\pm)}\big)^{6}} = \frac{(p+1)(5p^2+4p+1)}{p^3}
\end{equation}
These distributions for Fourier coefficients are specifically for prime spins. All Fourier coefficients for non-prime (composite) spins are fully determined in terms of these by Hecke relations because Maas cusp forms are eigenfunctions of Hecke operators:
\begin{equation}
    T_m \nu_{n,\pm}(\tau) = a_m^{(n,\pm)} \, \nu_{n,\pm}(\tau)  \qquad \text{where} \qquad 
    T_m f(\tau) = \frac{1}{\sqrt{m}} \sum_{\substack{a,b,d:\\ad=m\\  0 \leq b \leq d-1}}  f\left(\frac{a\tau+ b}{d} \right)  \,.
\end{equation}
This implies immediately:
\begin{equation}
\label{eq:HeckeDef}
\boxed{\;
    a_m^{(n,\pm)} a_{m'}^{(n,\pm)} = \sum_{\substack{ \ell | (m,m') \\ \ell > 0 }} a^{(n,\pm)}_{\frac{mm'}{\ell^2}} \;}
\end{equation}
for example, if $p,p'$ are prime we get the important multiplicative relation: $a^{(n,\pm)}_p a^{(n,\pm)}_{p'} = a^{(n,\pm)}_{pp'} + \delta_{pp'}$ (see Lemma \ref{lemma:hecke} for more relations of this type).

The Hecke relations allow us to construct the non-prime Fourier coefficients from the prime ones. This in turn implies that the variances (`normalization factors' ${\cal N}_m^\pm$) of the distributions of Fourier coefficients for non-prime spins follow from higher moments of the prime distributions. We give a few examples:
\begin{equation}
\begin{split}
    a_4^{(n,\pm)} = \big( a_2^{(n,\pm)} \big)^2 - 1 \quad &\Rightarrow \quad {\cal N}_4^\pm \equiv \overline{\big(a_4^{(n,\pm)}\big)^2 } = \overline{\big( a_2^{(n,\pm)} \big)^4}  - 2 \, \overline{ \big( a_2^{(n,\pm)} \big)^2} +1 \\
     a_6^{(n,\pm)} =  a_2^{(n,\pm)}a_3^{(n,\pm)} \;\; \quad &\Rightarrow \quad {\cal N}_6^\pm \equiv \overline{\big(a_6^{(n,\pm)}\big)^2 } = \overline{\big(a_2^{(n,\pm)}\big)^2}\;\overline{ \big(a_3^{(n,\pm)}\big)^2 } \\
     a_8^{(n,\pm)} = \big(a_2^{(n,\pm)}\big)^3 - 2 \, a_2^{(n,\pm)} \quad &\Rightarrow \quad {\cal N}_8^\pm \equiv \overline{\big( a_8^{(n,\pm)} \big)^2} = \overline{\big(a_2^{(n,\pm)} \big)^6} - 4 \; \overline{\big(a_2^{(n,\pm)} \big)^4} + 4 \; \overline{\big(a_2^{(n,\pm)} \big)^2}   
\end{split}
\label{eq:HeckeExamples}
\end{equation}
\begin{table}
\renewcommand{\arraystretch}{1.1}
{\small
  \begin{center}
    \begin{tabular}{c||c|c|c|c|c|c}
    \textbf{spin $m$} & $\overline{\big( a_m^{(n,\pm)} \big)}$ & $\overline{\big( a_m^{(n,\pm)} \big)^2}$
& $\overline{\big( a_m^{(n,\pm)} \big)^3}$ & $\overline{\big( a_m^{(n,\pm)} \big)^4}$ & $\overline{\big( a_m^{(n,\pm)} \big)^5}$ &$\overline{\big( a_m^{(n,\pm)} \big)^6}$ 
\\
         \hline\hline 
      2 & 0 & $\frac{3}{2} = 1.5$ & 0 & $\frac{15}{4} = 3.75$ & 0 & $\frac{87}{8} = 10.88..$  
      \\ 
      3 & 0 & $\frac{4}{3} = 1.33..$ & 0& $\frac{28}{9} = 3.11..$ & 0 & $\frac{232}{27} = 8.59..$ 
      \\ 
      4 & $\frac{1}{2} = 0.5$ & $\frac{7}{4} = 1.75$ & $\frac{25}{8} = 3.16..$ & $\frac{127}{16} = 7.94..$ & $\frac{601}{32} = 18.78..$ & $\frac{3055}{64} = 47.73..$ 
      \\
      5 & 0 & $\frac{6}{5} = 1.2$& 0 & $\frac{66}{25}=2.64$ & 0 & $\frac{876}{125} = 7.01..$ 
      \\ 
      6 & 0 & $2$ & 0 & $\frac{35}{3} =11.67..$ & 0 & $\frac{841}{9} = 93.44..$ 
      \\ 
      7 & 0 & $\frac{8}{7}= 1.14..$ & 0 & $\frac{120}{49}=2.45..$ & 0 & $\frac{2192}{343} = 6.39..$ 
      \\ 
      8 & 0 & $\frac{15}{8} = 1.88..$ & 0 & $\frac{831}{64}=12.98..$ & 0 & $\frac{67935}{512} = 132.7..$  
      \\ 
      9 & $\frac{1}{3} = 0.33..$ & $\frac{13}{9} = 1.44..$  & $\frac{61}{27}=2.26..$  & $\frac{469}{81} = 5.79..$ & $\frac{3181}{243} = 13.09..$ & $\frac{23857}{729} = 32.73..$ 
      \\ 
    \end{tabular}
  \end{center}
  }
    \caption{Exact values of the moments of the distributions of Fourier coefficients. For prime $m$, these follow from \eqref{eq:momentsFormula}. For composite $m$, they are constructed from prime moments using Hecke relations.}
    \label{tab:table1}
\end{table}
\begin{table}
\renewcommand{\arraystretch}{1.4}
  \begin{center}
    \begin{tabular}{c||c|c|c|c|c|c}
    \textbf{spin $m$}& $\overline{\big( a_m^{(n,\pm)} \big)}$ & $\overline{\big( a_m^{(n,\pm)} \big)^2}$
& $\overline{\big( a_m^{(n,\pm)} \big)^3}$& $\overline{\big( a_m^{(n,\pm)} \big)^4}$& $\overline{\big( a_m^{(n,\pm)} \big)^5}$ & $\overline{\big( a_m^{(n,\pm)} \big)^6}$ 
\\
         \hline \hline
      2 & $\substack{0.01.. (+)\\-0.01.. (-)}$ & $\substack{1.46.. (+)\\1.47.. (-)}$ & $\substack{0.02.. (+)\\-0.02.. (-)}$ & $\substack{3.56.. (+)\\3.62..(-)}$& $\substack{0.09.. (+)\\-0.08.. (-)}$  & $\substack{10.14..(+)\\10.35..(-)}$  
      \\ 
      3& $\substack{0.01.. (+)\\-0.01.. (-)}$  & $\substack{1.27..(+)\\1.30..(-)}$ & $\substack{0.03.. (+)\\-0.03.. (-)}$ & $\substack{2.87..(+)\\2.95..(-)}$ & $\substack{0.10.. (+)\\-0.09.. (-)}$ & $\substack{7.69..(+)\\7.97..(-)}$ 
      \\ 
      4 & $\substack{0.46.. (+)\\0.47.. (-)}$ & $\substack{1.65..(+)\\1.68..(-)}$ & $\substack{2.82.. (+)\\2.91.. (-)}$ & $\substack{7.10..(+)\\7.31..(-)}$ & $\substack{16.42.. (+)\\16.97.. (-)}$ & $\substack{41.08..(+)\\42.53..(-)}$ 
      \\
      5 & $\substack{0.01.. (+)\\-0.01.. (-)}$ & $\substack{1.13..(+)\\1.16..(-)}$ & $\substack{0.03.. (+)\\-0.03.. (-)}$ & $\substack{2.38..(+)\\2.45..(-)}$& $\substack{0.10.. (+)\\-0.09.. (-)}$  & $\substack{6.08..(+)\\6.28..(-)}$ 
      \\ 
      6 & $\substack{0.01.. (+)\\-0.01.. (-)}$ & $\substack{1.84..(+)\\1.89..(-)}$ & $\substack{0.11.. (+)\\-0.10.. (-)}$ & $\substack{9.99..(+)\\10.37..(-)}$ & $\substack{1.39.. (+)\\-1.21.. (-)}$ & $\substack{74.60..(+)\\77.76..(-)}$ 
      \\ 
      7 & $\substack{0.01.. (+)\\-0.01.. (-)}$ & $\substack{1.07..(+)\\1.09..(-)}$ & $\substack{0.03.. (+)\\-0.03.. (-)}$ & $\substack{2.19..(+)\\2.25..(-)}$ & $\substack{0.10.. (+)\\-0.09.. (-)}$ & $\substack{5.47..(+)\\5.65..(-)}$ 
      \\ 
      8 & $\substack{0.01.. (+)\\-0.01.. (-)}$ & $\substack{1.72..(+)\\1.76..(-)}$ & $\substack{0.11.. (+)\\-0.11.. (-)}$ & $\substack{10.86..(+)\\11.28..(-)}$ & $\substack{1.19.. (+)\\-1.19.. (-)}$ &  $\substack{104.5..(+)\\108.9..(-)}$ 
      \\ 
      9 & $\substack{0.27.. (+)\\0.30.. (-)}$ & $\substack{1.32..(+)\\1.36..(-)}$ & $\substack{1.90.. (+)\\2.00.. (-)}$ & $\substack{4.85..(+)\\5.08..(-)}$ & $\substack{10.52.. (+)\\11.10.. (-)}$ &  $\substack{25.72..(+)\\27.17..(-)}$ 
      \\ 
    \end{tabular}
  \end{center}
    \caption{Numerical values of the moments of the distributions of Fourier coefficients, computed using the Fourier coefficients for cusp forms with eigenvalue $R_n^\pm < 400$ (separately for even and odd parity).}
    \label{tab:table2}
\end{table}
We give exact analytical and numerical values for some of these moments in tables \ref{tab:table1} and \ref{tab:table2}. 
The composite spins $m=4$ and $m=8$ are special cases of a general result, see Lemma \ref{lemma:averages}.
The numerical values for the second moments  are within a few percent of the theoretical values. This error increases for higher moments due to the limited number of cusp forms available numerically. The numerical results for odd forms are consistently slightly better because we have more of them available.

\begin{figure}
    \centering
\includegraphics[width=.7\textwidth]{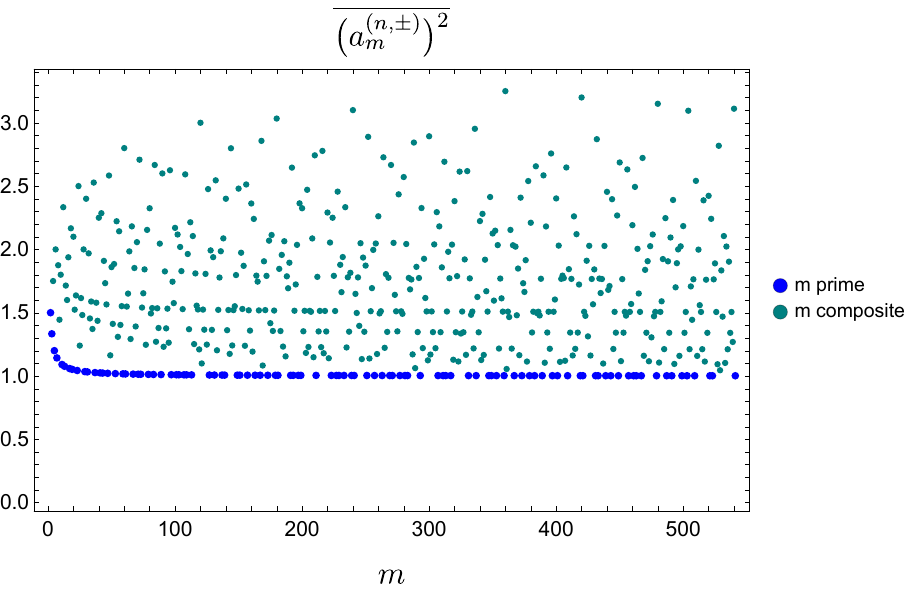}
    \caption{Exact variance of the Fourier coefficients as a function of $m$, up to the 100th prime. The prime coefficients have variance $1+\frac{1}{m}$, while the composite primes have much more complicated behaviour (which is determined by the Hecke relations and higher moments of the distribution of prime Fourier coefficients); in this range, all variances are  ${\cal O}(1)$, but as $m$ grows, the maximum possible value grows as well.}
    \label{fig:variance}
\end{figure}

More generally, computing just the variances ${\cal N}^\pm_m$ for {\it all} non-prime $m$ requires knowledge of {\it all} moments of the distributions of prime coefficients. Since the distributions are bounded, their moments determine the distributions fully. In other words, knowledge of {\it all} variances ${\cal N}_m^\pm$ is equivalent to complete knowledge of all the prime distributions \eqref{eq:coefficient_mu2}.
We show the variance of the Fourier coefficients for a large number of spins in figure \ref{fig:variance}. 

\vspace{5pt}
\section{Hecke relations, cusp form norms, and $L$-functions}
\label{app:norms}

In this appendix we provide some mathematical details relating to the norms of cusp forms and their relation to objects in analytic number theory. In order to be pedagogical, we provide step-by-step proofs of some important statements. Most of the general definitions and results can be found in the literature, see, for example, \cite{motohashi1997,shimura,sankaranarayanan,hoffstein}.

\subsection{Hecke relations and $L$-functions}

We first note some useful relations between Fourier coefficients of prime power spins.

\begin{lemma}
\label{lemma:hecke}
{\it Let $p,p_1,\ldots p_r$ be distinct primes. Then:
\begin{equation}
    \begin{split}
       (i)& \quad a_{p_1^{k_1}\cdots p_r^{k_r}}^{(n,\pm)} = a_{p_1^{k_1}}^{(n,\pm)} \cdots a_{p_r^{k_r}}^{(n,\pm)}  \\
       (ii)& \quad a^{(n,\pm)}_{p^{k}}  = a^{(n,\pm)}_{p^{k-1}} \, a^{(n,\pm)}_{p} -  (1-\delta_{k,1}) \, a^{(n,\pm)}_{p^{k-2}}  \\
    (iii)&\quad a_{p^{k}}^{(n,\pm)} = \frac{1}{2^k}\sum_{\ell=0}^{\lfloor k/2\rfloor}  {k+1\choose2\ell+1} \sum_{r=0}^\ell  {\ell \choose \,r\,} (-4)^{r}\big(a_p^{(n,\pm)}\big)^{k-2r} \\
(iv)& \quad \big(a_{p^{k}}^{(n,\pm)}\big)^2 = \frac{1}{2^{2k+1}} \sum_{\ell=0}^{k}  \left[ {2k+2\choose 2\ell+2} + (-1)^\ell {k+1 \choose \ell+1} \right] \sum_{r=0}^{\ell}{\ell \choose \,r\,} (-4)^r \, \big(a_{p}^{(n,\pm)}\big)^{2(k-r)}
    \end{split}
\end{equation}
}
\end{lemma}
\noindent{\it Proof:} $(i)$ and $(ii)$ follow immediately from the Hecke algebra \eqref{eq:HeckeDef}. $(iii)$ follows by viewing $(ii)$ as a recursion relation and solving it in terms of $a_p^{(n,\pm)}$. $(iv)$ follows from squaring and simplifying $(iii)$.$\qed$\\

Let us define the following Hecke $L$-functions for any of the cusp forms, defined by its Fourier coefficients:
\begin{equation}
    L^{(n,\pm)}(s) \equiv \sum_{m\geq 1} \frac{a_m^{(n,\pm)}}{m^s} \qquad (\text{Re}(s)>1)\,.
    \label{eq:Ldef}
\end{equation}
These $L$-functions are absolutely convergent in an $s$-half plane and they admit an analytic continuation to an entire function on the whole complex plane (see, e.g., \cite{sankaranarayanan}).

\begin{lemma}
    \label{lemma:Lfunc}
{\it The $L$-function \eqref{eq:Ldef} admits an Euler product representation:}
\begin{equation}
    L^{(n,\pm)}(s)  
    = \prod_{p\; \text{prime}} \frac{1}{ 1 - a_p^{(n,\pm)} \, p^{-s} + p^{-2s}}\,.
\end{equation}
\end{lemma}
\noindent{\it Proof:} 
Note that, as a consequence of the Hecke relations, we have
\begin{equation}
   \left[ 1 - a_p^{(n,\pm)} \, p^{-s} + p^{-2s} \right] \sum_{k\geq 0} a_{p^{k}}^{(n,\pm)}\,p^{-ks}
   =
   1 \,.
\end{equation}
The Euler product can then be written as a sum using Lemma \ref{lemma:hecke}(i):
\begin{equation}
    \begin{split}
    \prod_{p\; \text{prime}} \frac{1}{ 1 - a_p^{(n,\pm)} \, p^{-s} + p^{-2s}}
    &= 
    \prod_{p\; \text{prime}} \left( \sum_{k\geq 0} a_{p^{k}}^{(n,\pm)}\,p^{-ks} \right) = \sum_{m\geq 1} \frac{a_m^{(n,\pm)}}{m^s} \,.\qquad\qed
    \end{split}
\end{equation}
\newline

To make contact with the cusp form norms, we now consider the `symmetric square $L$-function', defined as\footnote{ More generally, for $\alpha_p,\beta_p \in \mathbb{C}$ satisfying
\begin{equation}
    1 - a_p^{(n,\pm)} \, p^{-s} + p^{-2s} = \left( 1 - \alpha_p p^{-s} \right) \left( 1 - \beta_p p^{-s}\right) \,,
\end{equation}
i.e., $\alpha_p + \beta_p = a_p^{(n,\pm)}$ and $\alpha_p \beta_p = 1$, the Ramanujan-Petersson conjecture asserts $|\alpha_p| = |\beta_p|=1$.
The symmetric $\ell$-th power $L$-function is then an automorphic function \cite{newton2021symmetric}, defined as 
\begin{equation}
    L_{\nu^\ell}^{(n,\pm)}(s)  
    = \prod_{p \text{ prime}} \,\prod_{k=0}^\ell \frac{1}{\left( 1 - \alpha_p^{\ell-2k} p^{-s}\right)} \,.
\end{equation}
The symmetric square $L$-function is the special case $\ell=2$. We suspect that $\ell$-th power $L$-functions play a role in the computation of higher moments of the CFT partition function.
}
\begin{equation}
    L_{\nu\times\nu}^{(n,\pm)}(s) \equiv \zeta(2s) \sum_{m\geq 1} \frac{a_{m^2}^{(n,\pm)}}{m^s} \qquad (\text{Re}(s)>1)\,.
\end{equation}
\begin{lemma}
    \label{lemma:Lsym}
\textit{The symmetric square $L$-function admits an Euler product representation:}
\begin{equation}
    L_{{\nu\times\nu}}^{(n,\pm)}(s)  
    = \prod_{p \text{ prime}} \frac{1}{1-\big(a_p^{(n,\pm)}\big)^2 \big( p^{-s} - p^{-2s} \big) + \left(p^{-s} - p^{-2s} - p^{-3s}\right)}
\end{equation}
\end{lemma}
\noindent{\it Proof:} The proof is the same as for Lemma \ref{lemma:Lfunc}, but starting from the observation that the following product simplifies:
\begin{equation}
   \left[1-\big(a_p^{(n,\pm)}\big)^2 \big( p^{-s} - p^{-2s} \big) + \left(p^{-s} - p^{-2s} - p^{-3s}\right)\right] \sum_{k\geq 0} a_{p^{2k}}^{(n,\pm)}\,p^{-ks}
   =
   1 - p^{-2s} \,,
\end{equation}
and recalling that $\prod_{p} (1-p^{-2s})^{-1} = \zeta(2s)$. $\qed$\\

\begin{figure}
    \centering
\includegraphics[width=.7\textwidth]{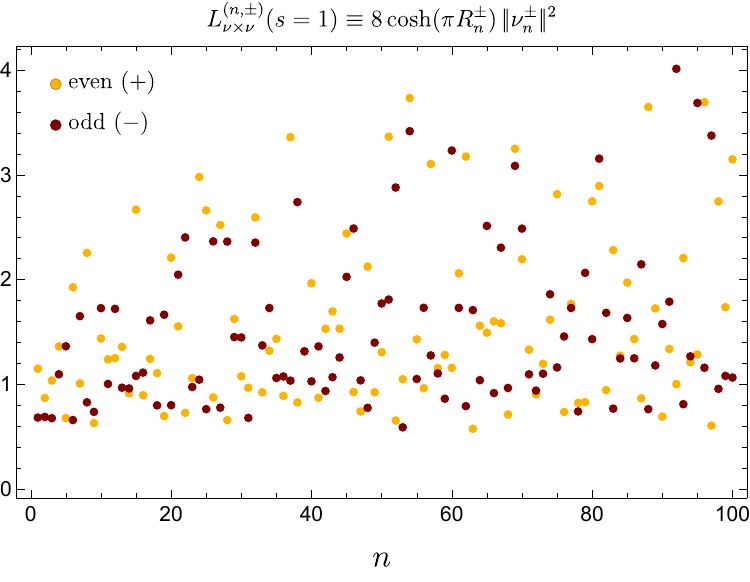}
    \caption{Plot of the even  and odd cusp form norms, rescaled by $8\cosh(\pi R_n^\pm)$. The norms themselves decay exponentially: $|\!|\nu_1^+|\!|^2 \approx 4.54 \times 10^{-20},\ldots , |\!|\nu_{100}^+|\!|^2 \approx 1.42\times 10^{-90}$ and $|\!|\nu_1^-|\!|^2 \approx 1.67 \times 10^{-14},\ldots , |\!|\nu_{100}^-|\!|^2 \approx 2.86\times 10^{-79}$. This is equivalently computable through the symmetric square $L$-function.}
    \label{fig:norms}
\end{figure}

\begin{tcolorbox}[width=\textwidth,colback=white]
\begin{theorem}
    \label{thm:norms}
{\it The norms of the cusp forms satisfy:}
\begin{equation}
    |\!|\nu_{n,\pm}|\!|^2 \equiv (\nu_{n,\pm} , \nu_{n,\pm}) = \frac{1}{8 \cosh(\pi R_n^\pm)} \, 
    L^{(n,\pm)}_{\nu\times\nu}(1) \,.
\end{equation}
\end{theorem}
\end{tcolorbox}
\noindent{\it Proof:} (See, e.g., references \cite{blomer2019symplectic,blomerSecondMomentTheory2019}.) We compute the norm using the Rankin-Selberg trick, i.e., note that a constant expression can be computed as the residue at $s=1$ with an Eisenstein series, which in turn allows for unfolding of the fundamental domain:
\begin{equation}
    \begin{split}
    |\!|\nu_{n,\pm}|\!|^2 
    &= \frac{\pi}{3}\, \text{Res}_{s=1} \big( |\nu_{n,\pm}(\,\cdot\,)|^2,\,E_s(\,\cdot\,) \big)
    \\
    &= \frac{\pi}{3}\,\text{Res}_{s=1}\int_{\cal F} \frac{dxdy}{y^2}\, |\nu_{n,\pm}(x+iy)|^2 \, E_s(x+iy) \\
    &=\frac{\pi}{3}\,\text{Res}_{s=1} \int_0^\infty dy \, y^{s-2} \, \left( \int_{-\frac{1}{2}}^{\frac{1}{2}} dx \, |\nu_{n,\pm}(x+iy)|^2 \right)\\
    &= \frac{\pi}{6}\, \text{Res}_{s=1}  \, \sum_{m\geq 1} \big( a_m^{(n,\pm)} \big)^2 \int_0^\infty dy \, y^{s-1}\,\big( K_{iR_n^\pm}(2\pi m y)\big)^2 \\
    &= \frac{\pi}{6}\, \text{Res}_{s=1}  \, \sum_{m\geq 1} \big( a_m^{(n,\pm)} \big)^2 \, \frac{\Gamma\left( \frac{s}{2} + i R_n^\pm \right)\Gamma\left( \frac{s}{2} - i R_n^\pm \right) \Gamma\left( \frac{s}{2}\right)^2}{8(\pi m)^s \Gamma(s)} \\
    &= \frac{\pi^2}{48\cosh(\pi R_n^\pm)}\, \text{Res}_{s=1} \sum_{m\geq 1}  \frac{\big( a_m^{(n,\pm)} \big)^2}{ m^s} \,.
    \end{split}
\end{equation}
To evaluate the residue, we note that the sum in the last line is related to a Rankin-Selberg zeta function and has an Euler product formula (\cite{sankaranarayanan}, Lemma 3.1 with $k=1$),
\begin{equation}
    \zeta(2s)\sum_{m\geq 1} \frac{\big( a_m^{(n,\pm)} \big)^2}{ m^s} = \zeta^2(s)\prod_{p \text{ prime}} \frac{1}{1+2p^{-s}-\big(a_p^{(n,\pm)}\big)^2p^{-s}+p^{-2s}} \,.
\end{equation}
This function is known to have a simple pole at $s=1$. Then,
\begin{equation}
    \begin{split}
    \zeta(2s) \sum_{m\geq 1}  \frac{\big( a_m^{(n,\pm)} \big)^2}{ m^s}
    &= \prod_{p \text{ prime}} \frac{1}{(1-p^{-s})^2\left(1+2p^{-s}-\big(a_p^{(n,\pm)}\big)^2p^{-s}+p^{-2s}\right)} \\
    &= \prod_{p \text{ prime}} \frac{1}{(1-p^{-s})\left(1-\big(a_p^{(n,\pm)}\big)^2 \big( p^{-s} - p^{-2s} \big) + \left(p^{-s} - p^{-2s} - p^{-3s}\right)\right)} \\
    &=  \zeta(s)L_{\nu\times\nu}^{(n,\pm)}(s)\,.
    \end{split}
\end{equation}
Taking the residue at $s=1$ of both sides and using $\text{Res}_{s=1}\zeta(s)=1$ yields
\begin{equation}
    |\!|\nu_{n,\pm}|\!|^2  = \frac{\pi^2}{48 \cosh(\pi R_n^\pm)} \sum_{m\geq 1} \frac{a_{m^2}^{(n,\pm)} }{m}=\frac{1}{8 \cosh(\pi R_n^\pm)} \, 
    L^{(n,\pm)}_{\nu\times\nu}(1)\,. \qquad \qed
\end{equation}

\subsection{Statistical averages and moments of $L$-functions}

Having reviewed some basic facts about the cusp form norms and the distribution of their Fourier coefficients, we can now state some of the crucial properties that hold after statistical averaging over $n$. Let us first state the following useful
\begin{lemma}
    \label{lemma:averages}
    {\it Let $k \in \mathbb{N}$ and $p$ be prime. Then the average over $n$ yields the following results:}
\begin{equation}
\begin{split}
 (i)& \qquad    \overline{\big(a_{p^k}^{(n,\pm)}\big)^{2}} = \sum_{\ell=0}^k p^{-\ell} = \frac{p-p^{-k}}{p-1} \\
 (ii)& \qquad \overline{\big(a_{p^{k+1}}^{(n,\pm)}a_{p^{k-1}}^{(n,\pm)}\big)} = \sum_{\ell=1}^k p^{-\ell} = \frac{1-p^{-k}}{p-1} \\
 (iv) & \qquad \overline{ \big( a_{p^k}^{(n,\pm)} \big)^2 \big( a_{p}^{(n,\pm)} \big)^2} = \frac{2(p+1) - p^{-k}(p+2+p^{-1})}{p-1}  \\
 (iii) & \qquad \overline{\big(a_p^{(n,\pm)}\big)^{2(k+1)} } = \frac{(p+1)^2}{p} \; \overline{\big(a_p^{(n,\pm)}\big)^{2k} } - (p+1) \, \frac{(2k)!}{k!(k+1)!} 
\end{split}
\label{eq:variancePrimePowerFormula2}
\end{equation}
\end{lemma}
{\it Proof:} To prove $(i)$, we start with Lemma \ref{lemma:hecke}$(iv)$ and evaluate its average using the moments of the distributions for prime Fourier coefficients \eqref{eq:momentsFormula}:
{\small
\begin{equation}
    \begin{split}    \overline{\big(a_{p^k}^{(n,\pm)}\big)^{2}} 
    &=
    \frac{p}{2^{2k+1}(p+1)} \sum_{\ell=0}^{k} \sum_{r=0}^{\ell} \left[ {2k+2\choose 2\ell+2} + (-1)^\ell {k+1 \choose \ell+1} \right] \\
    &\qquad \times {\ell \choose \,r\,} (-4)^r \, 
    \frac{(2(k-r))!}{(k-r)!(k-r+1)!}\, {}_2F_1\left(1, k-r+\frac{1}{2}, k-r+2 , \frac{4p}{(p+1)^2} \right)
    \\
    &=
    \frac{p}{2^{2k+1}(p+1)} \sum_{\ell=0}^{k} \sum_{r=0}^{\ell} \left[ {2k+2\choose 2\ell+2} + (-1)^\ell {k+1 \choose \ell+1} \right] \\
    &\qquad \times {\ell \choose \,r\,} (-4)^r \, 
    \frac{(2(k-r))!}{(k-r)!(k-r+1)!}\sum_{q= 0}^\infty \frac{\Gamma\left(k-r+\frac{1}{2}+q\right)\Gamma(k-r+2)}{\Gamma\left(k-r+\frac{1}{2}\right)\Gamma(k-r+2+q)} \left( \frac{4p}{(p+1)^2} \right)^q\\
    &=
    \frac{p}{p+1} \sum_{q=0}^{\infty} \sum_{\ell=0}^{k} \left[ {2k+2\choose 2\ell+2} + (-1)^\ell {k+1 \choose \ell+1} \right] \frac{(-1)^{k+q} \, \Gamma\left(\ell+\frac{3}{2}\right)}{\Gamma\left(\ell-k-q+\frac{1}{2}\right)\Gamma\left(k+q+2\right)}\left( \frac{4p}{(p+1)^2} \right)^q \\
    &=
    \frac{p}{p+1} \sum_{q=0}^\infty  \frac{(2q)!}{2^{2q}} \left[\frac{1}{(q!)^2} - \frac{\Theta(q-k)}{(q+k+1)!(q-k-1)!}  \right] \left( \frac{4p}{(p+1)^2} \right)^q \\
    &=  \frac{p}{p+1} \left[ \frac{p+1}{p-1} - \frac{p+1}{p^{k+1}(p-1)}\right]
    =\frac{p-p^{-k}}{p-1} \,, 
    \end{split}
\end{equation}
}\normalsize
where $\Theta(n)=1$ if $n>0$ and vanishes otherwise.
The second result, $(ii)$, can be proven in a similar fashion, using Lemma \ref{lemma:hecke}$(ii)$ to simplify. To prove $(iii)$, we use Lemma \ref{lemma:hecke}$(ii)$ to calculate as follows:
\begin{equation}
    \begin{split}
        \overline{ \big( a_{p^k}^{(n,\pm)} \big)^2 \big( a_{p}^{(n,\pm)} \big)^2} 
        &= \overline{\big(a_{p^{k+1}}^{(n,\pm)}+ (1-\delta_{k,0}) \, a_{p^{k-1}}^{(n,\pm)} \big)^2} \\
        &= \sum_{\ell=0}^{k+1} p^{-\ell} + \left(1-\delta_{k,0}\right)\left[\sum_{\ell=0}^{k-1} p^{-\ell} + 2 \sum_{\ell=1}^k p^{-\ell}\right] \\
        & =\frac{2(p+1) - p^{-k}(p+2+p^{-1})}{p-1} \,. 
    \end{split}
\end{equation}
Finally, to prove $(iv)$, we use \eqref{eq:momentsFormula} and the series representation of the hypergeometric function:
\begin{equation}
    \begin{split}
\overline{\big(a_p^{(n,\pm)}\big)^{2(k+1)} }
&= \frac{p}{p+1} \frac{(2k+2)!}{(k+1)!(k+2)!}  \sum_{q=0}^\infty \frac{\Gamma\left(k+\frac{3}{2}+q\right) \Gamma(k+3)}{\Gamma\left(k+\frac{3}{2} \right) \Gamma(k+3+q)} \left( \frac{4p}{(p+1)^2}\right)^q \\
&= \frac{4p}{p+1} \frac{(2k)!}{k!(k+1)!} \sum_{q=1}^\infty \frac{\Gamma\left(k+\frac{1}{2}+q\right) \Gamma(k+2)}{\Gamma\left(k+\frac{1}{2} \right) \Gamma(k+2+q)} \left( \frac{4p}{(p+1)^2}\right)^{q-1} \\
&= \frac{(p+1)^2}{p} \left[\overline{\big(a_p^{(n,\pm)}\big)^{2k} } - \frac{p}{p+1}\,\frac{(2k)!}{k!(k+1)!} \right] \,. \qquad\qed
    \end{split}
\end{equation}\\

\paragraph{Corollary:} {\it For any $k\in \mathbb{N}$ and $p$ prime,}
\begin{equation}
    \overline{\big( a_{p^{k}}^{(n,\pm)}\big)^2\,\left[ 1-\big(a_p^{(n,\pm)}\big)^2 \big( p^{-1} - p^{-2} \big) + ( p^{-1} - p^{-2} - p^{-3} )  \right] }=1-p^{-2}\,.
\end{equation}
\noindent{\it Proof:} 
Follows immediately from Lemma \ref{lemma:averages}$(i)$ and $(iii)$. $\qed$\\

Finally, the central property needed in our analysis of the gravity amplitude concerns the interplay of the moments of distributions of Fourier coefficients and the cusp form norms:
\begin{tcolorbox}[width=\textwidth,colback=white]
\begin{theorem}
\label{thm:main}
{\it Let $m\geq 1$ be any integer spin. Then the statistical averaging over different cusp forms indexed by $n$ yields:}
\begin{equation}
\overline{\big( a_{m}^{(n,\pm)}\big)^2 \left(8\cosh(\pi R_n^\pm) |\!|\nu_{n,\pm}|\!|^2\right)^{-1}} = 
    \overline{\big( a_{m}^{(n,\pm)}\big)^2\, \left( L_{\nu\times\nu}^{(n,\pm)}(1) \right)^{-1}} =\frac{6}{\pi^2} \,.
    \quad
\end{equation}
\end{theorem}
\end{tcolorbox}
\noindent{\it Proof:}
The first equality follows from Theorem \ref{thm:norms}. To prove the second equality, write the $L$-function in terms of its Euler product:
\begin{equation}
\begin{split}
   &\overline{\big( a_{m}^{(n,\pm)}\big)^2\, \left( L_{\nu\times\nu}^{(n,\pm)}(1) \right)^{-1}}\\
   &\qquad= \overline{\big( a_{m}^{(n,\pm)}\big)^2\,\prod_{p\text{ prime}} \left[ 1-\big(a_p^{(n,\pm)}\big)^2 \big( p^{-1} - p^{-2} \big) + \big( p^{-1} - p^{-2} - p^{-3} \big) \right]} \\
  &\qquad  = \prod_{p\text{ prime}} \left( 1- p^{-2}  \right)= \frac{1}{\zeta(2)} = \frac{6}{\pi^2}\,,
\end{split}
\end{equation}
where we applied the Corollary of Lemma \ref{lemma:averages} factor by factor after decomposing $a_m^{(n,\pm)}$ into  factors of Fourier coefficients of prime powers (Lemma \ref{lemma:hecke}$(i)$):
\begin{equation}
m=p_1^{k_1} \cdots p_r^{k_r} \qquad \Rightarrow \qquad
\big(a_m^{(n,\pm)}\big)^2 = \left(a_{p_1^{k_1}}^{(n,\pm)} \right)^2\cdots \left(a_{p_r^{k_r}}^{(n,\pm)}\right)^2 \,.\qquad \qed
\end{equation}

\subsection{Derivation of the arithmetic kernel $f^{(n,\pm)}$}
\label{app:fDerivation}

In the main text we verified that the arithmetic kernel \eqref{eq:KernelDef} has the required properties to produce a ramp in all spin sectors. We also outlined how to derive it, but provide more details here. The derivation essentially also shows that it is unique, up to modifications, which are invisible to our averaging condition \eqref{eq:fCond}.

\paragraph{Construction of $f^{(n,\pm)}$:}
The goal is to find a function $f^{(n,\pm)}$ which satisfies \eqref{eq:fCond}.
Since the Fourier coefficients have multiplicativity properties determined by Hecke relations, it is convenient to begin by decomposing $a_m^{(n,\pm)}$ into coefficients with prime-power index:
\begin{equation}
    m = p_1^{k_1}\cdots p_r^{k_r} \qquad \Rightarrow \qquad  a_{p_1^{k_1}\cdots p_r^{k_r}}^{(n,\pm)} = a_{p_1^{k_1}}^{(n,\pm)} \cdots a_{p_r^{k_r}}^{(n,\pm)}
    \label{eq:multF}
\end{equation}
where $p_i$ are distinct primes. Let us therefore first find a function $f^{(n,\pm)}_p$ which is fine tuned to Fourier coefficients with prime-power index $p^k$ such that:
\begin{equation}
    \overline{ \big( a_{p^k}^{(n,\pm)} \big)^2 f^{(n,\pm)}_p} = 1 \quad \text{ for all }  k\geq 0 \,.
    \label{eq:fCondPrime}
\end{equation}
It is important to note that $a_{p^k}^{(n,\pm)}$ is fully determined by powers of $a_p^{(n,\pm)}$, see Lemma \ref{lemma:hecke}$(iv)$.
In order for a condition such as \eqref{eq:fCondPrime} to hold, the function $f^{(n,\pm)}_p$ must balance different moments of the distribution of $a_p^{(n,\pm)}$. This is captured by an ansatz of the following form:
\begin{equation}
    f^{(n,\pm)}_p = \sum_{r\geq 0} c_{p,r} \, \big( a_p^{(n,\pm)} \big)^{2r}\,.
\end{equation}
We only need even powers of the Fourier coefficients because any odd powers will have vanishing expectation value. We also do not need any Fourier coefficients with spin other than $p$ because these are distributed independent of $a_p^{(n,\pm)}$, so they can be absorbed into $c_{p,r}$ as far as the averaged \eqref{eq:fCondPrime} is concerned.
The condition \eqref{eq:fCondPrime} then amounts to an infinite number of constraints on $c_{p,r}$. For example:
\begin{equation}
    \begin{split}
    &k=0:\qquad 1 \stackrel{!}{=} \overline{f^{(n,\pm)}_p} = \sum_{r\geq 0} c_{p,r}  \overline{\big( a_p^{(n,\pm)} \big)^{2r}}
    \\
    &k=1:\qquad 1 \stackrel{!}{=} \overline{\big(a_{p}^{(n,\pm)}\big)^2 \,f^{(n,\pm)}_p} = \sum_{r\geq 0} c_{p,r}  \overline{\big( a_p^{(n,\pm)} \big)^{2r+2}}
      \\
    &k=2:\qquad 1 \stackrel{!}{=} \overline{\big(a_{p^2}^{(n,\pm)}\big)^2 \,f^{(n,\pm)}_p} 
    = \overline{\left( \big(a_{p}^{(n,\pm)}\big)^2 - 1 \right)^2 \,f^{(n,\pm)}_p} 
    = -1 +\sum_{r\geq 0} c_{p,r}  \overline{\big( a_p^{(n,\pm)} \big)^{2r+4}} 
      \\
    &k=3:\qquad 1 \stackrel{!}{=} \overline{\big(a_{p^3}^{(n,\pm)}\big)^2 \,f^{(n,\pm)}_p} 
    = \overline{\left( \big(a_{p}^{(n,\pm)}\big)^3 - 2 a_p^{(n,\pm)} \right)^2 \,f^{(n,\pm)}_p} 
    = -4 +\sum_{r\geq 0} c_{p,r}  \overline{\big( a_p^{(n,\pm)} \big)^{2r+6}} 
    \end{split}
\end{equation}
and so on. Iterating this process, one finds for general $k$:
\begin{equation}
\label{eq:system}
     \sum_{r\geq 0} c_{p,r}  \overline{\big( a_p^{(n,\pm)} \big)^{2(k+r)}} = \frac{(2k)!}{k!(k+1)!} \,,
\end{equation}
where the r.h.s.\ is the $k$-th Catalan number. 
Using a general recursion relation of the moments of Fourier coefficients (Lemma \ref{lemma:averages}$(iv)$), we can write the r.h.s.\ as follows:
\begin{equation}
     \sum_{r\geq 0} c_{p,r}  \overline{\big( a_p^{(n,\pm)} \big)^{2(k+r)}} = 
     \frac{p+1}{p} \, \overline{\big( a_p^{(n,\pm)} \big)^{2k}} - \frac{1}{p+1} \, \overline{\big( a_p^{(n,\pm)} \big)^{2k+2}}
     \,.
\end{equation}
It is now obvious to see that a simple solution exists for all $k$:
\begin{equation}
    c_{p,0} = \frac{p+1}{p} \,,\qquad c_{p,1} = - \frac{1}{p+1} \,,\qquad c_{p,r>1} = 0 \,.
\end{equation}
To see that this is the {\it only} solution, note that the equations \eqref{eq:system} form a linear system. Therefore, the existence of any other solution $c'_{p,r}$ would mean that there exist coefficients $\tilde{c}_{p,r}\equiv c_{p,r} - c'_{p,r}$ such that
\begin{equation}
     \sum_{r\geq 0} \tilde{c}_{p,r}  \overline{\big( a_p^{(n,\pm)} \big)^{2(k+r)}} = 
     0
\end{equation}
for all $k$.
This can be written as an infinite list of equations labelled by $k$, which we call
\begin{equation}
    {\cal E}_{p,k} \equiv \sum_{r\geq k} \tilde{c}_{p,r-k}  \overline{\big( a_p^{(n,\pm)} \big)^{2r}} = 
     0\,.
\end{equation}
This can be thought of as an (infinite dimensional) triangular matrix acting on the vector of moments of Fourier coefficients.
If $\tilde{c}_{p,0} \neq 0$, we can form a linear combination which cancels all terms but one:
\begin{equation}
   0 = {\cal E}_{p,0} - \frac{\tilde{c}_{p,1}}{\tilde{c}_{p,0}} \, {\cal E}_{p,1} -  \left( \frac{\tilde{c}_{p,2}}{\tilde{c}_{p,0}} - \frac{\tilde{c}_{p,1}^2}{\tilde{c}_{p,0}^2}\right) {\cal E}_{p,2} - \ldots = \tilde{c}_{p,0}  \,.
\end{equation}
This contradicts the assumption, so we must have $\tilde{c}_{p,0}=0$. Next, if $\tilde{c}_{p,1} \neq 0$ we could form a similar linear combination
\begin{equation}
    0 = {\cal E}_{p,0} - \frac{\tilde{c}_{p,2}}{\tilde{c}_{p,1}} \, {\cal E}_{p,1} - \left( \frac{\tilde{c}_{p,3}}{\tilde{c}_{p,1}} - \frac{\tilde{c}_{p,2}^2}{\tilde{c}_{p,1}^2}\right) {\cal E}_{p,2} - \ldots = \tilde{c}_{p,1} \, \overline{\big(a_p^{(n,\pm)}\big)^2}\,,
\end{equation}
which is again contradictory and thus implies $\tilde{c}_{p,1} = 0$. Continuing this way, we must have $\tilde{c}_{p,k} = 0$ for all $k$. Therefore, there does not exist any solution different from $c_{p,k}$.\\

To summarize, we have shown that
\begin{equation}
    f^{(n,\pm)}_p = \frac{p+1}{p} - \frac{1}{p+1} \, \big( a_p^{(n,\pm)} \big)^2 
\end{equation}
solves \eqref{eq:fCondPrime} and is unique as far as our ansatz is concerned.\footnote{There are ways to modify $f^{(n,\pm)}_p$ that are invisible to the statistical averaging. For example, one can add odd powers of $a_p^{(n,\pm)}$ with arbitrary coefficients, as these will vanish in the evaluation of \eqref{eq:fCond}. See main text for more comments.} This function will give a ramp in all spin sectors of the form $m=p^k$. From the multiplicative property of the Fourier coefficients, \eqref{eq:multF}, it is then clear how to construct the function that will yield a linear ramp for all spins $m=p_1^{k_1} \cdots p_r^{k_r}$; indeed, we simply construct it as
\begin{equation}
\begin{split}
    f^{(n,\pm)} &=  \prod_{p \text{ prime}} \left[ \frac{p+1}{p} - \frac{1}{p+1} \, \big( a_p^{(n,\pm)} \big)^2 \right] \\
    &= \prod_{p \text{ prime}} \left( 1 - p^{-2} \right)^{-1} \times \left[ 1 - \big( a_p^{(n,\pm)} \big)^2 \left( p^{-1} - p^{-2} \right) + \left( p^{-1} - p^{-2} - p^{-3} \right) \right] \\
    &= \frac{\pi^2}{6}\frac{1}{L_{\nu\times\nu}^{(n,\pm)}(1)} \,,
\end{split}
\end{equation}
where we used $\zeta(2) = \prod_{p\text{ prime}} (1-p^{-2})^{-1} = \frac{\pi^2}{6}$ and we used the symmetric square $L$-function associated with the cusp form $\nu_{n,\pm}$, see Lemma \ref{lemma:Lsym}.
It is a meromorphic function in $s$ with a potential pole at $s=1$ \cite{shimura} (in our case there is no pole, so we can simply evaluate at $s=1$).\footnote{ Note that there would be a pole at $s=1$ if there was some prime Fourier coefficient with $a_p^{(n,\pm)}=\pm 2$. It is unproven but widely believed to be true that such a Fourier coefficient does not exist (Ramanijan-Petersson conjecture) \cite{sarnakStatisticalPropertiesEigenvalues1987}.}
This completes our derivation of the arithmetic kernel $f^{(n,\pm)}$.

\subsection{$L$-function for Eisenstein series}
\label{app:LfuncCont}

It is natural to define the following continuous family of $L$-functions for the Eisenstein series $E_{\frac{1}{2}+i\alpha}(x,y)$ in terms of their Fourier coefficients (c.f., \eqref{eq:aAlphaDef}):\footnote{The factor $\frac{1}{2}$ is unconventional, but will make the following discussion more convenient.}
\begin{equation}
    L^{(\alpha)}_E(s) \equiv \frac{1}{2}\sum_{m\geq 1} \frac{a_m^{(\alpha)}}{m^s} \,,\qquad a_m^{(\alpha)} = \frac{2\sigma_{2i\alpha}(m)}{m^{i\alpha}} \qquad\qquad (\text{Re}(s)>1)\,.
\end{equation}

\begin{lemma}
The meromorphic continuation of the Eisenstein series $L$-function is
\begin{equation}
    L^{(\alpha)}_E(s) =  \zeta(s+i\alpha) \zeta(s-i\alpha) \,.
\end{equation}
\end{lemma}
\noindent{\it Proof:} We expand the zeta-functions formally in the domain where they converge:
\begin{equation}
    \begin{split}
    \zeta(s+i\alpha) \zeta(s-i\alpha) &= \sum_{n_1 \geq 1}\sum_{n_2 \geq 1} \frac{1}{n_1^{s+i\alpha} n_2^{s-i\alpha}} = \sum_{m\geq 1}\, \sum_{\substack{n_1,n_2:\\n_1 n_2 = m}} \frac{1}{n_1^{s+i\alpha} n_2^{s-i\alpha}} \\
    &= \sum_{m\geq 1} \frac{1}{m^{s+i\alpha}} \, \sigma_{2i\alpha}(m) = L_E^{(\alpha)}(s) \qquad \qed
    \end{split}
\end{equation}
\\

Note in particular:
\begin{equation}
  L_E^{(2\alpha)}(s=1) = \cosh(\pi \alpha)\Lambda(i\alpha)\Lambda(-i\alpha) \,, \qquad \Lambda(s) = \pi^{-s} \Gamma(s) \zeta(2s)\,,
\end{equation}
which is similar to the symmetric square $L$-function for cusp forms evaluated at $s=1$. See \cite{Haehl:2023mhf} for more details on this connection.

\section{Effects of chaos across different spin sectors}
\label{sec:ramp-independence}

In this appendix we show (and review) that the existence of a ramp (or plateau) in a given spin sector is generically not sufficient to conclude the existence of a ramp (or plateau) in another spin sector. 
We previously showed this for the Eisenstein series in \cite{Haehl:2023tkr}, and only briefly review those results here. We mainly focus here on extending this result to the Maass cusp form spectrum.

\vspace{5pt}
\subsection{Signatures of a spin $m=0$ ramp at spin $(m_1,m_2)$}
\label{sec:spin0imprint}

Recall the simple form of correlations $\langle {\color{colHighlight} z_{\frac{1}{2}+i\alpha_1} z_{\frac{1}{2}+i\alpha_2} } \rangle$ that correspond to a linear ramp in the $m=0$ sector, \eqref{eq:spin0Ramp}. Since these correlations also enter into the spectral form factor for all $m>0$, we can ask about their {\it imprint} onto higher spin sectors. We previously found in \cite{Haehl:2023tkr} (numerically) for the contribution to the spin $(m_1,m_2)$ spectral form factor due to the existence of a ramp at spin 0:\footnote{The factor 2 relative to \cite{Haehl:2023tkr} is for the same reason as in \eqref{eq:ZPZPuniv}.}
\begin{equation}
\label{eq:ZZresNum}
\big\langle \widetilde{Z}^{m_1}_\text{P,cont.}(y_1) \widetilde{Z}^{m_2}_\text{P,cont.}(y_2)
    \big\rangle \supset_{_0} 2\,\lambda_{m_1} \, \delta_{m_1 m_2} \; e^{-2\pi (m_1 y_1+m_2y_2)} \, \sqrt{\frac{y_1 y_2}{y_1+y_2}}  + \ldots 
    \quad\;\; (y_i \gg 1)
\end{equation}
where ``$\supset_{_0}$'' means that we only consider the contribution to the left hand side that is implied by the existence of a ramp at spin $m=0$. The first few spin-dependent prefactors in this expression are 
\begin{equation}
    \begin{split}
        \lambda_1 = 0.761.. \,,\quad
        \lambda_2 = 0.644.. \,,\quad
        \lambda_3 = 0.613.. \,,\quad
        \lambda_4 = 0.532.. \,,\quad
        \lambda_5 = 0.548.. \,,\;\; \text{etc.} 
    \end{split}
\end{equation}
Details can be found in \cite{Haehl:2023tkr}. Crucially, since \eqref{eq:ZZresNum} is strictly subleading to the ramp \eqref{eq:ZPZPuniv}, the form of the spin 0 correlations advocated in \eqref{eq:spin0Ramp} is consistent by itself and does not affect the slope or existence of ramps in any other spin sector.

\vspace{5pt}
\subsection{Signatures of a spin $m$ ramp at spin $(m_1,m_2)$: Eisenstein series}
\label{sec:spin1imprint}

Let us now assume the existence of a linear ramp in the Eisenstein spectrum at spin $m$. From \eqref{eq:z-m-result-early}, we would infer the following imprint of a spin $m$ ramp onto the spin $(m_1,m_2)$ sector:
{\small
 \begin{equation}
  \begin{split}
      &\big\langle \widetilde{Z}_\text{P,cont.}^{m_1}(y_1)\widetilde{Z}_\text{P,cont.}^{m_2}(y_2) \big\rangle\\
      &\quad
     \supset_{_m} \frac{\sqrt{y_1y_2}}{\pi^2} \iint d\alpha_1 d\alpha_2 \,  \big\langle {\color{colHighlight}z_{\frac{1}{2}+i\alpha_1} z_{\frac{1}{2}+i\alpha_2} }\big\rangle_{\text{spin }m\text{ ramp}} \, \frac{\sigma_{2i\alpha_1}(m_1)\sigma_{2i\alpha_2}(m_2)}{m_1^{i\alpha_1}m_2^{i\alpha_2}\Lambda(-i\alpha_1)\Lambda(-i\alpha_2)} \, K_{i\alpha_1}(2\pi m_1y_1) K_{i\alpha_2}(2\pi m_2y_2) \\
     &\quad
     = \frac{2\sqrt{y_1y_2}}{\pi^2} \int d\alpha \,   \alpha\tanh(\pi \alpha) \,  \frac{m^{2i\alpha}\sigma_{2i\alpha}(m_1)\sigma_{2i\alpha}(m_2)}{(m_1m_2)^{i\alpha} \sigma_{2i\alpha}(m)^2} \, K_{i\alpha}(2\pi m_1y_1) K_{i\alpha}(2\pi m_2y_2)
  \end{split}
  \label{eq:ZmZmspin1ImprintTest}
 \end{equation}
 }\normalsize
 where the notation ``$\supset_{_m}$'' means that we consider the contribution to the spectral form factor that is implied by the existence of a ramp in the spin $m$ sector.
 For $m=1$, this expression is particularly simple. Its numerical evaluation gives \cite{Haehl:2023tkr}
   \begin{equation}
  \label{eq:Zm1Zm1result}
      \big\langle \widetilde{Z}_\text{P,cont.}^{m_1}(y_1)\widetilde{Z}_\text{P,cont.}^{m_2}(y_2) \big\rangle
     \supset_{_{m=1}}  \sigma_0(m_1) \times\delta_{m_1m_2} \, \frac{1}{\pi}  \,\frac{y_1y_2}{y_1+y_2} \,  e^{-2\pi m_1 ( y_1 +  y_2)}
 \end{equation}
 with the divisor function giving the following count:
\begin{equation}
    \sigma_0(1) = 1\, , \quad \sigma_0(2) = 2\, ,\quad \sigma_0(3) = 2\, , \quad \sigma_0(4) = 3\, , \quad \sigma_0(5) = 2\, , \;\; \text{etc.}
\end{equation}
So, unlike for spin 0, the higher spin ramps do imprint onto the slope of ramps in other spin sectors. This is analogous to the situation with the `naive' ansatz for the spin $m$ ramp in the cusp form case discussed in the main text, see \eqref{eq:znznApproxDef}. It would be interesting to analyze if the naive ansatz for ramps in the Eisenstein sector can be improved, or if the above analysis hints at a deeper inconsistency with ramps for $m>0$ being encoded in Eisenstein series at all.

\vspace{5pt}
\subsection{Signatures of a spin $m$ ramp at spin $(m_1,m_2)$: Maass cusp forms}
\label{sec:spin1imprintMaass}

\begin{figure}
    \centering
\includegraphics[width=.48\textwidth]{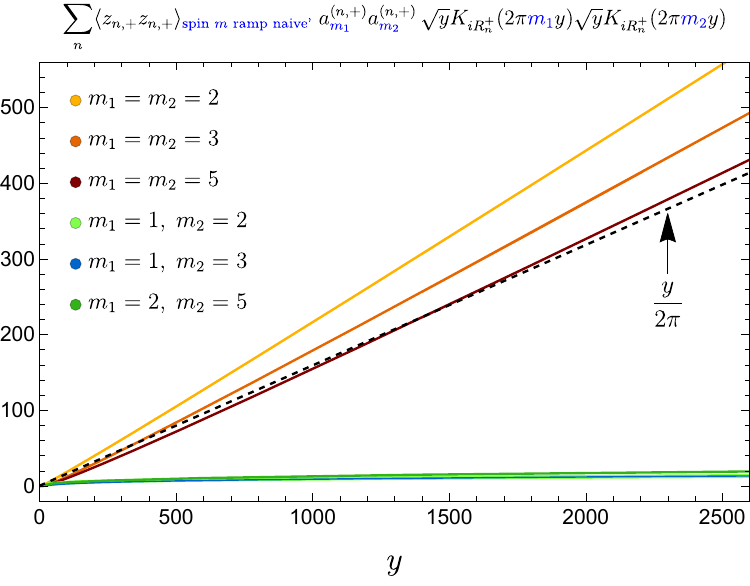}$\,$
\includegraphics[width=.48\textwidth]{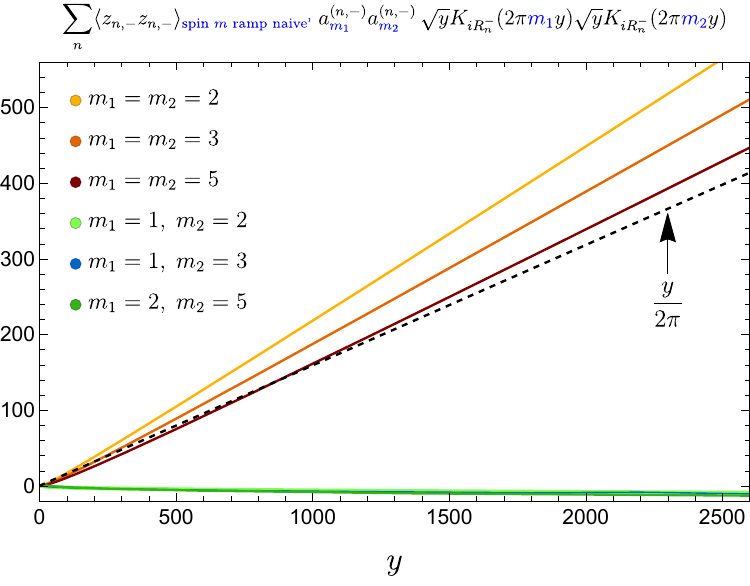}
    \caption{Numerical evaluation of \eqref{eq:ZmZmdiscImprint}: we show the imprint of a ramp in the spin $m=1$ sector onto the spin $(m_1,m_2)$ sector of the spectral form factor, assuming the naive form of correlations \eqref{eq:z-m-R-result-disc-naive-app} which is not engineered to have information about any other spin sector. While the asymptotic contribution has the correct linear $y$-dependence (for $m_1=m_2$), it does not have the correct slope to account for all the information encoded in a ramp.}
    \label{fig:imprint}
\end{figure}

The numerical analysis of the previous subsection can be generalized straightforwardly to the case of cusp forms. To this end, we adapt the calculation \eqref{eq:ZmZmspin1ImprintTest}: let us assume that the spin $m$ spectrum of Maass cusp forms contains a linear ramp. As discussed in the main text, the statistical averaging over cusp form data, which is automatic in the large $y_i$ limit, means that there are different choices of correlations which would all yield a linear ramp in some given spin sector. Ultimately, we found \eqref{eq:summaryRamp} by demanding a ramp with the correct slope in {\it every} spin sector. That is, we demanded that the imprint of any spin sector is the same on any other spin sector. In this appendix we analyze the consequences of working with less fine-tuned spectral correlations that are engineered to only describe RMT statistics in a fixed spin sector. In particular, consider the most naive ansatz, obtained by taking \eqref{eq:z-m-R-result-disc} and simply dividing out the Fourier coefficients:
\begin{equation}
 \big\langle {\color{colHighlight}z_{n_1,\pm}\,z_{n_2,\pm}}\big\rangle_{\text{spin }m \text{ ramp naive'}} 
 \approx \frac{1}{a_{m}^{(n_1,\pm)} a_{m}^{(n_2,\pm)}} \, \frac{2R_{n_1}^\pm \tanh(\pi R_{n_1}^\pm)}{ \pi^2\,\bar{\mu}_\pm(R_{n_1}^\pm)}\, \delta_{n_1n_2}   \,.
\label{eq:z-m-R-result-disc-naive-app}
\end{equation}
Such a correlation implies that the spin $(m_1,m_2)$ sector must contain the following term:
\begin{equation}
\begin{split}
  &\big\langle \widetilde{Z}_{\text{P,disc.,}\pm}^{m_1}(y_1)\widetilde{Z}_{\text{P,disc.,}\pm}^{m_2}(y_2) \big\rangle \\
  &\qquad \supset_{_m}  \sum_{n_1,n_2} \langle {\color{colHighlight}z_{n_1,\pm}\, z_{n_2,\pm}} \rangle_{\text{spin }m\text{ ramp naive'}} \; a_{m_1}^{(n_1,\pm)} a_{m_2}^{(n_2,\pm)}  \, \sqrt{y_1}K_{iR_{n_1}^\pm}(2\pi m_1 y_1)\sqrt{y_2} K_{iR_{n_2}^\pm}(2\pi m_2 y_2) \\
  &\qquad \approx \sum_{n} \frac{2R_n^\pm \, \tanh(\pi R_n^\pm)}{\pi^2 \bar{\mu}_\pm(R_n^\pm)} \, \frac{a_{m_1}^{(n,\pm)} a_{m_2}^{(n,\pm)}}{\big(a_m^{(n,\pm)}\big)^2}  \, \sqrt{y_1}K_{iR_{n}^\pm}(2\pi m_1 y_1) \sqrt{y_2}K_{iR_{n}^\pm}(2\pi m_2 y_2)\,.
\end{split}
\label{eq:ZmZmdiscImprint}
\end{equation}
This can be evaluated numerically, which yields similar results as in the case of Eisenstein series discussed in the previous subsection: 
\begin{equation}
\begin{split}
  \big\langle \widetilde{Z}_{\text{P,disc.,}\pm}^{m_1}(y_1)\widetilde{Z}_{\text{P,disc.,}\pm}^{m_2}(y_2) \big\rangle \supset_{_m}  \eta_{m,m_1}^\pm \times\delta_{m_1m_2} \, \frac{1}{\pi}  \,\frac{y_1y_2}{y_1+y_2} \,  e^{-2\pi m_1 ( y_1 +  y_2)} \qquad (y_i \gg 1)\,,
\end{split}
\label{eq:ZmZmdiscImprint2}
\end{equation}
where the spin-dependent coefficient $\eta_{m,m_1}^\pm$ is generally different from 1. This is illustrated in figure \ref{fig:imprint}. 
From the curves in that figure, we find the following numerical fit:
\begin{equation}
\text{num. fit:}\quad \left\{
  \begin{aligned}
    \eta^+_{1,1} &= 1 \,,\quad \eta^+_{1,2} = 1.44..  \,,\quad \eta^+_{1,3} = 1.25..  \,,\quad \eta^+_{1,4} = 1.61..   \,,\quad \eta^+_{1,5} = 1.10..  \,,\;\; \ldots\\
    \eta^-_{1,1} &= 1 \,,\quad \eta^-_{1,2} = 1.46..  \,,\quad \eta^-_{1,3} = 1.28..  \,,\quad \eta^-_{1,4} = 1.65..   \,,\quad \eta^-_{1,5} = 1.13..  \,,\;\; \ldots
  \end{aligned}    
  \right.
\end{equation}
This matches within $\sim\! 10\%$ with the theoretical expectation based on statistical averaging, namely $\eta^\pm_{m,m_1} = {\cal N}^\pm_{m_1}/{\cal N}^\pm_{m}$, which follows after replacing squares of Fourier coefficients by their variances in \eqref{eq:ZmZmdiscImprint}:\footnote{ Computing ${\cal N}^\pm_{m_1}/{\cal N}^\pm_{m}$ using the finite number of cusp forms available to us, i.e., using \eqref{eq:variancesNum}, yields agreement within $\sim\! 2\%$. This shows that a still much larger number of cusp forms is required in order to get very close to the theoretical values for $\eta^\pm_{m,m_1}$.}
\begin{equation}
\text{theoretical values:}\qquad
    \eta^\pm_{1,1} = 1 \,,\quad \eta^\pm_{1,2} = \frac{3}{2}  \,,\quad \eta^\pm_{1,3} = \frac{4}{3}  \,,\quad \eta^\pm_{1,4} = \frac{7}{4}   \,,\quad \eta^\pm_{1,5} = \frac{5}{4}  \,,\;\; \ldots
\end{equation}

The fact that the prefactor $\eta^\pm_{m,m_1}$ in \eqref{eq:ZmZmdiscImprint2} is not 1 means that the ramp at spin $(m_1,m_2)$ is not fully encoded in the ramp at spin $m$. Random matrix universality in one spin sector therefore does not imply random matrix universality in a different spin sector -- as is consistent with general expectations in the theory of quantum chaos. Instead, one must fine-tune the approximation \eqref{eq:z-m-R-result-disc-naive-app} in a way that is informed by cusp form data in all other spin sectors. This is achieved by the arithmetic kernel $f^{(n,\pm)}$ discussed in the main text.

\subsection{Independence of the plateaus}
\label{sec:plateau-independence}

\begin{figure}
    \centering
\includegraphics[width=.75\textwidth]{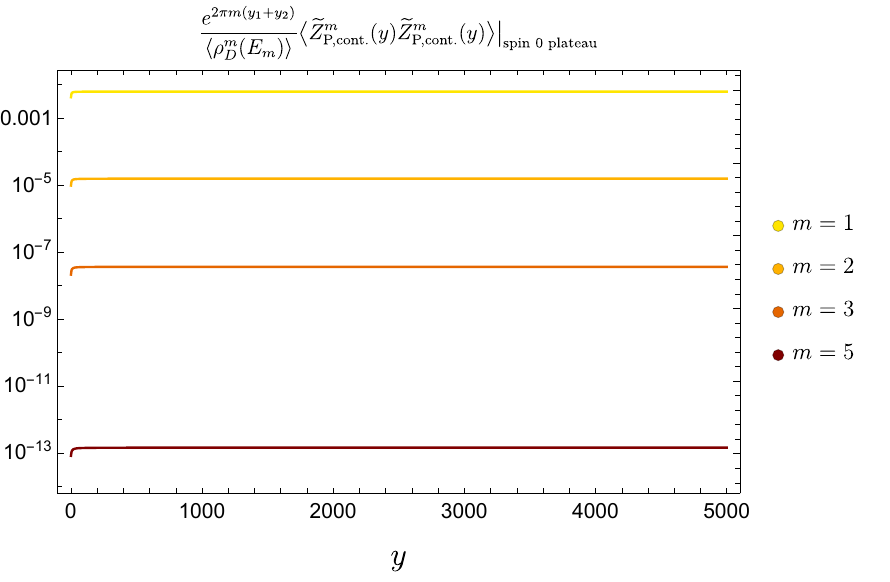}
    \caption{Plot of the imprint of the spin $0$ plateau on other spin sectors (note the log scaling of the $y$-axis). In order to get a result that is $c$-independent and compare with the true spin $m$ plateau \eqref{eq:plateau-sff}, we normalize the imprint by $\langle \rho_D(E_m) \rangle e^{-2\pi m (y_1+y_2)}$; at large $c$, $\langle \rho_D(E_0) \rangle/\langle \rho_D(E_m) \rangle\approx \frac{1}{2} e^{-2\pi m}$. With this normalization, the true spin $m$ plateau becomes equal to $1/2$ for $y_1=y_2$. The results shown therefore amount to a small constant $\sim {\cal O}(e^{-2\pi m})$.}
    \label{fig:spin0-plateau-imprint}
\end{figure}

Similar to the case of the ramps analyzed above and in \cite{Haehl:2023tkr}, here we discuss  the numerical imprint of a plateau in one spin sector onto other sectors. We begin with the imprint of a spin 0 plateau onto the spin $(m_1,m_2)$ sector:
{\small
 \begin{equation}
  \begin{split}
      &\big\langle \widetilde{Z}_\text{P,cont.}^{m_1}(y_1)\widetilde{Z}_\text{P,cont.}^{m_2}(y_2) \big\rangle\\
      &\quad
     \supset_{m=0}^{\plt} \frac{\sqrt{y_1y_2}}{4\pi^2} \iint d\alpha_1 d\alpha_2 \,  \big\langle{\color{colHighlight} z_{\frac{1}{2}+i\alpha_1} z_{\frac{1}{2}+i\alpha_2} }\big\rangle_{\text{spin }0\text{ plateau}} \, \frac{a_{m_1}^{(\alpha_1)}a_{m_2}^{(\alpha_2)}}{\Lambda(-i\alpha_1)\Lambda(-i\alpha_2)} \, K_{i\alpha_1}(2\pi m_1y_1) K_{i\alpha_2}(2\pi m_2y_2) \\
     &\quad
     = \frac{\sqrt{y_1y_2}}{\pi^2} \int d\alpha \,  2 i \pi^2 \frac{1}{\sinh(\pi \alpha)} \,  \frac{\sigma_{2i\alpha}(m_1)\sigma_{-2i\alpha-2}(m_2)}{m_1^{i\alpha}m_2^{-i\alpha-1} \Lambda(-i\alpha) \Lambda(i\alpha+1) } \, K_{i\alpha}(2\pi m_1y_1) K_{-i\alpha-1}(2\pi m_2y_2)
  \end{split}
  \label{eq:ZmZmspin0PlateauImprint}
 \end{equation}
}\normalsize
The result, as shown in figure \ref{fig:spin0-plateau-imprint}, is
\begin{equation}
      \big\langle \widetilde{Z}_\text{P,cont.}^{m_1}(y_1)\widetilde{Z}_\text{P,cont.}^{m_2}(y_2) \big\rangle
     \supset_{m=0}^{\plt} \lambda_m^{\text{(p)}} \, \langle \rho_D^m(E_m) \rangle e^{-2\pi |m|(y_1+y_2)} \qquad (y_i \gg 1)\,,
  \label{eq:ZmZmspin0PlateauNumeric}
 \end{equation}
where the spin $m$ coefficient $\lambda_m^{\text{(p)}}$ is a small, ${\cal O}(e^{-2\pi m})$ coefficient:
\begin{equation}
    \begin{split}
        \lambda_1^{\text{(p)}} \approx 6.04 \cross 10^{-3} \,,\quad
        \lambda_2^{\text{(p)}} \approx 1.51 \cross 10^{-5} \,,\quad
        \lambda_3^{\text{(p)}} \approx 3.49 \cross 10^{-8} \,,\quad
        \lambda_5^{\text{(p)}} \approx 1.38 \cross 10^{-13} \,,\; \ldots 
    \end{split}
\end{equation}
Thus the spin 0 plateau produces a small constant in the spin $m$ sector. Since the plateau also goes to a constant for $y_i\gg 1, \, y_1/y_2=$ fixed, this represents a subleading correction to the true spin $m$ plateau. This is a similar situation to the imprint of the spin 0 ramp; however, here we find that it is subleading due to the {\it coefficient}, rather than the {\it functional form}.

\begin{figure}
    \centering
\includegraphics[width=.75\textwidth]{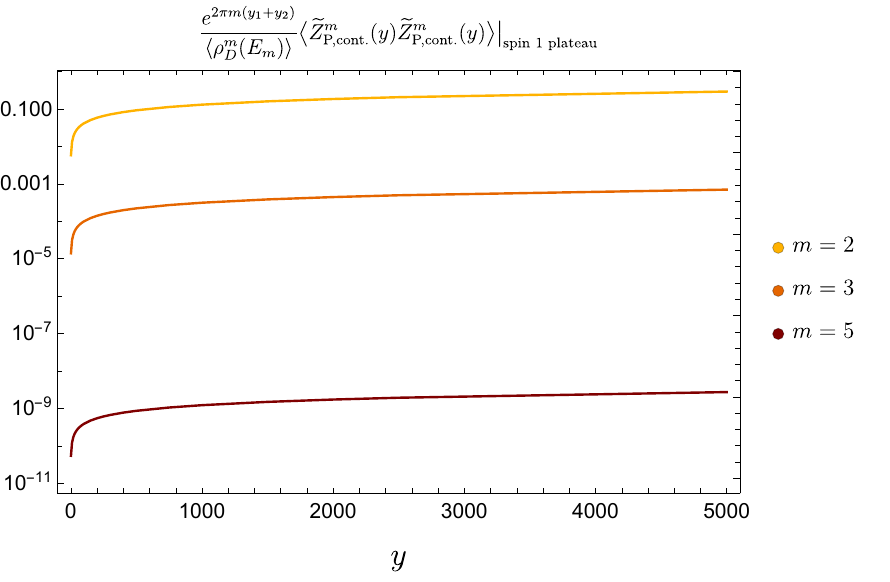}
    \caption{Plot of the imprint of the spin $1$ plateau on other spin sectors for $y_1=y_2=y$. We again normalize by $\langle \rho_D(E_m) \rangle e^{-2\pi m (y_1+y_2)}$; at large $c$, $\langle \rho_D(E_1) \rangle/\langle \rho_D(E_m) \rangle\approx  e^{-2\pi (m-1)}$. The result is a function that grows like $\sqrt{y}$; creating a similar plot for a fixed $y_2$ yields the functional form in \eqref{eq:ZmZmspin1PlateauNumeric}.}
    \label{fig:spin1-plateau-imprint}
\end{figure}

The imprint of a spin 1 plateau, through a similar calculation, is as follows (see figure \ref{fig:spin1-plateau-imprint}):
\begin{equation}
      \big\langle \widetilde{Z}_\text{P,cont.}^{m_1}(y_1)\widetilde{Z}_\text{P,cont.}^{m_2}(y_2) \big\rangle
     \supset_{m=1}^{\plt} \mu_m^{\text{(p)}}\sqrt{\frac{y_1 + y_2}{2}} \langle \rho_D^m(E_m) \rangle \frac{\sqrt{y_1 y_2}}{y_1+y_2}e^{-2\pi |m|(y_1+y_2)} \quad\;\; (y_i\gg 1)
  \label{eq:ZmZmspin1PlateauNumeric}
 \end{equation}
where
\begin{equation}
    \begin{split}
        \mu_2^{\text{(p)}} \approx 8.13..\cross 10^{-3} \,,\quad
        \mu_3^{\text{(p)}} \approx 1.97.. \cross 10^{-5} \,,\quad
        \mu_5^{\text{(p)}} \approx 7.78.. \cross 10^{-11} \,,\; \ldots 
    \end{split}
\end{equation}
We obtain a function that dominates over the plateau at large $y_i$. This is a similar situation to the ramp, where the imprint of the spin 1 ramp dominated over the true spin $m$ ramp; however, here we find that it dominates due to the {\it functional form}, rather than the {\it coefficient}. It would obviously be interesting to study the implications of this further.

\subsection{Comments on the plateau and the cusp forms}\label{sec:plateau-cusp-forms}
When trying to find an expression for the spectral decomposition of the plateau into the cusp forms, we can apply the logic of Section \ref{cusp}, i.e., use arithmetic chaos and the continuous approximation. Using \eqref{eq:alphaPlateau-cont}, this would immediately give:
\small{
\begin{equation}
\label{eq:disc-plateau-correlations}
    \langle {\color{colHighlight2}z^{m_1}_{n_1,\pm}z^{m_1}_{n_2,\pm}}\rangle_{\text{spin }m\text{ plateau}} \stackrel{?}{\approx} -\frac{4m}{\pi \bar{\mu}(R_{n_1}^\pm)\bar{\mu}(R_{n_2}^\pm)} \langle \rho_D^m(E_m) \rangle {\cal D}\left( \frac{1}{\left(R_{n_1}^\pm-R_{n_2}^\pm\right)^2}+\frac{1}{\left(R_{n_1}^\pm+R_{n_2}^\pm\right)^2}\right)\delta_{m_1 m_2} \, .
\end{equation}
}\normalsize
These correlations should then produce a plateau in the spectral form factor. Unfortunately, \eqref{eq:disc-plateau-correlations} is not as well suited for numerical analysis as the ramp. The reason is that the factor $(R_{n_1}^\pm \pm R_{n_2}^\pm)^{-2}$ decays for large $R_n^\pm$, meaning that the integrand is peaked at small values of $R_n^\pm$. This is in contrast to the ramp, where we instead had the factor $R_{n_1}^\pm \tanh (\pi R_{n_1}^\pm)$ which leads to an integrand peaked at large values of $R_n^\pm$.

However we do expect that as $y_i$ and $R_{n_i}^\pm$ increase, the continuous approximation \eqref{eq:disc-plateau-correlations} becomes better. The reason is that in the continuous approximation, the region of $R_{n_i}^\pm$ where the integrand has support increases as $y_i\rightarrow \infty$. Thus, even though the correlations are peaked at small $R_{n_i}$, \eqref{eq:disc-plateau-correlations} should reproduce the plateau at sufficiently large $y_i$. We do not have access to enough cusp form data to demonstrate this, and we leave \eqref{eq:disc-plateau-correlations} as a conjecture.

\vspace{10pt}
\bibliographystyle{JHEP}
\newpage

\providecommand{\href}[2]{#2}\begingroup\raggedright\endgroup

\end{document}